\documentclass{article}

\usepackage[main, final]{neurips_2026}

\usepackage[utf8]{inputenc} 
\usepackage[T1]{fontenc}    
\usepackage{hyperref}       
\usepackage{url}            
\usepackage{booktabs}       
\usepackage{amsfonts}       
\usepackage{nicefrac}       
\usepackage{microtype}      
\usepackage{xcolor}         
\usepackage{algorithm}
\usepackage{amsmath}
\usepackage{algpseudocode}
\usepackage{graphicx}
\usepackage{amsmath}
\usepackage{mathtools}
\usepackage{float}
\usepackage{multirow}
\usepackage{wrapfig}
\usepackage{amsthm}
\newtheorem{theorem}{Theorem}
\newtheorem{remark}{Remark}

\definecolor{burntorange}{rgb}{0.8, 0.33, 0.0}
\definecolor{carrotorange}{rgb}{0.93, 0.57, 0.13}
\definecolor{darkorange}{rgb}{1.0, 0.55, 0.0}
\definecolor{deepcarrotorange}{rgb}{0.91, 0.41, 0.17}
\usepackage{natbib}
\usepackage{enumitem}
\newcommand{\ours}{\textsc{ZeUS}}

\title{Efficient and Accurate Zero Shot Generation of Symmetric Protein Complexes}

\author{
    Rory Gao$^{1}$ \quad
    Yuanzhou Chen$^{1}$ \quad
    Prajit Rajkumar$^{2}$ \quad
    Wei Wang$^{1}$ \\
    $^{1}$Department of Computer Science, University of California, Los Angeles \\
    $^{2}$Skaggs School of Pharmacy and Pharmaceutical Sciences, University of California, San Diego \\
    \texttt{gaorory@cs.ucla.edu, adrianchen@cs.ucla.edu} \\
    \texttt{prajkumar@ucsd.edu, weiwang@cs.ucla.edu}
}

\begin{document}

\maketitle

\begin{abstract}
Symmetric organization is a fundamental principle of biomolecular complex assembly. The same principle underlies engineered applications, from vaccine platforms to nanomaterials, making symmetric \textit{de novo} generation a core capability for computational protein design. We establish a theoretical foundation equating generation of an assembly with underlying symmetry group $G$ to generation within a quotient space over $G$. This directly allows $\mathrm{SE}(3)$-equivariant models to generate symmetric structures with minimal distribution shift. 
With this insight we introduce \textbf{\ours{}} (\textbf{Ze}ro-shot \textbf{U}nified \textbf{S}ymmetrization), a framework that converts any pretrained $\mathrm{SE}(3)$-equivariant backbone generative model into an efficient generator of symmetric assemblies with no retraining. 
\ours{} operates on a single \emph{canonical subunit} and utilizes \emph{phantom subunits} to preserve internal model operations, reducing memory and runtime cost by $\mathcal{O}(|G|)$ for finite groups $G$, compared to a full-complex forward pass. We implement \ours{} for IPA-based models (FrameDiff, FoldFlow) and axial-attention-based RFdiffusion, showing high generation efficiency with minimal degradation to generation quality. 
We then apply \ours{} to generate symmetric complexes of over $10{,}000$ residues with exact, untruncated attention, a scale previously intractable with any model.

\end{abstract}
\section{Introduction}
\label{sec:intro}


Symmetry is a defining organizational principle of biological macromolecular assemblies. From proteasomes to the nuclear pore complex, the most functionally consequential large-scale protein architectures in nature adopt regular, symmetric quaternary structures, and over $\sim40\%$ of proteins in the Protein Data Bank~\citep{berman2000protein} exhibit some form of symmetry~\citep{duarte2022exploring}. 
Symmetry is also a naturally attractive design strategy for large protein architectures, as an assembly of \textit{k} identical subunits is encoded by a single gene and self-organizes under more favourable free-energy minimization ~\citep{zandi2004origin}. 
The same logic motivates engineered symmetric design, which underpins applications from vaccine platforms and targeted drug delivery to enzyme cascades and biomaterials. Therefore, \textit{de novo} generation of symmetric protein complexes is a core capability for computational protein design. 
However, training data for symmetric structures is sparse. For example, dimers are well represented in the PDB, but higher order assemblies, such as tetrahedral complexes, number at most a few hundred structures~\citep{rcsb_symmetry_stats}, making training and finetuning on symmetric structures difficult. 
We argue in favour of leveraging the generative capacity of existing models for the task of symmetric protein design, since modern protein backbone generators~\citep{watson2023de, ingraham2023illuminating} are already highly capable of producing diverse, designable protein structures.

In this work we introduce \textbf{\ours{}} (\textbf{Ze}ro-shot \textbf{U}nified \textbf{S}ymmetrization), a zero shot generation framework that converts any $\mathrm{SE}(3)$-equivariant backbone generative model into an efficient symmetric assembly generator with no retraining and no modification of model weights.

A central observation of \ours{} is that \emph{symmetric generation is naturally a generative problem in a quotient space over the symmetry group}. A protein assembly with $k$-fold symmetry group $G$ with $|G|=k$ is fully determined by its single representative subunit, and the full complex is the orbit of that subunit under the group action. A symmetric assembly within an ambient configuration space $\mathcal{X}$ is therefore equivalent to its subunit assembly within the quotient space $\mathcal{X}/G$, and a generative model for symmetric assemblies is precisely a generative model within this quotient. 

Our work makes the following contributions:

\begin{enumerate}
    \item \textbf{Quotient space formulation.} First, we establish that symmetric protein generation can be interpreted as generation in the quotient space $\mathcal{X}/G$. This is further realized as generation on a single canonical subunit, made aware of its copies defined by group actions. 
    \item \textbf{\ours{}, an efficient zero shot symmetrization algorithm.} We introduce \ours, which is a framework that allows any backbone diffusion or flow-matching model that satisfies SE(3)-equivariance to generate protein complexes of arbitrary finite symmetry $G$, including cyclic $(C_n)$, dihedral ($D_n$), and polyhedral $(T, O, I)$, without requiring any additional training or fine-tuning.
    \item \textbf{Empirical validation of \ours.} We implement \ours{} for two architecture classes: IPA-based models (FrameDiff, FoldFlow) and the axial-attention-based RFdiffusion. We ensure that our method retains generation quality compared to the baseline full-complex inference by evaluating generation designability (scTM, scRMSD), diversity (pairwiseTM), and novelty (pdbTM) across cyclic groups with separate distributional comparisons for polyhedral groups. \ours \space generates distributionally similar designs well within one standard deviation for all metrics compared to the baseline. 
    \item \textbf{Scaling symmetric complex generation.} We show theoretically and demonstrate empirically, a significant time and memory reduction of $\mathcal{O}(k)$ during \ours{}'s forward pass, allowing us to tractably generate assemblies that exceed 10{,}000 total residues in length, an order of magnitude inaccessible to prior methods on a single GPU. We demonstrate this with generation of $C_{50}$, $D_{50}$, $H$, and $I$ groups, all diverse and unique. 
\end{enumerate}

\section{Background and Related Work}

\paragraph{Bottom-up vs.\ top-down design.}
Approaches to symmetric protein design can be broadly characterized as
\emph{bottom-up} or \emph{top-down}. Bottom-up methods
~\citep{bale2016accurate, bethel2023precisely,divine2021designed, huddy2024blueprinting, shen2024de} design modular, reusable building blocks
that are subsequently assembled into a symmetric complex. Top-down methods
instead design the entire assembly at once, optionally conditioned on
high-level design goals. Modern diffusion- and flow-matching-based generative
models fall in this category.

\paragraph{Generative models for protein design.}
Top-down generative models can be grouped into three categories: 
\emph{Backbone-only} models generate residue frames that are then passed to an inverse-folding network such as ProteinMPNN~\citep{dauparas2022robust} to recover an amino acid sequence; representative
examples include RFdiffusion ~\citep{watson2023de}, FrameDiff~\citep{yim2023se}, FoldFlow~\citep{bose2024se3}, and FrameFlow~\citep{yim2024improved}. 
\emph{Backbone--sequence co-design} models jointly
generate structure and sequence: FoldFlow2~~\citep{huguet2024sequence}, MultiFlow~~\citep{campbell2024generative}, and ProteinGenerator~\citep{Lisanza2025}. 
Finally, \emph{All-atom} models additionally model sidechain atoms; this includes APM~\citep{chen2025all}, RFdiffusion3~\citep{butcher2025novo}, Protpardelle~\citep{chu2024all}, and Chroma~\citep{ingraham2023illuminating}. 

We restrict our study to backbone generative models, which form the largest and most mature family of SE(3)-equivariant protein generators and are the natural setting for our quotient-space construction. 
To our knowledge, no backbone or all-atom generative model has been retrained or fine-tuned specifically for symmetric generation.
\paragraph{Existing methods for symmetric backbone generation.}
Among generative models, only Chroma~\citep{ingraham2023illuminating} and RFdiffusion~\citep{watson2023de} have been applied to symmetric complex generation. 
Both treat symmetric design as full-complex generation with explicit re-symmetrization: at each denoising step, the network performs a forward pass over all residues in the assembly, and the predicted update from a designated canonical subunit is propagated to all other subunits according to the target symmetry group. This approach has two limitations. 
First, the forward pass scales with the size of the full assembly, making it prohibitively expensive in time and memory for higher order symmetries.
Second, intermediate model representations are not constrained to be symmetric. Identical subunits are processed independently within the
network, with symmetry imposed only on the coordinate update. In effect, these methods do not \emph{leverage} the symmetry of the
problem; they merely \emph{enforce} it.

\begin{figure}[t]
    \centering
    \includegraphics[width=1.0\textwidth]{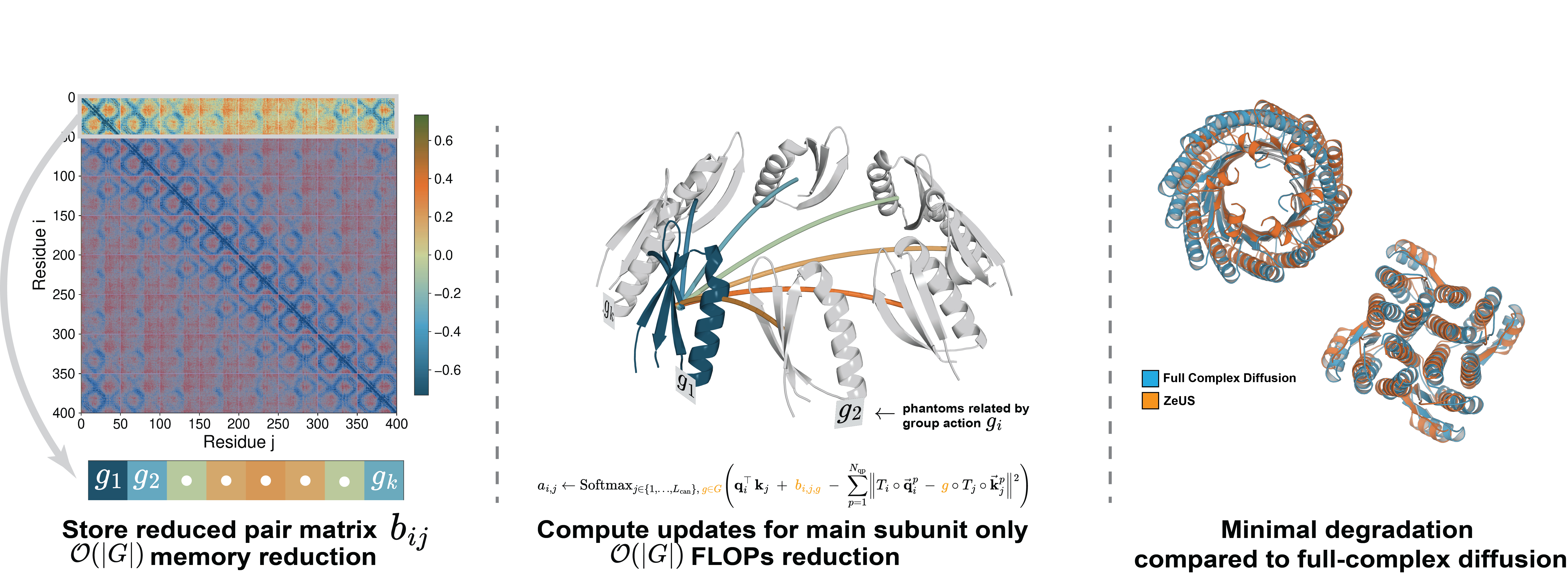} 
    \caption{Overview of \ours{}. \underline{Left:} Pairwise matrices under symmetric input are degenerate. \ours{} realizes $\mathcal{O}(k)$ memory savings by storing the reduced block row representation during model forward. \underline{Middle:} $\mathcal{O}(k)$ compute savings are realized by computing symmetry-aware attentional updates to the main/canonical subunit only. \underline{Right:} \ours{} has minimal output degradation compared to full-complex diffusion, with a mean difference in scRMSD of 0.131 $\AA$ over all evaluated models and symmetries.}
    \vspace{-10pt}
    \label{fig:overview} 
\end{figure}
\section{Theoretical Foundations}
\label{sec:theory}

This section develops the theoretical framework underlying \ours{}.
Section~\ref{sec:notation} establishes notation for backbone frames, symmetry groups, and the quotient space of symmetric assemblies.
Section~\ref{sec:equivalence} contains our core theoretical contribution, that full-complex generation is equivalent to quotient-space generation, which is realized practically through the use of \emph{phantom} copies of a canonical subunit.
Detailed proofs and extended discussions are deferred to Appendix~\ref{app:theory}.

\subsection{Notation and Preliminaries}
\label{sec:notation}

\paragraph{Backbone models and equivariance.} Following \citet{jumper2021highly, yim2023se}, an $N$-residue protein backbone is parameterized as $\mathbf{T} = [T_1,\ldots,T_N] \in \mathrm{SE}(3)^N$, where each frame $T_n = (r_n, \mathbf{x}_n)$ decomposes into a rotation $r_n \in \mathrm{SO}(3)$ and a translation $\mathbf{x}_n \in \mathbb{R}^3$ at the $\mathrm{C}_\alpha$ position of residue $n$.
The group $\mathrm{SE}(3)$ acts on frames by left multiplication: 
$T_0 \cdot T_n \coloneqq (r_0 r_n,\; r_0 \mathbf{x}_n + \mathbf{x}_0)$.
We denote $\mathrm{SE}(3)^N_0$ the centered submanifold satisfying $\sum_n \mathbf{x}_n = \mathbf{0}$, which supports diffusion processes in practice \citep{yim2023se}.

A generative model $f_\theta$ is \emph{$\mathrm{SE}(3)$-equivariant} if $f_\theta(T_0 \cdot \mathbf{T}) = T_0 \cdot f_\theta(\mathbf{T})$ for any global rigid transform $T_0 \in \mathrm{SE}(3)$, with the action extending element-wise. 
Correspondingly, all internal scalar representations, such as pairwise distances, attention logits, or node embeddings projected to invariant features, are invariant and satisfy $s(T_0 \cdot \mathbf{T}) = s(\mathbf{T})$.
This is realized by the IPA module in \citet{jumper2021highly, yim2023se, bose2024se3}, and enforced throughout RoseTTAFold in \citet{watson2023de}.

In addition, consider a multi-chain protein as the generative target of $f_\theta$: 
$\mathbf{T} = \text{concat} [ \mathbf{T}_1, \cdots, \mathbf{T}_m ]$, 
where $\mathbf{T}_1, \cdots, \mathbf{T}_m$ are individual amino acid chains. 
We say $f_\theta$ satisfies \emph{chain-permutation equivariance} if reordering the input chains leaves the output assembly unchanged up to the same reordering, 
i.e. $f_\theta(P_\pi \cdot \mathbf{T}) = P_\pi \cdot f_\theta(\mathbf{T})$ for any $\pi \in S_m$, 
where $P_\pi$ permutes the $m$ chain blocks while preserving residue order within each chain, 
i.e. $P_\pi \cdot \text{concat} [ \mathbf{T}_1, \cdots, \mathbf{T}_m ] = \text{concat} [ \mathbf{T}_{\pi(1)}, \cdots, \mathbf{T}_{\pi(m)} ]$.
This is achieved by enforcing internal node and pair representations to satisfy 
$h(P_\pi \cdot \mathbf{T}) = P_\pi \cdot h(\mathbf{T})$ and $s(P_\pi \cdot \mathbf{T}) = P_\pi \cdot s (\mathbf{T}) \cdot P_\pi^\top$, respectively.
In practice, some models break chain-permutation-equivariance and thus we discuss the effect of the non-strict equivalence  in Section~\ref{app:permutation}.

\paragraph{Symmetry groups.}
Let $G$ be a subgroup of $\mathrm{SE}(3)$ acting on $\mathbb{R}^3$. We consider two classes:
\begin{itemize} [leftmargin=*, topsep=0pt]
\item \emph{Finite point groups}: cyclic $C_n$, dihedral $D_n$, and polyhedral $T$, $O$, $I$ (with orders $12$, $24$, $60$, respectively).
\item \emph{Infinite discrete groups}: translational lattices, screw-axis groups, and their semi-direct products with point groups; generating periodic chains, planar lattices, helical filaments, and space-group structures. Experiments in this class are limited since backbone models fall highly out-of-distribution. 
\end{itemize}
We write $G = \{g_1 = \mathrm{Id}, g_2, \ldots, g_k\}$ with $|G| = k$ for finite groups, and in practice approximate infinite group orbits by a finite neighborhood as discussed in Appendix~\ref{app:diminishing}.

\vspace{-7pt}
\paragraph{Canonical subunit, phantom copies, and the quotient space.}
We refer to a specific single representative of an assembly as the \emph{canonical subunit} $\mathbf{T}^{(1)} \in \mathrm{SE}(3)^N_0$, and to its image $g_i \cdot \mathbf{T}^{(1)}$ under each non-identity group element $g_i$ as a \emph{phantom copy}.
The corresponding $G$-symmetric assembly is the orbit
\begin{equation}
  \mathbf{T}^{\mathrm{sym}} = \bigl[g_1 \cdot \mathbf{T}^{(1)},\; \ldots,\; g_k \cdot \mathbf{T}^{(1)}\bigr] \in \mathrm{SE}(3)^{kN}.
  \label{eq:orbit}
\end{equation}
For the sake of notational clarity, we mostly consider the case where $|G| = k < \infty$. 
The key observation is that a $G$-symmetric assembly is completely and uniquely determined by the group $G$ together with a single canonical subunit.
This establishes a bijection between $G$-symmetric assemblies and a single subunit existing within the \emph{quotient space} $\mathrm{SE}(3)^{kN}_0 / G$, where \emph{equivalent elements under symmetric group actions are collapsed into a single copy}. 
The next section builds upon this intuition, that generating a $G$-symmetric assembly is equivalent to sampling over $\mathrm{SE}(3)^{N}_0 / G$, a space of dimension $k$ times smaller than the ambient $\mathrm{SE}(3)^{kN}$ of full-complex generation.

\subsection{Equivalence of Generation Strategies}
\label{sec:equivalence}

Applying our quotient space view, we propose three strategies for generating symmetric structures: 
\begin{enumerate}[leftmargin=22pt, topsep=0pt]
  \item[(I)] \textbf{Full-complex generation}: a model operates on the full $kN$-residue assembly $\mathbf{T}^{\mathrm{sym}}$, updating all copies simultaneously. With augmented symmetric updates, this is the strategy taken by current methods \citep{watson2023de,ingraham2023illuminating}.
  \item[(II)] \textbf{Quotient-space generation}: a model operates directly on $\mathrm{SE}(3)^N_0 / G$, taking equivalence classes as inputs and outputs. This is the theoretical ideal as it removes all symmetry redundancy by construction and assumes parameterized access to the quotient manifold.
  \item[(III)] \textbf{Phantom-copy generation}: a model operates on a single canonical representative $\mathbf{T}^{(1)} \in \mathrm{SE}(3)^N_0$, while consulting its phantom copies $\{g_i \cdot \mathbf{T}^{(1)}\}_{i=2}^k$ as auxiliary geometric context. This is the \ours{} approach.
\end{enumerate}

The two relationships of interest are the descent (I)$\Rightarrow$(II), where a full-assembly model induces a well-defined operator on the quotient space, and the expansion (II)$\Rightarrow$(III), where this quotient-space operator is constructed concretely through phantom copies.
We now discuss both relationships.

\vspace{-7pt}
\paragraph{Descent from full-complex to quotient-space.}
$\mathrm{SE}(3)$-equivariance ensures $G$-symmetric inputs yield $G$-symmetric outputs, and the action of $f_\theta$ on a $G$-symmetric input is fully determined by what it does to one canonical subunit.
This in turn means that $f_\theta$ descends to a well-defined map on equivalence classes: 

\begin{theorem}[Descent of full-complex models to quotient space]
\label{thm:descent}
Let $f_\theta : \mathrm{SE}(3)^{kN} \to \mathrm{SE}(3)^{kN}$ be $\mathrm{SE}(3)$-equivariant and chain-permutation-equivariant.
Then there exists a unique map
\[
  f_\theta^{/G} : \mathrm{SE}(3)^N_0 / G \to \mathrm{SE}(3)^N_0 / G
\]
such that, for every canonical subunit $\mathbf{T}^{(1)}$ with assembly $\mathbf{T}^{\mathrm{sym}} = [g_1 \cdot \mathbf{T}^{(1)}, \ldots, g_k \cdot \mathbf{T}^{(1)}]$ and every $i \in [k]$,
\begin{equation}
  f_\theta(\mathbf{T}^{\mathrm{sym}}) \big|_{(i,n)} \;=\; g_i \cdot f_\theta^{/G}\bigl([\mathbf{T}^{(1)}]\bigr) \big|_n.
  \label{eq:descent}
\end{equation}
\end{theorem}

\paragraph{What this quotient space model looks like.}
Theorem~\ref{thm:descent} guarantees that an idealized quotient-space model $f_\theta^{/G}$ exists.
We now describe these operations directly, working component-by-component in quotient space.
First, operations in ambient space can act on residues (node features), on pairs of residues (edge features, attention matrices), and more generally on $r$-tuples of residues. 
In the quotient space, however, each node is an equivalence class with $k$ ambient representatives, so an operation that consumes a pair of quotient-space nodes, for example, corresponds to $k$ distinct types of pairs in ambient space (one per relative group element $g \in G$; see Figure~\ref{fig:overview}) and must produce $k$ outputs to capture all of them. 
Every primitive of $f_\theta$ admits such a quotient-space counterpart, and $f_\theta^{/G}$ is the composition of these counterparts. 
The precise statement of the operation typology and its completeness across $\mathrm{SE}(3)$-equivariant architectures is given in Appendix~\ref{app:proof_descent_expansion}.

\vspace{-7pt}
\paragraph{Expansion from quotient-space to ambient phantom-copy generation.}
The quotient-space description above operates on equivalence classes, which are hard to parameterize for existing architectures.
Expansion recovers a concrete computation by translating each quotient-space operation back into the ambient space. 
For example, an edge in quotient space can be represented by $k$ pair of residues, between a canonical representative in $\mathbf{T}^{(1)}$ and a phantom copy in $g \cdot \mathbf{T}^{(1)}$, where $g$ cycles through $G$. 
This expansion preserves the architecture of $f_\theta$ while focusing memory and compute on the canonical representative. 
We formalize the construction as an operator $f_\theta^{\mathrm{ph}}$ and state the resulting equivalence as the following theorem.

\begin{theorem}[Phantom-copy realization]
\label{thm:expansion}
The phantom-copy operator $f_\theta^{\mathrm{ph}} : \mathrm{SE}(3)^N_0 \to \mathrm{SE}(3)^N_0$, constructed by per-operation ambient-space expansion of $f_\theta^{/G}$ as described above, satisfies
\begin{equation}
  f_\theta^{/G}\bigl([\mathbf{T}^{(1)}]\bigr) \;=\; \bigl[\, f_\theta^{\mathrm{ph}}(\mathbf{T}^{(1)}) \,\bigr]
  \label{eq:expansion}
\end{equation}
for every representative $\mathbf{T}^{(1)} \in \mathrm{SE}(3)^N_0$.
\end{theorem}

\begin{remark}
Beyond edge features, each primitive component of $f_\theta^{/G}$ has an ambient-space expansion centered around the canonical chain: a triplet operation becomes $k^2$ ambient triplets, and aggregations sum these contributions back via group-indexed \emph{\texttt{einsum}}'s. 
The composition of these expansions is exactly $f_\theta^{\mathrm{ph}}$.
A precise statement of the operation typology and an inductive proof for tuples are given in Appendix~\ref{app:proof_descent_expansion}, while empirical deployments are presented in Section~\ref{sec:methods}.
\end{remark}

Together, Theorem~\ref{thm:descent} and Theorem~\ref{thm:expansion} give the practical equivalence underlying \ours{}, reducing full-complex generation $f_\theta(\mathbf{T}^{\mathrm{sym}})$ to phantom-copy generation $f_\theta^{\mathrm{ph}}(\mathbf{T}^{(1)})|_n$. 

\vspace{-7pt}
\paragraph{Local geometry preservation under quotient.}
A central implication of Theorem~\ref{thm:descent} is that a model trained in $\mathrm{SE}(3)^N_0$ can be applied, zero shot, to generate $G$-symmetric assemblies.
The quotient space $\mathrm{SE}(3) / G$, where each residue of the projection of a $G$-symmetric assembly lives, is a manifold whose local geometry is (a.e.) identical to that of the ambient $\mathrm{SE}(3)$.
The distribution shift therefore lies only in global geometry, and any short-range structural priors the model has internalized transfer without loss.

\vspace{-7pt}

\section{Methods}
\label{sec:methods}




We translate Theorems~\ref{thm:descent} and~\ref{thm:expansion} into concrete computational savings, guided by the central observation that \emph{every internal representation of an $\mathrm{SE}(3)$-equivariant model, evaluated on a $G$-symmetric assembly, is equivariant under the $G$-action, and is therefore duplicated $k$ times in the full-complex forward pass}. Recovering this redundancy yields a uniform $\mathcal{O}(k)$ reduction in both memory and compute, applied across every layer and every operation type as they now act on the canonical subunit alone. 

\subsection{Memory Savings: Reduction of Phantom Copies}
\label{sec:reductions}

\begin{itemize}[leftmargin=*, topsep=2pt, itemsep=2pt]
  \item \textbf{Node features} reduce from $kN$ to $N$: scalar (invariant) components of the phantom copies are identical to those of the canonical subunit, and equivariant (frame-valued) components are recovered as the group-transformed canonical embeddings under $g_i$.
  \item \textbf{Edge features} reduce from $k^2 N^2$ to $k N^2$: by the same principle, only the $k$ distinct relative-group blocks incoming to the canonical subunit need be maintained. Phantom-to-phantom edges are never required, since each admits a counterpart involving the canonical chain via $g_i$. 
\end{itemize}

\subsection{Compute Savings: Symmetry-Guided Expansion of Canonical Chain}

The runtime counterpart to Section~\ref{sec:reductions} is to compute updates only for the canonical subunit, which by Theorems~\ref{thm:descent} and~\ref{thm:expansion} is exact provided all attention-like operations remain \emph{phantom-aware}. We describe the most salient modifications required for each of the two architecture classes targeted by \ours{}. Full details of additional modifications to architectures can be found in Appendix~\ref{app:algorithms}.

\paragraph{IPA architectures (FrameDiff, FoldFlow).}
We modify the IPA layer to efficiently pass messages from phantom residues into the canonical subunit. 
To compute the attention softmax over all $k$ group transforms of the $kN$ residues, we only need queries from the canonical subunit, while keys need to come from all copies. 
Invariant scalar keys are tied across copies, but point-valued keys are equivariant and must be transformed by the group action. Dropping constants for brevity, the attention logit becomes
\begin{equation}
a_{i,j} \leftarrow \mathrm{Softmax}_{j \in \{1,\dots,kN\},\ {g \in G}} \left(
   \mathbf{q}_i^{\top}\,\mathbf{k}_j
   \;+\; {b_{i,j}}
   \;-\; \textstyle\sum_{p}
      \bigl\| T_i \circ \vec{\mathbf{q}}_i^{p} \,-\, {g} \circ T_j \circ \vec{\mathbf{k}}_j^{p} \bigr\|^2
   \right),
\end{equation}
where $i$ ranges over the residues in the canonical subunit and $j$ ranges over all residues in the complex.

\paragraph{Axial-attention architectures (RFdiffusion).}
Pair-track axial attention~\citep{watson2023de} is the most memory-intensive block in the RoseTTAFold / RFdiffusion architecture. 
Standard Axial Attention initiates its sequences on the full $(kN) \times (kN)$ pair-wise tensor and thus requires all $k^2 N^2$ elements. 
As discussed in Section~\ref{subsec:pair_block}, however, the canonical block-row contains all distinct pair interactions, as every phantom-centered block is a block-column permutation. 
We replace axial attention with an implementation that consumes only the canonical block-row and computes the full pair update via group-indexed shuffles, never materializing the $(kN) \times (kN)$ tensor:
\begin{equation}
a_{i,j} = \mathrm{Softmax}_{j}\left(
{\textstyle \sum_{g \in G}}\, 
\left[
(\pi_g (\mathbf{Q}))_{i}^{\top}\, 
(\pi_g (\mathbf{K}))_{1+ ((j-1) \ (modN))}
\right]
+\; b_{i,j}
\right),
\end{equation}
where $i$ ranges over the residues in the canonical subunit and $j$ ranges over all residues in the complex. $\mathbf{Q}$ and $\mathbf{K}$ are the block-row pair query and key matrices. $\pi_g (\mathbf{X})$ denotes the column permutation operation of a block-row $\mathbf{X}$ by the group permutation matrix determined through $g$. 
The output is exactly equivalent to the full axial update, and computation never exceeds $\mathcal{O}(k N^2)$ pair memory.

\subsection{Block Structure of Pair-wise Interaction Matrices}
\label{subsec:pair_block}
Here we expand upon the discussion on axial-attention by studying pair-wise feature matrices. 
Quotient to phantom equivalence implies a specific exploitable structure in any pair-wise interaction matrix.
Indexing residues as $(i,n)$ for chain-$i \in [k]$ and position $n \in [N]$, we arrange a $kN \times kN$ interaction matrix $A$ in $N \times N$ blocks $A^{[ij]}$ collecting entries from chain-$j$'s residues to chain-$i$'s.

\begin{theorem}[Block Structure of Interaction Matrices]
\label{thm:attention_block}Let $A$ be a pair-wise interaction matrix, featured within an $\mathrm{SE}(3)$-equivariant model evaluated on a $G$-symmetric assembly, whose entries are all $\mathrm{SE}(3)$-invariant pairwise features. Then for any $i, j \in [k]$,
$A^{[ij]} \;=\; A^{[1,\, \mathrm{idx}(g_i^{-1} g_j)]}$, 
where $\mathrm{idx}: G \to [k]$ is the inverse of group element indexing. 
\end{theorem}

By the block-permutation theorem, the block-row $i$ associated with
$g \in G$ satisfies $A^{[i, :]} = \pi_g(A^{[1,:]})$.

A proof is given in Appendix~\ref{app:proof_attention}. 
Theorem~\ref{thm:attention_block} implies a block-wise permutation pattern for pair-wise interaction matrices specific to the symmetry group $G$, where each row of blocks is a permutation of the first row. 
For abelian groups, consecutive rows are cyclic shifts of the first row; for non-abelian groups, the permutation is determined by the group multiplication table.
A visualization is in Figure~\ref{fig:overview}, and more extensive visualizations are in Appendix~\ref{app:block}. 
\section{Experiments}
\label{sec:experiments}

\subsection{Symmetric Generation Quality}

By Theorem~\ref{thm:descent}, \ours{} produces samples equivalent to full-complex symmetric diffusion under pure $\mathrm{SE}(3)$-invariant attention and chain-permutation equivariance. We now empirically characterize the \emph{equivalence} of generation quality across architectures, symmetry groups, and design metrics and the \emph{efficiency} of the observed runtime and memory savings. 

\paragraph{Setup.}
We benchmark three backbone models across two architectural classes: IPA-based FrameDiff and FoldFlow, and axial attention based RFdiffusion. For each model 
we unconditionally generate $200$ designs at five cyclic symmetries ($C_2, C_3, C_4, C_6, C_8$), and $100$ designs at $C_{12}$ with subunit length $N = 50$, yielding total complex sizes from 
$100$ to $600$ residues. 
In practice, model design choices can violate chain-permutation equivariance. FrameDiff and FoldFlow were trained on single chain input. We implement symmetric multi chain inference that respects this invariance. RFdiffusion is not chain-permutation equivariant and thus we implement a full-complex diffusion ablation that enforces chain-permutation equivariance during inference, but without any \ours{} reductions. We compare \ours{} with both the default RFdiffusion full-complex inference and this ablated inference to distinguish the effect of symmetry reduction from that of modifying the pretrained model. In addition, we discuss the theoretical implications in Appendix ~\ref{app:permutation}

Each configuration is run with full-complex diffusion and with ZeUS, and an additional chain-permutation ablation for RFdiffusion. All experiments were run on a single NVIDIA A100-40GB GPU. 

\paragraph{Metrics.} We report three classes of generation-quality metrics, following prior work.

\textbf{Designability.} For each backbone design we predict 8 sequences with ProteinMPNN~\citep{dauparas2022robust}, refold each with ESMFold~\citep{lin2023evolutionary}, and measure the self-consistency RMSD (scRMSD) and TM-score (scTM) against the original backbone. A backbone is \emph{designable} if its best scRMSD falls below $2\,\text{\AA}$ or its best scTM exceeds $0.5$. We report the fraction of designable structures together with the mean best scTM and scRMSD over all refolded designs. 
FrameDiff and FoldFlow are not trained on multi-chain structures, so their baseline designability rates on symmetric assemblies are modest. The mean scTM and scRMSD complement the fraction designable by enabling a more sensitive comparison.

\textbf{Novelty}. We measure novelty as the maximum TM-score against the PDB for designable complexes only, retrieved via Foldseek~\citep{van2024fast}. A lower TM-score means lower structural similarity to the PDB and is thus desirable. We report it at both the subunit and full-complex levels, since these capture independent aspects of complex novelty.

\textbf{Diversity}. Following FoldFlow ~\citep{bose2024se3}, we report the mean pairwise TM-score among designable complexes only, computed at the full-complex level.

\textbf{Efficiency.} We report peak GPU memory during a denoising step (\texttt{torch.cuda. max\_memory\_allocated}) and average wall-clock time per denoising step.
\begin{table*}[h!]
\centering
\small
\setlength{\tabcolsep}{6pt}
\caption{
RFdiffusion Designability (frac. designable, scRMSD) quality across cyclic symmetry groups.
Full, Ablation, and ZeUS denote full-complex inference, the chain-invariance
ablation, and \ours{} inference, respectively.
The $\Delta$ columns report the difference
$\Delta=\mathrm{ZeUS}-\mathrm{Ablation}$.
}
\label{tab:rfdiffusion_design_quality}
\begin{tabular}{l cccc cccc}
\toprule
& \multicolumn{4}{c}{Designability (\%) $\uparrow$}
& \multicolumn{4}{c}{mean best scRMSD (\AA) $\downarrow$} \\
\cmidrule(lr){2-5} \cmidrule(lr){6-9}
Symmetry
& full & ablation & ZeUS & {\boldmath$\Delta$} \textbf{(pp)}
& full & ablation & ZeUS & {\boldmath$\Delta$} \textbf{(\AA)} \\
\midrule
$C_{2}$ & 97.0 & 96.5 & 95.5 & {\boldmath$-1.0$} & 1.04 & 0.90 & 0.89 & {\boldmath$-0.01$} \\
$C_{3}$ & 82.0 & 76.0 & 76.0 & {\boldmath$0.0$} & 1.69 & 1.94 & 1.76 & {\boldmath$-0.18$} \\
$C_{4}$ & 67.0 & 61.5 & 54.5 & {\boldmath$-7.0$} & 2.43 & 2.39 & 2.72 & {\boldmath$+0.33$} \\
$C_{6}$ & 57.0 & 51.0 & 47.0 & {\boldmath$-4.0$} & 3.53 & 3.83 & 3.98 & {\boldmath$+0.16$} \\
$C_{8}$ & 48.5 & 41.0 & 46.5 & {\boldmath$+5.5$} & 4.70 & 5.04 & 5.10 & {\boldmath$+0.06$} \\
$C_{12}$ & 51.0 & 36.0 & 40.0 & {\boldmath$+4.0$} & 5.68 & 6.04 & 5.91 & {\boldmath$-0.12$} \\
\bottomrule
\end{tabular}
\end{table*}

\begin{table*}[t]
\centering
\small
\setlength{\tabcolsep}{6pt}
\caption{
Complex-level diversity and novelty of designable structures.
Lower values indicate greater diversity and novelty, respectively.
The $\Delta$ columns report the difference
$\Delta=\mathrm{ZeUS}-\mathrm{Full}$.
}
\label{tab:complex_diversity_novelty_deltas}
\begin{tabular}{ll ccc ccc}
\toprule
& & \multicolumn{3}{c}{diversityTM $\downarrow$}
& \multicolumn{3}{c}{pdbTM (novelty) $\downarrow$} \\
\cmidrule(lr){3-5} \cmidrule(lr){6-8}
Model & Symmetry & full & ZeUS & {\boldmath$\Delta$} & full & ZeUS & {\boldmath$\Delta$} \\
\midrule
\multirow{6}{*}{FrameDiff} & $C_{2}$ & 0.223 & 0.231 & {\boldmath$+0.008$} & 0.157 & 0.132 & {\boldmath$-0.025$} \\
 & $C_{3}$ & 0.226 & 0.202 & {\boldmath$-0.024$} & 0.097 & 0.078 & {\boldmath$-0.019$} \\
 & $C_{4}$ & 0.237 & 0.232 & {\boldmath$-0.005$} & 0.103 & 0.068 & {\boldmath$-0.035$} \\
 & $C_{6}$ & 0.260 & 0.286 & {\boldmath$+0.026$} & 0.059 & 0.044 & {\boldmath$-0.015$} \\
 & $C_{8}$ & 0.299 & 0.292 & {\boldmath$-0.007$} & 0.047 & 0.026 & {\boldmath$-0.022$} \\
 & $C_{12}$ & 0.374 & -- & \textbf{--} & 0.033 & 0.020 & {\boldmath$-0.013$} \\
\midrule
\multirow{6}{*}{FoldFlow} & $C_{2}$ & 0.249 & 0.247 & {\boldmath$-0.001$} & 0.126 & 0.130 & {\boldmath$+0.003$} \\
 & $C_{3}$ & 0.214 & 0.215 & {\boldmath$+0.001$} & 0.069 & 0.075 & {\boldmath$+0.005$} \\
 & $C_{4}$ & 0.219 & 0.220 & {\boldmath$+0.001$} & 0.058 & 0.046 & {\boldmath$-0.012$} \\
 & $C_{6}$ & 0.263 & 0.263 & {\boldmath$-0.001$} & 0.054 & 0.040 & {\boldmath$-0.015$} \\
 & $C_{8}$ & 0.320 & 0.283 & {\boldmath$-0.037$} & 0.021 & 0.035 & {\boldmath$+0.014$} \\
 & $C_{12}$ & 0.395 & 0.368 & {\boldmath$-0.027$} & 0.017 & 0.018 & {\boldmath$+0.001$} \\
\midrule
\multirow{6}{*}{RFdiffusion} & $C_{2}$ & 0.252 & 0.260 & {\boldmath$+0.008$} & 0.037 & 0.038 & {\boldmath$+0.001$} \\
 & $C_{3}$ & 0.229 & 0.225 & {\boldmath$-0.004$} & 0.034 & 0.031 & {\boldmath$-0.002$} \\
 & $C_{4}$ & 0.255 & 0.243 & {\boldmath$-0.011$} & 0.024 & 0.026 & {\boldmath$+0.002$} \\
 & $C_{6}$ & 0.259 & 0.244 & {\boldmath$-0.015$} & 0.017 & 0.017 & {\boldmath$0.000$} \\
 & $C_{8}$ & 0.271 & 0.268 & {\boldmath$-0.003$} & 0.013 & 0.012 & {\boldmath$-0.001$} \\
 & $C_{12}$ & 0.304 & 0.306 & {\boldmath$+0.001$} & 0.010 & 0.006 & {\boldmath$-0.004$} \\
\bottomrule
\end{tabular}
\end{table*}

\begin{figure}[h!]
    \centering
    \includegraphics[width=1.0\textwidth]{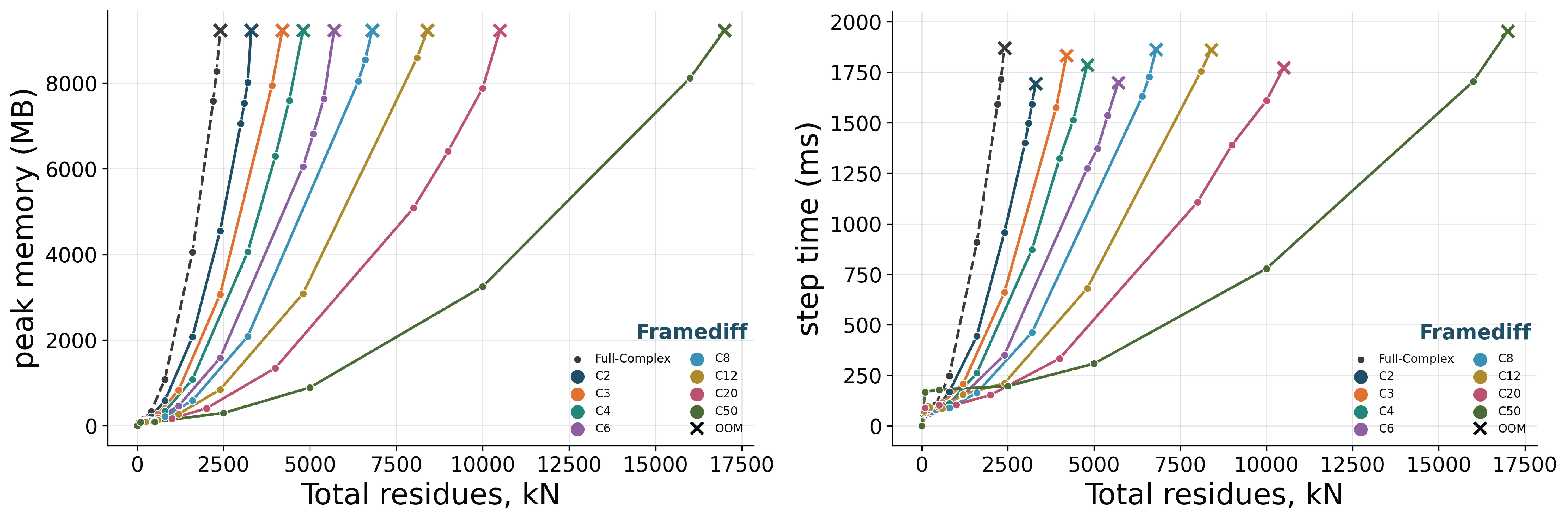} 
    \caption{Left: GPU VRAM usage. Right: forward pass wall-clock time for the full-complex baseline compared to \ours{}, at varying cyclic group sizes. Models are run until out-of-memory, with the largest $N_{\text{res}}$ that fits in memory marked.}
    \label{fig:time_memory}
\end{figure}

\paragraph{Equivalence across architectures and symmetries.}
Table~\ref{tab:rfdiffusion_design_quality} shows that \ours{} achieves equivalent design quality to the chain-permutation equivariant RFdiffusion ablation, with an mean signed scRMSD difference of only $0.04\,\text{\AA}$. While there is some degradation in both designability and scRMSD when moving from full-complex diffusion to the ablation, this degradation is attributable to the ablation itself rather than \ours{}. Our results show that \ours{} preserves the design quality of the corresponding ablated model.

Similarly, Table~\ref{tab:complex_diversity_novelty_deltas} shows that diversity and novelty remain largely unchanged between \ours{} and full-complex diffusion across all evaluated models, with mean signed differences of $-0.0053$ in diversityTM and $-0.0076$ in pdbTM, respectively. Together, these results are consistent with the equivalence established in Theorem~\ref{thm:descent}, across different architectures and symmetry groups.

Table~\ref{tab:all_combined} in Appendix~\ref{app:experimental} provides comprehensive results across all models and symmetries. Designability, diversity, and novelty agree within $1\sigma$ between both methods across all six cyclic symmetries.

\paragraph{Memory and runtime scale linearly in $k$.}
Figure~\ref{fig:time_memory} shows that for a fixed memory budget, the total number of residues \ours{} can fit in memory grows as $\sqrt{k}$, relative to the baseline. At $C_{50}$, this enables generation of complexes up to $\sim\!16{,}000$ residues, compared to $\sim\!2{,}400$ under the full-complex baseline. The runtime curve follows the same pattern, with the per-step advantage of \ours{} growing with $k$. Conversely this shows the $\mathcal{O}(k)$ linear reduction predicted in Section~\ref{sec:reductions} when $N$ is kept fixed.  

\paragraph{Polyhedral symmetries.}
We do not report designability metrics for polyhedral generations as self-consistency-based designability is currently unreliable for polyhedral assemblies. Current established folding models struggle to correctly predict the structure of large symmetric complexes~\citep{zhu2023evaluation}. 
This implies that we cannot determine whether a given design under $T$, $O$, or $I$ symmetry is genuinely undesignable, or whether the folding model itself is the bottleneck. 
Nevertheless we report distributional comparisons of polyhedral generations with the baseline full-complex and \ours{}, available in Appendix~\ref{app:polyhedral}. 
Finally, \ours{} results remain distributionally similar for additional metrics such as steric clashes, subunit-level novelty, and secondary structure composition, available in Appendix~\ref{app:additionalmetrics}.

\subsection{Large Scale Design}

\begin{wrapfigure}{r}{0.4\textwidth}
    \vspace{-10pt}
    \centering
    \includegraphics[width=0.4\textwidth]{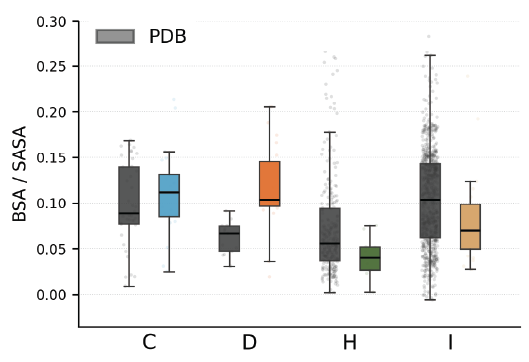} 
    \caption{Normalized buried surface area of \ours{} generations compared to PDB proteins matched by symmetry.}
    \vspace{-15pt}
    \label{fig:bsa}
\end{wrapfigure}

Generation of large-scale biomolecular complexes (>$\sim$1000 residues) is an active frontier of de novo design \citep{herzik2019high, bryant2022predicting}. At this size, existing generative models exhibit substantially higher failure rates and are unable to calculate full attention on all residues on a single GPU ~\citep{lin2026fold}. \ours{}'s quotient space reformulation makes this regime tractable to compute exact, untruncated attention over the full symmetric complex. 
To showcase this, we use \ours{} on FrameDiff to generate large $C_{50}$, $D_{50}$, $I$ (Icosahedral), and $H$ (Helical) complexes, at sizes of 10{,}000, 10{,}000, 6000, and 4500 total residues, respectively.
\begin{figure}[t]
    \centering
    \includegraphics[width=1.0\textwidth]{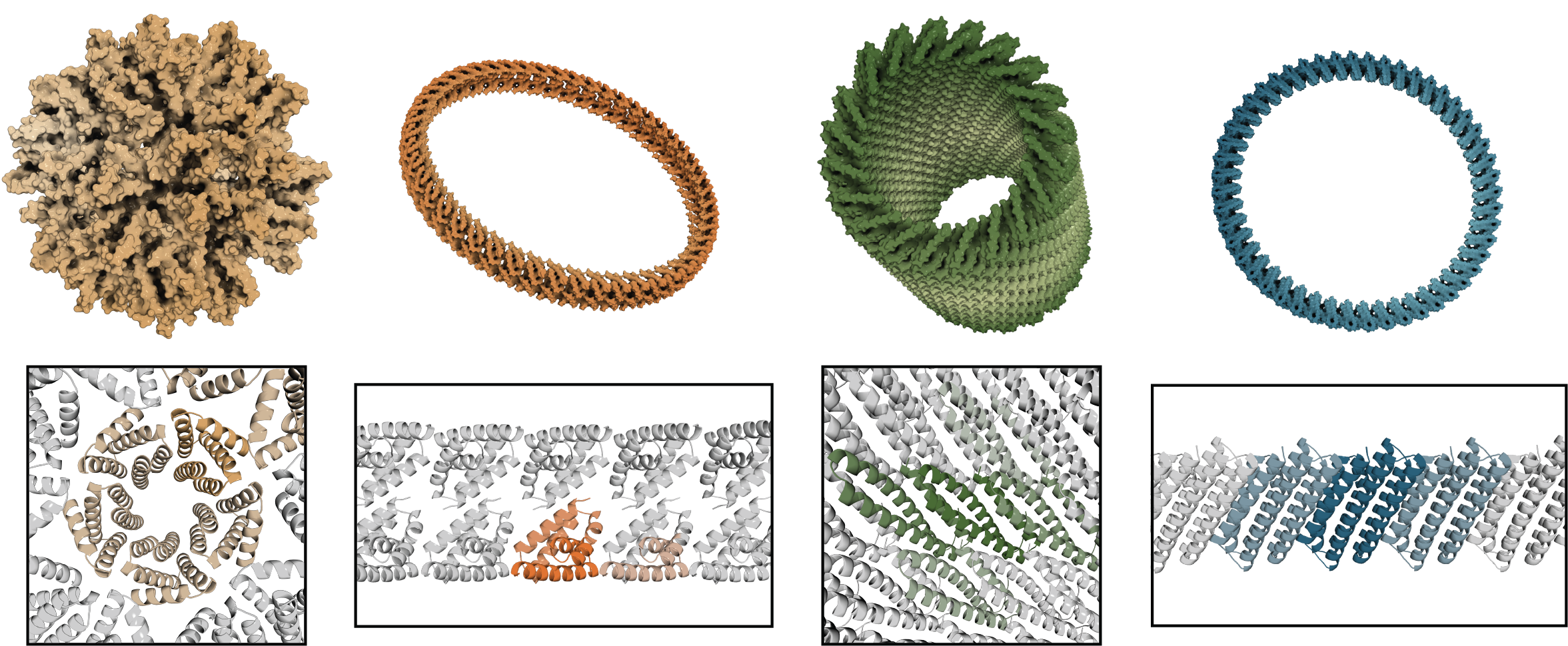} 
    \caption{Generated $I$, $D_{50}$, $H$, $C_{50}$,  designs with $N_{res}=6000, 10000, 4500, 10000$ respectively, displayed along with their interchain contacts. Contacts respect the imposed symmetry in each case.}
    \vspace{-10pt}
    \label{fig:large_scale} 
\end{figure}

The generated designs are structurally coherent, with interchain interfaces that respect the imposed symmetry and characteristic motifs emerging per symmetry class. Helical assemblies are dominated by $\alpha$-helices aligned along the tilt axis, mirroring both natural helical proteins and prior \textit{de novo} helical designs~\citep{shen2018novo}, while dihedral interfaces respect both the cyclic and stacking components of $D_{50}$.

To probe the global plausibility of our generated complexes, we calculate the buried surface area of each subunit protein normalized by the unbound solvent-accessible surface area as a proxy for interchain contact. We find that the property distributions of our large-scale designs closely overlap with natural PDB proteins in each symmetry class. In addition to being globally plausible, the designs themselves remain geometrically distinct from any PDB structure as measured by radius and present reasonable alternative large-scale geometries (closest PDB matches reported in Appendix~\ref{app:closest_pdb}). We note that these are backbone-only structures, and assessing designability is intractable for any such proposed structure at this scale ~\citep{bryant2022predicting}. Our contribution here is demonstrating that existing backbone generators can produce coherent symmetric complexes at a previously unattainable scale; improving validity in this regime is left to future work.

\section{Conclusions, Limitations, and Future Work}
\label{sec:conclusion}

We introduced \ours{}, a method that converts any $\mathrm{SE}(3)$-equivariant and chain-permutation-equivariant backbone generative model into a symmetric assembly generator with lossless $\mathcal{O}(k)$ reductions in memory and runtime. The central insight is to formulate symmetric backbone design as a generative problem within a quotient space. We implement this insight for the two dominant backbone architecture classes, IPA-based models (FrameDiff, FoldFlow) and the axial-attention-based RFdiffusion. Empirically, \ours{} matches the designability, diversity, and novelty of full-complex symmetric inference across all settings tested while achieving the predicted $\mathcal{O}(k)$ savings. The same reduction unlocks generation of symmetric complexes an order of magnitude larger than feasible under any explicit-symmetrization method, without truncation or approximation. We acknowledge that the equivalence of transfer is not as strong when the base model is not fully compliant, nevertheless we anticipate that fine-tuning on symmetric designs and tightening the $\mathrm{SE}(3)$-equivariance and chain-permutation equivariance of existing architectures will be natural next steps toward even more accurate and scalable symmetric complex design.



\begin{ack}

This work was partially supported by the National Institutes
of Health (OT2OD038003, U54OD036472, U54DK097771), the National
Science Foundation (2312501, 2106859), and Amazon, NEC, and the Easton Center. 
\end{ack}

\newpage
\clearpage
{
\small
\bibliographystyle{plainnat}
\bibliography{citation}

@Article{watson2023de,
author={Watson, Joseph L.
and Juergens, David
and Bennett, Nathaniel R.
and Trippe, Brian L.
and Yim, Jason
and Eisenach, Helen E.
and Ahern, Woody
and Borst, Andrew J.
and Ragotte, Robert J.
and Milles, Lukas F.
and Wicky, Basile I. M.
and Hanikel, Nikita
and Pellock, Samuel J.
and Courbet, Alexis
and Sheffler, William
and Wang, Jue
and Venkatesh, Preetham
and Sappington, Isaac
and Torres, Susana V{\'a}zquez
and Lauko, Anna
and De Bortoli, Valentin
and Mathieu, Emile
and Ovchinnikov, Sergey
and Barzilay, Regina
and Jaakkola, Tommi S.
and DiMaio, Frank
and Baek, Minkyung
and Baker, David},
title={De novo design of protein structure and function with RFdiffusion},
journal={Nature},
year={2023},
month={Aug},
day={01},
volume={620},
number={7976},
pages={1089-1100},
issn={1476-4687},
doi={10.1038/s41586-023-06415-8},
url={https://doi.org/10.1038/s41586-023-06415-8}
}

@inproceedings{yim2023se,
author = {Yim, Jason and Trippe, Brian L. and De Bortoli, Valentin and Mathieu, Emile and Doucet, Arnaud and Barzilay, Regina and Jaakkola, Tommi},
title = {SE(3) diffusion model with application to protein backbone generation},
year = {2023},
publisher = {JMLR.org},
booktitle = {Proceedings of the 40th International Conference on Machine Learning},
articleno = {1672},
numpages = {39},
location = {Honolulu, Hawaii, USA},
series = {ICML'23}
}

@inproceedings{
bose2024se3,
title={{SE}(3)-Stochastic Flow Matching for Protein Backbone Generation},
author={Avishek Joey Bose and Tara Akhound-Sadegh and Guillaume Huguet and Kilian Fatras and Jarrid Rector-Brooks and Cheng-Hao Liu and Andrei Cristian Nica and Maksym Korablyov and Michael M. Bronstein and Alexander Tong},
booktitle={The Twelfth International Conference on Learning Representations},
year={2024},
url={https://openreview.net/forum?id=kJFIH23hXb}
}

@article{
dauparas2022robust,
author = {J. Dauparas  and I. Anishchenko  and N. Bennett  and H. Bai  and R. J. Ragotte  and L. F. Milles  and B. I. M. Wicky  and A. Courbet  and R. J. de Haas  and N. Bethel  and P. J. Y. Leung  and T. F. Huddy  and S. Pellock  and D. Tischer  and F. Chan  and B. Koepnick  and H. Nguyen  and A. Kang  and B. Sankaran  and A. K. Bera  and N. P. King  and D. Baker },
title = {Robust deep learning–based protein sequence design using ProteinMPNN},
journal = {Science},
volume = {378},
number = {6615},
pages = {49-56},
year = {2022},
doi = {10.1126/science.add2187},
URL = {https://www.science.org/doi/abs/10.1126/science.add2187},
eprint = {https://www.science.org/doi/pdf/10.1126/science.add2187},
}

@article{van2024fast,
  title={Fast and accurate protein structure search with Foldseek},
  author={Van Kempen, Michel and Kim, Stephanie S and Tumescheit, Charlotte and Mirdita, Milot and Lee, Jeongjae and Gilchrist, Cameron LM and S{\"o}ding, Johannes and Steinegger, Martin},
  journal={Nature biotechnology},
  volume={42},
  number={2},
  pages={243--246},
  year={2024},
  publisher={Nature Publishing Group US New York}
}

@article{lin2023evolutionary,
  title={Evolutionary-scale prediction of atomic-level protein structure with a language model},
  author={Zeming Lin and Halil Akin and Roshan Rao and Brian Hie and Zhongkai Zhu and Wenting Lu and Nikita Smetanin and Robert Verkuil and Ori Kabeli and Yaniv Shmueli and Allan dos Santos Costa and Maryam Fazel-Zarandi and Tom Sercu and Salvatore Candido and Alexander Rives},
  journal={Science},
  volume={379},
  number={6637},
  pages={1123--1130},
  year={2023},
  publisher={American Association for the Advancement of Science}
}

@Article{jumper2021highly,
author={Jumper, John
and Evans, Richard
and Pritzel, Alexander
and Green, Tim
and Figurnov, Michael
and Ronneberger, Olaf
and Tunyasuvunakool, Kathryn
and Bates, Russ
and {\v{Z}}{\'i}dek, Augustin
and Potapenko, Anna
and Bridgland, Alex
and Meyer, Clemens
and Kohl, Simon A. A.
and Ballard, Andrew J.
and Cowie, Andrew
and Romera-Paredes, Bernardino
and Nikolov, Stanislav
and Jain, Rishub
and Adler, Jonas
and Back, Trevor
and Petersen, Stig
and Reiman, David
and Clancy, Ellen
and Zielinski, Michal
and Steinegger, Martin
and Pacholska, Michalina
and Berghammer, Tamas
and Bodenstein, Sebastian
and Silver, David
and Vinyals, Oriol
and Senior, Andrew W.
and Kavukcuoglu, Koray
and Kohli, Pushmeet
and Hassabis, Demis},
title={Highly accurate protein structure prediction with AlphaFold},
journal={Nature},
year={2021},
month={Aug},
day={01},
volume={596},
number={7873},
pages={583-589},
issn={1476-4687},
doi={10.1038/s41586-021-03819-2},
url={https://doi.org/10.1038/s41586-021-03819-2}
}

@article{
bale2016accurate,
author = {Jacob B. Bale  and Shane Gonen  and Yuxi Liu  and William Sheffler  and Daniel Ellis  and Chantz Thomas  and Duilio Cascio  and Todd O. Yeates  and Tamir Gonen  and Neil P. King  and David Baker },
title = {Accurate design of megadalton-scale two-component icosahedral protein complexes},
journal = {Science},
volume = {353},
number = {6297},
pages = {389-394},
year = {2016},
doi = {10.1126/science.aaf8818},
URL = {https://www.science.org/doi/abs/10.1126/science.aaf8818},
eprint = {https://www.science.org/doi/pdf/10.1126/science.aaf8818},
}

@article{zandi2004origin,
  title={Origin of icosahedral symmetry in viruses},
  author={Zandi, Roya and Reguera, David and Bruinsma, Robijn F and Gelbart, William M and Rudnick, Joseph},
  journal={Proceedings of the National Academy of Sciences},
  volume={101},
  number={44},
  pages={15556--15560},
  year={2004},
  publisher={National Academy of Sciences}
}

@article{bryant2022predicting,
  title={Predicting the structure of large protein complexes using AlphaFold and Monte Carlo tree search},
  author={Bryant, Patrick and Pozzati, Gabriele and Zhu, Wensi and Shenoy, Aditi and Kundrotas, Petras and Elofsson, Arne},
  journal={Nature communications},
  volume={13},
  number={1},
  pages={6028},
  year={2022},
  publisher={Nature Publishing Group UK London}
}

@article{yim2024improved,
  title={Improved motif-scaffolding with SE(3) flow matching},
  author={Yim, Jason and Campbell, Andrew and Mathieu, Emile and Foong, Andrew YK and Gastegger, Michael and Jim{\'e}nez-Luna, Jos{\'e} and Lewis, Sarah and Satorras, Victor Garcia and Veeling, Bastiaan S and No{\'e}, Frank and Barzilay, Regina and Jaakkola, Tommi S},
  journal={arXiv},
  year={2024}
}

@article{chu2024all,
  title={An all-atom protein generative model},
  author={Chu, Alexander E and Kim, Jinho and Cheng, Lucy and El Nesr, Gina and Xu, Minkai and Shuai, Richard W and Huang, Po-Ssu},
  journal={Proceedings of the National Academy of Sciences},
  volume={121},
  number={27},
  pages={e2311500121},
  year={2024},
  publisher={National Academy of Sciences}
}

@article{campbell2024generative,
  title={Generative flows on discrete state-spaces: Enabling multimodal flows with applications to protein co-design},
  author={Campbell, Andrew and Yim, Jason and Barzilay, Regina and Rainforth, Tom and Jaakkola, Tommi},
  journal={arXiv},
  year={2024}
}

@article{huguet2024sequence,
  title={Sequence-augmented SE(3)-flow matching for conditional protein backbone generation},
  author={Huguet, Guillaume and Vuckovic, James and Fatras, Kilian and Thibodeau-Laufer, Eric and Lemos, Pablo and Islam, Riashat and Liu, Cheng-Hao and Rector-Brooks, Jarrid and Akhound-Sadegh, Tara and Bronstein, Michael and Tong, Alexander and Bose, Avishek Joey},
  journal={Advances in neural information processing systems},
  volume={37},
  pages={33007--33036},
  year={2024}
}

@article{shen2018novo,
  title={De novo design of self-assembling helical protein filaments},
  author={Hao Shen and Jorge A. Fallas and Eric Lynch and William Sheffler and Bradley Parry and Nicholas Jannetty and Justin Decarreau and Michael Wagenbach and Juan Jesus Vicente and Jiajun Chen and Lei Wang and Quinton Dowling and Gustav Oberdorfer and Lance Stewart and Linda Wordeman and James De Yoreo and Christine Jacobs-Wagner and Justin Kollman and David Baker},
  journal={Science},
  volume={362},
  number={6415},
  pages={705--709},
  year={2018},
  publisher={American Association for the Advancement of Science}
}

@misc{rcsb_symmetry_stats,
  author       = {{RCSB Protein Data Bank}},
  title        = {{PDB} Statistics: Homo-oligomers Global Assembly Symmetries per Year},
  howpublished = {\url{https://www.rcsb.org/stats/symmetry/growth}},
  note         = {Accessed: 2026-05-07},
  year         = {2025}
}

@article{berman2000protein,
  title={The protein data bank},
  author={Berman, Helen M and Westbrook, John and Feng, Zukang and Gilliland, Gary and Bhat, Talapady N and Weissig, Helge and Shindyalov, Ilya N and Bourne, Philip E},
  journal={Nucleic acids research},
  volume={28},
  number={1},
  pages={235--242},
  year={2000},
  publisher={Oxford University Press}
}

@article{duarte2022exploring,
  title={Exploring protein symmetry at the RCSB Protein Data Bank},
  author={Duarte, Jose M and Dutta, Shuchismita and Goodsell, David S and Burley, Stephen K},
  journal={Emerging topics in life sciences},
  volume={6},
  number={3},
  pages={231--243},
  year={2022},
  publisher={Portland Press Ltd.}
}

@article{shrake1973environment,
  title={Environment and exposure to solvent of protein atoms. Lysozyme and insulin},
  author={Shrake, Andrew and Rupley, John A},
  journal={Journal of molecular biology},
  volume={79},
  number={2},
  pages={351--371},
  year={1973},
  publisher={Elsevier}
}

@article{lovsdorfer2013statistical,
  title={Statistical analysis of sizes and shapes of virus capsids and their resulting elastic properties},
  author={Lo{\v{s}}dorfer Bo{\v{z}}i{\v{c}}, An{\v{z}}e and {\v{S}}iber, Antonio and Podgornik, Rudolf},
  journal={Journal of biological physics},
  volume={39},
  number={2},
  pages={215--228},
  year={2013},
  publisher={Springer}
}

@article{chen2025all,
  title={An all-atom generative model for designing protein complexes},
  author={Ruizhe Chen and Dongyu Xue and Xiangxin Zhou and Zaixiang Zheng and Xiangxiang Zeng and Quanquan Gu},
  journal={arXiv preprint arXiv:2504.13075},
  year={2025}
}

@article{butcher2025novo,
  title={De novo Design of All-atom Biomolecular Interactions with RFdiffusion3},
  author={Jasper Butcher and Rohith Krishna and Raktim Mitra and Rafael I. Brent and Yanjing Li and Nathaniel Corley and Paul T. Kim and Jonathan Funk and Simon Mathis and Saman Salike and Aiko Muraishi and Helen Eisenach and Tuscan Rock Thompson and Jie Chen and Yuliya Politanska and Enisha Sehgal and Brian Coventry and Odin Zhang and Bo Qiang and Kieran Didi and Max Kazman and Frank DiMaio and David Baker},
  journal={bioRxiv},
  year={2025}
}

@article{herzik2019high,
  title={High-resolution structure determination of sub-100 kDa complexes using conventional cryo-EM},
  author={Mark A. Herzik and Mengyu Wu and Gabriel C. Lander},
  journal={Nature communications},
  volume={10},
  number={1},
  pages={1032},
  year={2019},
  publisher={Nature Publishing Group UK London}
}

@article{
divine2021designed,
author = {Robby Divine  and Ha V. Dang  and George Ueda  and Jorge A. Fallas  and Ivan Vulovic  and William Sheffler  and Shally Saini  and Yan Ting Zhao  and Infencia Xavier Raj  and Peter A. Morawski  and Madeleine F. Jennewein  and Leah J. Homad  and Yu-Hsin Wan  and Marti R. Tooley  and Franziska Seeger  and Ali Etemadi  and Mitchell L. Fahning  and James Lazarovits  and Alex Roederer  and Alexandra C. Walls  and Lance Stewart  and Mohammadali Mazloomi  and Neil P. King  and Daniel J. Campbell  and Andrew T. McGuire  and Leonidas Stamatatos  and Hannele Ruohola-Baker  and Julie Mathieu  and David Veesler  and David Baker },
title = {Designed proteins assemble antibodies into modular nanocages},
journal = {Science},
volume = {372},
number = {6537},
pages = {eabd9994},
year = {2021},
doi = {10.1126/science.abd9994},
URL = {https://www.science.org/doi/abs/10.1126/science.abd9994},
eprint = {https://www.science.org/doi/pdf/10.1126/science.abd9994},
}

@article{lin2026fold,
  title={Fold-CP: A Context Parallelism Framework for Biomolecular Modeling},
  author={Dejun Lin and Simon Chu and Vishanth Iyer and Youhan Lee and John St John and Kevin Boyd and Brian Roland and Xiaowei Ren and Guoqing Zhou and Zhonglin Cao and Polina Binder and Yuliya Zhautouskaya and Jakub Zakrzewski and Maximilian Stadler and Kyle Gion and Yuxing Peng and Xi Chen and Tianjing Zhang and Philipp Junk and Michelle Dimon and Paweł Gniewek and Fabian Ortega and McKinley Polen and Ivan Grubisic and Ali Bashir and Graham Holt and Danny Kovtun and Matthias Grass and Luca Naef and Rui Wang and Jian Peng and Anthony Costa and Saee Paliwal and Eddie Calleja and Timur Rvachov and Neha Tadimeti and Roy Tal and Emine Kucukbenli},
  journal={arXiv preprint arXiv:2603.14806},
  year={2026}
}

@article{zhu2023evaluation,
  title={Evaluation of AlphaFold-Multimer prediction on multi-chain protein complexes},
  author={Zhu, Wensi and Shenoy, Aditi and Kundrotas, Petras and Elofsson, Arne},
  journal={Bioinformatics},
  volume={39},
  number={7},
  pages={btad424},
  year={2023},
  publisher={Oxford University Press}
}

@Article{bethel2023precisely,
author={Bethel, Neville P.
and Borst, Andrew J.
and Parmeggiani, Fabio
and Bick, Matthew J.
and Brunette, T. J.
and Nguyen, Hannah
and Kang, Alex
and Bera, Asim K.
and Carter, Lauren
and Miranda, Marcos C.
and Kibler, Ryan D.
and Lamb, Mila
and Li, Xinting
and Sankaran, Banumathi
and Baker, David},
title={Precisely patterned nanofibres made from extendable protein multiplexes},
journal={Nature Chemistry},
year={2023},
month={Dec},
day={01},
volume={15},
number={12},
pages={1664-1671},
issn={1755-4349},
doi={10.1038/s41557-023-01314-x},
url={https://doi.org/10.1038/s41557-023-01314-x}
}

@Article{ingraham2023illuminating,
author={Ingraham, John B.
and Baranov, Max
and Costello, Zak
and Barber, Karl W.
and Wang, Wujie
and Ismail, Ahmed
and Frappier, Vincent
and Lord, Dana M.
and Ng-Thow-Hing, Christopher
and Van Vlack, Erik R.
and Tie, Shan
and Xue, Vincent
and Cowles, Sarah C.
and Leung, Alan
and Rodrigues, Jo{\~a}o V.
and Morales-Perez, Claudio L.
and Ayoub, Alex M.
and Green, Robin
and Puentes, Katherine
and Oplinger, Frank
and Panwar, Nishant V.
and Obermeyer, Fritz
and Root, Adam R.
and Beam, Andrew L.
and Poelwijk, Frank J.
and Grigoryan, Gevorg},
title={Illuminating protein space with a programmable generative model},
journal={Nature},
year={2023},
month={Nov},
day={01},
volume={623},
number={7989},
pages={1070-1078},
issn={1476-4687},
doi={10.1038/s41586-023-06728-8},
url={https://doi.org/10.1038/s41586-023-06728-8}
}

@Article{shen2024de,
author={Shen, Hao
and Lynch, Eric M.
and Akkineni, Susrut
and Watson, Joseph L.
and Decarreau, Justin
and Bethel, Neville P.
and Benna, Issa
and Sheffler, William
and Farrell, Daniel
and DiMaio, Frank
and Derivery, Emmanuel
and De Yoreo, James J.
and Kollman, Justin
and Baker, David},
title={De novo design of pH-responsive self-assembling helical protein filaments},
journal={Nature Nanotechnology},
year={2024},
month={Jul},
day={01},
volume={19},
number={7},
pages={1016-1021},
issn={1748-3395},
doi={10.1038/s41565-024-01641-1},
url={https://doi.org/10.1038/s41565-024-01641-1}
}

@Article{huddy2024blueprinting,
author={Huddy, Timothy F.
and Hsia, Yang
and Kibler, Ryan D.
and Xu, Jinwei
and Bethel, Neville
and Nagarajan, Deepesh
and Redler, Rachel
and Leung, Philip J. Y.
and Weidle, Connor
and Courbet, Alexis
and Yang, Erin C.
and Bera, Asim K.
and Coudray, Nicolas
and Calise, S. John
and Davila-Hernandez, Fatima A.
and Han, Hannah L.
and Carr, Kenneth D.
and Li, Zhe
and McHugh, Ryan
and Reggiano, Gabriella
and Kang, Alex
and Sankaran, Banumathi
and Dickinson, Miles S.
and Coventry, Brian
and Brunette, T. J.
and Liu, Yulai
and Dauparas, Justas
and Borst, Andrew J.
and Ekiert, Damian
and Kollman, Justin M.
and Bhabha, Gira
and Baker, David},
title={Blueprinting extendable nanomaterials with standardized protein blocks},
journal={Nature},
year={2024},
month={Mar},
day={01},
volume={627},
number={8005},
pages={898-904},
issn={1476-4687},
doi={10.1038/s41586-024-07188-4},
url={https://doi.org/10.1038/s41586-024-07188-4}
}

@article{Lisanza2025,
author={Lisanza, Sidney Lyayuga
and Gershon, Jacob Merle
and Tipps, Samuel W. K.
and Sims, Jeremiah Nelson
and Arnoldt, Lucas
and Hendel, Samuel J.
and Simma, Miriam K.
and Liu, Ge
and Yase, Muna
and Wu, Hongwei
and Tharp, Claire D.
and Li, Xinting
and Kang, Alex
and Brackenbrough, Evans
and Bera, Asim K.
and Gerben, Stacey
and Wittmann, Bruce J.
and McShan, Andrew C.
and Baker, David},
title={Multistate and functional protein design using RoseTTAFold sequence space diffusion},
journal={Nature Biotechnology},
year={2025},
month={Aug},
day={01},
volume={43},
number={8},
pages={1288-1298},
issn={1546-1696},
doi={10.1038/s41587-024-02395-w},
url={https://doi.org/10.1038/s41587-024-02395-w}
}

@inproceedings{jiao2024space,
  title={Space group constrained crystal generation},
  author={Jiao, Rui and Huang, Wenbing and Liu, Yu and Zhao, Deli and Liu, Yang},
  booktitle={International Conference on Learning Representations},
  volume={2024},
  pages={6836--6853},
  year={2024}
}

@article{kazeev2025wyckoff,
  title={Wyckoff transformer: Generation of symmetric crystals},
  author={Kazeev, Nikita and Nong, Wei and Romanov, Ignat and Zhu, Ruiming and Ustyuzhanin, Andrey and Yamazaki, Shuya and Hippalgaonkar, Kedar},
  journal={arXiv preprint arXiv:2503.02407},
  year={2025}
}

@article{chang2026space,
  title={Space group equivariant crystal diffusion},
  author={Chang, Rees and Pak, Angela and Guerra, Alex and Zhan, Ni and Richardson, Nick and Ertekin, Elif and Adams, Ryan},
  journal={Advances in Neural Information Processing Systems},
  volume={38},
  pages={72772--72805},
  year={2026}
}

@article{puny2025space,
  title={Space group conditional flow matching},
  author={Puny, Omri and Lipman, Yaron and Miller, Benjamin Kurt},
  journal={arXiv preprint arXiv:2509.23822},
  year={2025}
}

@article{wallace2024novo,
  title={De novo design of peptides that bind specific conformers of $\alpha$-synuclein},
  author={Wallace, Hailey M and Yang, Hyunjun and Tan, Sophia and Pan, Henry S and Yang, Rose and Xu, Junyi and Jo, Hyunil and Condello, Carlo and Polizzi, Nicholas F and DeGrado, William F},
  journal={Chemical Science},
  volume={15},
  number={22},
  pages={8414--8421},
  year={2024},
  publisher={The Royal Society of Chemistry}
}

@article{lin2026planar,
  title={Planar Symmetric Pattern Generation},
  author={Lin, Ning and Chen, Luxi and Chen, Huaguan and Cen, Jiacheng and Li, Chongxuan and Huang, Wenbing and Sun, Hao},
  journal={arXiv preprint arXiv:2606.02073},
  year={2026}
}
}








\newpage
\appendix

\section{Theoretical Details}
\label{app:theory}

\subsection{Proof of Theorems~\ref{thm:descent} and~\ref{thm:expansion}: Equivalence of Generation Strategies}
\label{app:proof_descent_expansion}

This appendix provides full proofs of Theorems~\ref{thm:descent} and~\ref{thm:expansion}.
Appendix~\ref{app:proof_descent} establishes the descent (I)$\Rightarrow$(II), proving that an $\mathrm{SE}(3)$-equivariant and chain-permutation-equivariant full-complex model induces a unique well-defined operator on the quotient space.
Appendix~\ref{app:operation_typology} introduces the operation typology used by both proofs, characterizing the primitives that compose any $\mathrm{SE}(3)$-equivariant graph network in terms of $r$-tuple operations on residue frames.
Appendix~\ref{app:proof_expansion} establishes the expansion (II)$\Rightarrow$(III) by constructing the phantom-chain operator primitive-by-primitive and proving its equivalence with $f_\theta^{/G}$ on canonical representatives.

\subsubsection{Setup and Notation}
\label{app:proof_setup}

Let $f_\theta : \mathrm{SE}(3)^{kN} \to \mathrm{SE}(3)^{kN}$ be a backbone model, where the input $\mathbf{T} = [\mathbf{T}_1, \cdots, \mathbf{T}_k]$ can be separated into $k$ distinct components of size $N$ (i.e., every chain belongs to a single component in our context).
We say $f_\theta$ is \emph{$\mathrm{SE}(3)$-equivariant} if $f_\theta(g \cdot \mathbf{T}) = g \cdot f_\theta(\mathbf{T})$ for every $g \in \mathrm{SE}(3)$, where the action is element-wise on the $kN$ residues.
We say $f_\theta$ is \emph{chain-permutation-equivariant} if $f_\theta(\sigma \cdot \mathbf{T}) = \sigma \cdot f_\theta(\mathbf{T})$ for every permutation $\sigma \in \mathfrak{S}_k$ of the $k$ component blocks of $\mathbf{T} \in \mathrm{SE}(3)^{kN}$.
Let $G = \{g_1 = \mathrm{Id}, g_2, \ldots, g_k\}$ be a finite symmetry group, and write $\mathbf{T}^{\mathrm{sym}} = (g_1 \cdot \mathbf{T}^{(1)}, \ldots, g_k \cdot \mathbf{T}^{(1)})$ for the $G$-symmetric assembly generated by a canonical subunit $\mathbf{T}^{(1)} \in \mathrm{SE}(3)^N_0$.
Equivalence classes $[\mathbf{T}^{(1)}]$ in $\mathrm{SE}(3)^N_0 / G$ are taken under the relation $\mathbf{T} \sim \mathbf{T}'$ if and only if $\mathbf{T}' = g \cdot \mathbf{T}$ for some $g \in G$.

\subsubsection{Proof of Theorem~\ref{thm:descent}: Descent}
\label{app:proof_descent}

The proof proceeds in three steps: first, we show that $f_\theta(\mathbf{T}^{\mathrm{sym}})$ is itself $G$-symmetric, allowing the equivalence class of any single chain to represent the entire output; second, we use this to construct $f_\theta^{/G}$ on equivalence classes; third, we verify that the construction is independent of the choice of representative and is the unique map satisfying \eqref{eq:descent}.

\paragraph{Step 1: $G$-symmetry of the output.}
The construction of $f_\theta^{/G}$ requires that $f_\theta(\mathbf{T}^{\mathrm{sym}})$ be $G$-symmetric, since taking the equivalence class of a single copy is well-defined only when all copies are $G$-related.
For any $g \in G$, the diagonal action of $g$ on $\mathbf{T}^{\mathrm{sym}}$ produces $g \cdot \mathbf{T}^{\mathrm{sym}} = (g g_1 \cdot \mathbf{T}^{(1)}, \ldots, g g_k \cdot \mathbf{T}^{(1)})$, which by closure of $G$ under multiplication is a permutation of the original orbit; concretely, $g \cdot \mathbf{T}^{\mathrm{sym}} = \pi_g \cdot \mathbf{T}^{\mathrm{sym}}$ where $\pi_g \in \mathfrak{S}_k$ is the permutation defined by $g g_i = g_{\pi_g(i)}$.
Applying $f_\theta$ to both sides and using $\mathrm{SE}(3)$-equivariance on the left and chain-permutation-equivariance on the right yields
\begin{equation}
  g \cdot f_\theta(\mathbf{T}^{\mathrm{sym}}) \;=\; \pi_g \cdot f_\theta(\mathbf{T}^{\mathrm{sym}}).
  \label{eq:descent_symmetry}
\end{equation}
Reading off the $i$-th block of each side, the $i$-th copy of the output equals $g$ applied to the $\pi_g^{-1}(i)$-th copy, so the output orbit is closed under $G$ and $f_\theta(\mathbf{T}^{\mathrm{sym}})$ is itself $G$-symmetric.

\paragraph{Step 2: Construction of $f_\theta^{/G}$.}
Define $f_\theta^{/G} : \mathrm{SE}(3)^N_0 / G \to \mathrm{SE}(3)^N_0 / G$ as follows.
Given an equivalence class $[\mathbf{T}^{(1)}]$, choose any representative $\mathbf{T}^{(1)}$, form the orbit $\mathbf{T}^{\mathrm{sym}}$, apply $f_\theta$ to obtain the updated assembly $f_\theta(\mathbf{T}^{\mathrm{sym}})$, and set $f_\theta^{/G}([\mathbf{T}^{(1)}]) = [f_\theta(\mathbf{T}^{\mathrm{sym}})|_{(1,\cdot)}]$, the equivalence class of the first canonical chain.
By Step 1 the output is $G$-symmetric, so the choice of which copy to read off is immaterial: any other copy yields the same equivalence class.

\paragraph{Step 3: Well-definedness and uniqueness.}
We verify that $f_\theta^{/G}$ is independent of the choice of representative $\mathbf{T}^{(1)}$ within $[\mathbf{T}^{(1)}]$.
Suppose $\mathbf{T}^{(1)\prime} = h \cdot \mathbf{T}^{(1)}$ for some $h \in G$.
The orbit built from $\mathbf{T}^{(1)\prime}$ is $\mathbf{T}^{\mathrm{sym}\prime} = (g_1 h \cdot \mathbf{T}^{(1)}, \ldots, g_k h \cdot \mathbf{T}^{(1)})$, which by closure of $G$ is a permutation $\sigma_h$ of the original orbit $\mathbf{T}^{\mathrm{sym}}$.
By chain-permutation-equivariance, $f_\theta(\mathbf{T}^{\mathrm{sym}\prime}) = \sigma_h \cdot f_\theta(\mathbf{T}^{\mathrm{sym}})$, and the first copy of $f_\theta(\mathbf{T}^{\mathrm{sym}\prime})$ is therefore some copy of $f_\theta(\mathbf{T}^{\mathrm{sym}})$, which by Step 1 belongs to the same equivalence class as the first.
The output equivalence class is therefore the same regardless of representative, establishing well-definedness.
Finally, equation~\eqref{eq:descent} follows directly from \eqref{eq:descent_symmetry} by reading off the $i$-th block: $f_\theta(\mathbf{T}^{\mathrm{sym}})|_{(i,n)} = g_i \cdot f_\theta(\mathbf{T}^{\mathrm{sym}})|_{(1,n)} = g_i \cdot f_\theta^{/G}([\mathbf{T}^{(1)}])|_n$.
Uniqueness of $f_\theta^{/G}$ satisfying \eqref{eq:descent} follows because the right-hand side determines the equivalence class of the canonical-copy output for every input class.
\qed

\subsubsection{Operation Typology for Equivariant Graph Networks}
\label{app:operation_typology}

The proof of Theorem~\ref{thm:expansion} requires a structural proposition on $f_\theta$: that it decomposes into a sequence of \emph{$\mathrm{SE}(3)$-equivariant primitive operations} on $r$-tuples of residue frames.
We make this proposition precise here.

\paragraph{Primitive operations.}
A \emph{primitive operation} consumes a finite collection of input features, each indexed by an ordered tuple of residue indices, and produces a single output feature indexed by an ordered tuple of residue indices.
For rigidity, we generally write the input arities as a multiset $\{r_{\mathrm{in}}^{(1)}, \ldots, r_{\mathrm{in}}^{(p)}\}$ and the output arity as $r_{\mathrm{out}}$, so that the operation maps an input collection
\begin{equation}
\label{eq:input_arities}
\bigg( x^{(1)}_{n^{(1)}_1 \cdots n^{(1)} _{r_{\mathrm{in}} ^{(1)}}}, \ldots, x^{(p)}_{n^{(p)}_1 \cdots n^{(p)}_{r_{\mathrm{in}}^{(p)}}} \bigg)
\end{equation}
to an output $y_{m_1 \cdots m_{r_{\mathrm{out}}}}$.
Each output index ranges independently over residues (a ``for each'' over the corresponding axis); 
indices that appear in some input tuple but not in the output tuple are \emph{contracted} by a pooling operation (sum, mean, max, etc.) over the residues in the missing axes, altogether definable as an \texttt{einsum}-like operation.

This typology, while abstruse, covers all standard primitives used in $\mathrm{SE}(3)$-equivariant backbone models, with mostly low tuple dimensionalities:
\begin{itemize} [leftmargin=*]
    \item A pointwise node map is the case $p = 1$, $r_{\mathrm{in}}^{(1)} = r_{\mathrm{out}} = 1$ with no contraction.
    \item An edge construction from a pair of node features is $p = 2$, $r_{\mathrm{in}}^{(1)} = r_{\mathrm{in}}^{(2)} = 1$, $r_{\mathrm{out}} = 2$, also without contraction: the output edge $y_{nm}$ is built from the node features $x^{(1)}_n$ and $x^{(2)}_m$ via an $\mathrm{SE}(3)$-equivariant function of frames $T^n$ and $T^m$.
    \item An edge-to-edge transformation is $p = 1$, $r_{\mathrm{in}}^{(1)} = r_{\mathrm{out}} = 2$.
    \item Standard attention is $p = 2$ with one input feature of arity $2$ (the attention logits or kernel $\alpha_{nm}$) and one of arity $1$ (the values $v_m$), and output arity $r_{\mathrm{out}} = 1$ (the aggregated node update); the index $n$ is preserved as a ``for each'' axis, and the index $m$ is contracted by attention-weighted summation $y_n = \sum_m \alpha_{nm} v_m$.
    \item Triplet operations such as those appearing in axial attention have $r_{\mathrm{in}} = 3$ for some input feature, and may either preserve the triplet structure ($r_{\mathrm{out}} = 3$) or collapse it to a pair or node by contraction.
\end{itemize}

\paragraph{Proposition.}
It is plain to see that all graph-based models operating on tensors can be decomposed into a string of these primitive operations. 
We therefore assume throughout that $f_\theta$ decomposes as a finite composition of such primitive operations.

This also proposes an important new perspective for Theorem~\ref{thm:descent}: the descent of the compound model $f_{\theta}$ onto quotient space $f_{\theta}^{/G}$ is the same as the descent and re-composition of each individual component in $f_{\theta}$. Proof immediately follows from the same line of reasoning on each $\mathrm{SE}(3)$-equivariant primitive operation. 

\subsubsection{Proof of Theorem~\ref{thm:expansion}: Expansion}
\label{app:proof_expansion}

We construct the phantom-copy operator $f_\theta^{\mathrm{ph}}$ by giving the ambient-space expansion of each primitive operation in $f_\theta^{/G}$, then prove the equivalence \eqref{eq:expansion} by induction on the composition.

\paragraph{Step 1: Group-indexed features in the quotient space.}
Before expanding individual primitives, we describe the data layout of features in the quotient-space picture and its ambient-space realization.
A feature of arity $r$ in the ambient space is indexed by an ordered tuple of residues $(n_1, \ldots, n_r)$.
On the quotient space, each residue index becomes an equivalence class with $k$ ambient representatives, so an $r$-tuple of equivalence classes corresponds to $k^r$ distinct ambient tuples; gauge-fixing the first index to a canonical representative leaves $k^{r-1}$ distinct ambient tuples, indexed by relative group elements in $G^{r-1}$.
Concretely, a quotient-space feature of arity $r$ is realized in the ambient space as a feature indexed by $(n_1, \ldots, n_r; g_2, \ldots, g_r)$, where the first $r$ indices range over canonical-copy residues and the remaining $r-1$ group indices specify which phantom-copy representatives provide the second through $r$-th frames in the implicit ambient tuple $(\mathbf{T}^{(1, n_1)}, g_2 \cdot \mathbf{T}^{(1, n_2)}, \ldots, g_r \cdot \mathbf{T}^{(1, n_r)})$.
We refer to this as the \emph{group-indexed representation} of the quotient-space feature, and write $\tilde{x}_{n_1 \cdots n_r}^{g_2, \ldots, g_r}$ for its components.

\paragraph{Step 2: Ambient-space expansion of a single primitive.}
Consider a primitive operation $P$ with input arities $\{r_{\mathrm{in}}^{(1)}, \ldots, r_{\mathrm{in}}^{(p)}\}$ and output arity $r_{\mathrm{out}}$ in $f_\theta$, taking inputs $x^{(1)}, \ldots, x^{(p)}$ and producing output $y$.
The corresponding quotient-space primitive $P^{/G}$ takes group-indexed inputs $\tilde{x}^{(i)}$ with $r_{\mathrm{in}}^{(i)} - 1$ group-element axes each, and produces a group-indexed output $\tilde{y}$ with $r_{\mathrm{out}} - 1$ group-element axes.
Its ambient-space expansion $P^{\mathrm{ph}}$ proceeds as follows:
\begin{enumerate}[leftmargin=18pt, topsep=2pt, itemsep=2pt]
  \item For each output residue tuple $(m_1, \ldots, m_{r_{\mathrm{out}}})$ and output group-index tuple $(h_2, \ldots, h_{r_{\mathrm{out}}})$, evaluate $P$ on the ambient representatives implied by these indices: specifically, instantiate each input $x^{(i)}$ at the ambient tuple of frames specified by the appropriate canonical-copy residues and group elements, gauge-fixed so that the first input index aligns with the canonical-copy output index $m_1$.
  \item Indices that appear in some input but not in the output are contracted by the pooling operation specified by $P$, summed (or otherwise reduced) over the residues in those axes.
  Crucially, indices that appear as group-element axes in some input but not in the output must also be contracted; this corresponds to summing over the corresponding group elements in $G$, weighted by the original pooling rule.
  \item The resulting value is stored at $\tilde{y}_{m_1 \cdots m_{r_{\mathrm{out}}}}^{h_2, \ldots, h_{r_{\mathrm{out}}}}$.
\end{enumerate}

We claim that for any $G$-symmetric input $\mathbf{T}^{\mathrm{sym}}$, the output of $P$ applied to the full assembly, restricted to output tuples whose first index lies in the canonical copy, equals the output of $P^{\mathrm{ph}}$ applied to the canonical representative.
The argument has two parts.

\emph{(a) Per-tuple equivariance.}
For any ambient input tuple of the form $(g_{i_1} \cdot \mathbf{T}^{(1, n_1)}, \ldots, g_{i_r} \cdot \mathbf{T}^{(1, n_r)})$ arising from the orbit of a canonical residue tuple under group elements $(g_{i_1}, \ldots, g_{i_r})$, $\mathrm{SE}(3)$-equivariance of $P$ gives
\begin{equation}
  P(g_{i_1} \cdot \mathbf{T}^{(1, n_1)}, \ldots, g_{i_r} \cdot \mathbf{T}^{(1, n_r)}) \;=\; g_{i_1} \cdot P(\mathbf{T}^{(1, n_1)}, g_{i_1}^{-1} g_{i_2} \cdot \mathbf{T}^{(1, n_2)}, \ldots, g_{i_1}^{-1} g_{i_r} \cdot \mathbf{T}^{(1, n_r)}).
\end{equation}
When the output first index is in the canonical copy ($i_1 = 1$, so $g_{i_1} = \mathrm{Id}$), this is exactly the value $P^{\mathrm{ph}}$ stores at the corresponding output index, with relative group elements $g_{i_1}^{-1} g_{i_j} = g_{i_j}$ providing the input-side phantom-copy assignments.

\emph{(b) Pooling consistency.}
For pooled axes, the full-assembly computation of $P$ on $\mathbf{T}^{\mathrm{sym}}$ contracts a pooled input index over all $kN$ residues in the assembly.
By the orbit structure of the input, this $kN$-residue sum decomposes into $k$ subsums, each over the $N$ residues of one copy.
The subsum corresponding to copy $i$ is, by part (a), exactly the value computed by $P^{\mathrm{ph}}$ at the corresponding canonical-residue tuple with that input axis assigned phantom-copy group element $g_i$.
The total full-assembly contraction is therefore the sum of $P^{\mathrm{ph}}$ values over all $k$ phantom-copy assignments of the contracted axis, which is exactly the group-element contraction performed by $P^{\mathrm{ph}}$ in step 2 of its expansion.
The same argument applies to all pooled input axes simultaneously, since the orbit structure factors over independent residue indices.

Together, parts (a) and (b) establish that the full-assembly output of $P$, restricted to canonical-copy output tuples, agrees with $P^{\mathrm{ph}}$ applied to the canonical representative.

\paragraph{Step 3: Composition.}
The phantom-copy operator $f_\theta^{\mathrm{ph}}$ is the composition of the per-primitive expansions $P_\ell^{\mathrm{ph}}$ in the same order as the primitives in $f_\theta = P_L \circ \cdots \circ P_1$.
We prove $f_\theta^{/G}([\mathbf{T}^{(1)}]) = [f_\theta^{\mathrm{ph}}(\mathbf{T}^{(1)})]$ by induction on $L$.

For $L = 1$, the equality follows directly from Step 2.
For the inductive step, suppose the equality holds for compositions of fewer than $L$ primitives.
Let $\mathbf{T}^{\mathrm{sym}, \ell}$ denote the intermediate assembly produced after the first $\ell$ primitives are applied to $\mathbf{T}^{\mathrm{sym}}$.
Each prefix composition $P_\ell \circ \cdots \circ P_1$ is itself $\mathrm{SE}(3)$-equivariant and chain-permutation-equivariant (these properties are preserved under composition of equivariant primitives), so by Theorem~\ref{thm:descent} applied to each prefix, every $\mathbf{T}^{\mathrm{sym}, \ell}$ is $G$-symmetric.
Step 2 applied to $P_L$ at the input $\mathbf{T}^{\mathrm{sym}, L-1}$ gives that the canonical-copy output of $P_L(\mathbf{T}^{\mathrm{sym}, L-1})$ equals $P_L^{\mathrm{ph}}$ applied to the canonical representative of $\mathbf{T}^{\mathrm{sym}, L-1}$.
By the inductive hypothesis, this canonical representative equals $(P_{L-1}^{\mathrm{ph}} \circ \cdots \circ P_1^{\mathrm{ph}})(\mathbf{T}^{(1)})$.
Composing, the canonical-copy output of $f_\theta(\mathbf{T}^{\mathrm{sym}})$ equals $(P_L^{\mathrm{ph}} \circ \cdots \circ P_1^{\mathrm{ph}})(\mathbf{T}^{(1)}) = f_\theta^{\mathrm{ph}}(\mathbf{T}^{(1)})$, completing the induction.
\qed

\paragraph{Remarks on the construction.}
The phantom-copy realization preserves the architecture of $f_\theta$ in the sense that each primitive $P^{\mathrm{ph}}$ uses the same parameters and computational structure as the original $P$, applied to a different input set.
The architectural changes are limited to bookkeeping: intermediate features carry $k^{r-1}$ group-indexed copies for arity-$r$ features, and pooling operations now span both residue and group-element axes.
Memory and compute reductions follow because the canonical-copy outputs are computed only once rather than $k$ times as in the full-complex forward pass.

\subsection{Proof of Theorem~\ref{thm:attention_block}: Block Structure of Pair-wise Interaction Attention Matrices}
\label{app:proof_attention}

We prove the block-structure equality $A^{[ij]} = A^{[1, \mathrm{idx}(g_i^{-1} g_j)]}$ stated in Theorem~\ref{thm:attention_block}, then expand on the resulting permutation patterns for abelian and non-abelian groups.

\paragraph{Setup.}
Let $A \in \mathbb{R}^{kN \times kN}$ denote a pair-wise interaction matrix featured in an $\mathrm{SE}(3)$-equivariant model, being evaluated on a $G$-symmetric assembly $\mathbf{T}^{\mathrm{sym}} = (g_1 \cdot \mathbf{T}^{(1)}, \cdots, g_k \cdot \mathbf{T}^{(1)})$.
We index residues as $(i, n)$ for copy $i \in [k]$ and position $n \in [N]$, with $A_{(i,n),(j,m)}$ denoting the logit from key residue $(j, m)$ to query residue $(i, n)$.
Arranging $A$ into $k \times k$ blocks of size $N \times N$, the $(i, j)$-th block $A^{[ij]}$ has entries $A^{[ij]}_{nm} = A_{(i,n),(j,m)}$.
Entries are $\mathrm{SE}(3)$-invariant scalars. 

\paragraph{Proof of Theorem~\ref{thm:attention_block}.}
By chain-permutation-equivariance, the permutation 
$\pi_g: i \rightarrow \mathrm{idx}(g \cdot g_i)$ acts equivariantly on the matrix $A$, so for a single entry: 
\begin{align*}
    A_{(i,n),(j,m)} (\mathbf{T}^{\text{sym}}) 
    &= A_{(i,n),(j,m)} (\pi_g \circ \mathbf{T}^{\text{sym}}) \\
    &= \pi_g \circ A_{(i,n),(j,m)} (\mathbf{T}^{\text{sym}})
    = A_{(\pi_{g}(i),n),(\pi_{g}(j),m)} (\mathbf{T}^{\text{sym}}), 
\end{align*}
where the first equation follows since the permutation $\pi_g$ acting on the $G$-symmetric assembly is equivalent to the geometric action of $g$, which keeps it intact. 
Taking $g = g_i^{-1}$, we have immediately
\begin{equation*}
    A_{(i,n),(j,m)} = A_{(1,n),(\mathrm{idx}(g_i^{-1} g_j), m)}, 
\end{equation*}
which is to say,
\begin{equation*}
  A^{[ij]}_{nm} = A^{[1,\, \mathrm{idx}(g_i^{-1} g_j)]}_{nm}. 
\end{equation*}
\qed

\paragraph{Permutation patterns: abelian and non-abelian groups.}
The block equality says that the $i$-th row of blocks of $A$ is a permutation of the first row, with the $j$-th block of the $i$-th row equal to the $\mathrm{idx}(g_i^{-1} g_j)$-th block of the first row.
This permutation pattern is fully determined by the group multiplication table of $G$ together with the chosen indexing $\mathrm{idx}$.

For abelian $G$ (cyclic groups $C_n$ and direct products thereof), the block at position $(i, j)$ depends only on $g_i^{-1} g_j$, which itself depends only on the index difference $j - i$ modulo $n$ (under the natural indexing $g_i = g^{i-1}$ for a generator $g$).
The block matrix is therefore a circulant: each row is a cyclic shift of the first row, with shift amount equal to the row index.
This is the pattern exploited by efficient circulant-matrix multiplications for cyclic-symmetric assemblies.

For non-abelian $G$ (dihedral $D_n$, polyhedral $T$, $O$, $I$), the permutation indexed by $g_i^{-1} g_j$ is the action of $g_i^{-1}$ on the indexing of $G$ via left multiplication, which is the regular representation of $G$.
The block matrix is therefore a $G$-equivariant matrix in the sense of group representation theory: it commutes with the action of $G$ on the row and column indices.
Concretely, the row indexed by $g_i$ is obtained from the first row by relabeling the column indices according to $j \mapsto \mathrm{idx}(g_i \cdot g_j)$ (the action of $g_i$ by left multiplication on the indexing of $G$).
For $C_n$ this reduces to the cyclic shift above; for $D_n$ the shifts are interleaved with reflections; for the polyhedral groups the permutations correspond to the rotational subgroup action on the orbit of the canonical copy.
A visualization of the permutation patterns for each group used in our experiments is provided in Appendix~\ref{app:block}.

\subsection{Detailed Feature Reductions}
\label{app:feature_reduction_full}

This appendix expands the per-component reductions sketched in the main text into a precise account of how each feature type stored by an $\mathrm{SE}(3)$-equivariant backbone model is realized in the phantom-copy operator $f_\theta^{\mathrm{ph}}$.
The general principle, formalized in Appendix~\ref{app:operation_typology}, is that an arity-$r$ feature on a $G$-symmetric assembly is determined by the canonical-residue tuple together with $r-1$ relative group elements; the phantom-copy realization stores only this gauge-fixed representation.
We instantiate the principle for each feature type appearing in FrameDiff/FoldFlow and RFdiffusion.

\paragraph{Node features (arity $r = 1$).}
Node features have no group-element axis ($k^{r-1} = 1$).
Scalar (invariant) node features are identical across all $k$ copies of a $G$-symmetric assembly by $\mathrm{SE}(3)$-invariance, so storing them only for the canonical subunit reduces the memory footprint from $kN \cdot D_h$ to $N \cdot D_h$, where $D_h$ is the node embedding dimension.
Frame-valued (equivariant) node features---residue frames $T^{(i, n)}$ themselves, predicted frame updates $\Delta T^{(i, n)}$---are tied across copies by $T^{(i, n)} = g_i \cdot T^{(1, n)}$, so only the canonical-copy frames $T^{(1, n)}$ are stored explicitly; phantom-copy frames are reconstructed on demand by applying $g_i$.
This reduction is exact and applies uniformly to every node-level tensor in the model, including IPA local-frame coordinates.

\paragraph{Edge features (arity $r = 2$).}
Edge features carry one group-element axis ($k^{r-1} = k$), corresponding to the relative group element between query and key residues.
The phantom-copy realization stores edge features as a tensor $\tilde{z}_{nm}^g$ for $n, m \in [N]$ and $g \in G$, with memory footprint $kN^2 D_z$ where $D_z$ is the edge embedding dimension.
Reading off the relevant blocks: $\tilde{z}_{nm}^{\mathrm{Id}}$ contains the intra-canonical-subunit edge from residue $m$ to residue $n$; $\tilde{z}_{nm}^{g_j}$ for $j \geq 2$ contains the inter-copy edge from phantom residue $(j, m)$ to canonical residue $(1, n)$.
Phantom-to-phantom edges (e.g., from $(j_1, m)$ to $(j_2, n)$ with $j_1, j_2 \geq 2$) are not stored explicitly; by the block-structure result of Theorem~\ref{thm:attention_block}, every such edge equals an edge in the canonical block-row indexed by group element $g_{j_2}^{-1} g_{j_1}$.
The compression from $k^2 N^2 D_z$ to $kN^2 D_z$ corresponds to retaining only the canonical block-row of the full $kN \times kN$ edge tensor.

\paragraph{Edge-feature initialization with self-conditioning.}
FrameDiff and FoldFlow initialize edge features at each diffusion step using a binned $\mathrm{C}_\alpha$ pairwise distance matrix from the previous self-conditioned prediction $\hat{\mathbf{x}}^{sc}$ \citep{yim2023se}.
On a $G$-symmetric assembly, this distance matrix has the same block structure as the attention matrix: $d^{(i, n), (j, m)} = d^{(1, n), (\mathrm{idx}(g_i^{-1} g_j), m)}$, since pairwise distances are $\mathrm{SE}(3)$-invariant.
The phantom-copy realization computes binned distances only for the canonical block-row, indexed by $g \in G$, and uses these as the initialization of $\tilde{z}_{nm}^g$.
The self-conditioned prediction $\hat{\mathbf{x}}^{sc}$ is itself $G$-symmetric in \ours{} by induction (the canonical-copy prediction at each diffusion step is propagated to phantom copies by explicit group action), so the resulting $\tilde{z}_{nm}^g$ matches the canonical block-row of the full-complex computation exactly.
At the first diffusion step, $\hat{\mathbf{x}}^{sc}$ is initialized to zero, consistent with the training protocol.

\paragraph{Attention logits and weights (arity $r = 2$).}
Attention logits inherit the same arity and group-axis structure as edge features.
The phantom-copy realization stores $\tilde{a}_{nm}^g$ for $n, m \in [N]$ and $g \in G$, with memory footprint $kN^2$ per attention head.
After softmax normalization, the attention weights $\tilde{\alpha}_{nm}^g$ inherit the same shape; aggregation of values to canonical-residue updates contracts the key residue index $m$ and the group-element axis $g$:
\begin{equation}
  v_n \;=\; \sum_{g \in G}\, \sum_{m=1}^{N} \tilde{\alpha}_{nm}^g \cdot \tilde{v}_m^g,
\end{equation}
which is precisely the einsum-style group-axis contraction described in the proof of Theorem~\ref{thm:expansion}.

\paragraph{Frame updates.}
Frame updates are pointwise frame-valued node maps and follow the node-feature reduction.
Predicted updates $\Delta T^{(1, n)}$ are computed for the canonical subunit only; phantom-copy updates are obtained as $\Delta T^{(i, n)} = g_i \cdot \Delta T^{(1, n)}$ during the assembly reconstruction at the end of each diffusion step.
This step is the practical mechanism that propagates the canonical-copy update to the full assembly, and it incurs $O(kN)$ rather than $O(kN \cdot \text{model-cost})$ work.

\section{Further Discussions}

\subsection{Extended Related Work}
\label{app:extended_related_work}

Representing symmetric structures through symmetry-inequivalent components
and their group orbits is a well-established construction used across
crystallography and protein design. In crystal generation,
DiffCSP++~\citep{jiao2024space} and
SGEquiDiff~\citep{chang2026space} model symmetry-inequivalent atoms under
Wyckoff constraints and recover symmetry-related copies through
space-group operations. Related approaches exploit the same symmetry
structure in different ways: WyFormer~\citep{kazeev2025wyckoff} generates
discrete descriptions of occupied Wyckoff sites that are subsequently
converted into crystal structures, while Space Group Conditional Flow
Matching~\citep{puny2025space} initializes symmetry-compatible atomic
configurations and preserves their Wyckoff constraints through an
equivariant flow over the full unit cell. In amyloid-binder design,
Wallace et al.~\citep{wallace2024novo} use the target fibril's screw symmetry
to replicate helical binder motifs along its surface and jointly optimize
binder--fibril and binder--binder interactions. Related symmetry
enforcement also appears in continuous pattern design:
Lin et al.~\citep{lin2026planar} construct continuous representations
invariant under any of the 17 wallpaper groups, with applications to
pattern, paper-cutting, topology, and material design.

\subsection{Violation of Permutation Invariance in Chain Embeddings}
\label{app:permutation}

The chain-permutation-equivariance assumption of Theorems~\ref{thm:descent} and~\ref{thm:expansion} is violated in practice by the relative-residue-index encoding $\phi(m-n)$ used to initialize edge features and chain offsets in RFdiffusion~\citep{watson2023de}. RFdiffusion's default inference splits the chains with $L$, before initializing edge features with relative sequential distances. In general, any such method that instantiates signed relative embeddings is not chain-permutation equivariant, as the swapping of chains $i$ and $j$ changes the sign of the relative-index encoding. This has the effect of breaking the equivalence claim of \ours{} with respect to pairwise interaction blocks.
Recall the block structure of the $kN \times kN$ pair-wise interaction matrix from Theorem~\ref{thm:attention_block}: 
under exact chain-permutation equivariance, $A^{[ij]} = A^{[1, g_i^{-1} g_j]}$ for all $i, j \in [k]$.
The intra-copy diagonal blocks $A^{[ii]}$ are computed exclusively from intra-copy edge features, and are therefore unaffected by the inter-copy offset $\Delta$. Hence they match exactly across all copies even in the presence of this violation.
The inter-copy off-diagonal blocks $A^{[ij]}$'s, with $i \neq j$, are affected by this discrepancy. The equality $A^{[ij]} = A^{[1, g_i^{-1} g_j]}$ then becomes approximate rather than exact.

\paragraph{Proposed fixes.}
Several possible fixes can restore exact chain-permutation equivariance, though they certainly require model retraining and lie outside the zero shot scope of this work.
The simplest fix is an out-of-range token, which replaces inter-copy relative indices with a fixed special token, which is assigned a learned embedding distinct from any in-range relative index. This is the approach adopted by some structure-prediction models for chain breaks, including AF-Multimer's chain-relative encoding \citep{jumper2021highly}.
A more advanced method may involve a symmetry-aware encoding that represents the relative group relation between the pair, designed to be invariant across chain permutations.
Both preserve the within-copy semantics of relative position while restoring exact chain-permutation equivariance, and we view this development as a natural direction for future work.

\subsection{Phantom-Copy Truncation for Infinite Symmetry Groups}
\label{app:diminishing}

For finite point groups, all $|G| = k$ phantom copies are retained and Theorems~\ref{thm:descent} and~\ref{thm:expansion} hold exactly outside the chain-embedding caveat of Appendix~\ref{app:permutation}.
For infinite groups such as lattice and helical symmetries, the orbit is unbounded and exact phantom-copy generation is impossible. We propose to implement distance-based truncation as a natural way to deal with this, motivated by the fact that many architectures already imbue distance priors. For example, IPA has a distance term that penalizes residue pairs quadratically.

\paragraph{Distance Truncation.}
Given a canonical subunit $\mathbf{T}^{(1)}$ and a possibly infinite symmetry group $G$, we truncate the phantom-copy set to those chains whose minimum residue--residue distance to the canonical chain falls below a hyperparameter $d_{\mathrm{max}}$, that is, $\{g \cdot \mathbf{T}^{(1)} : g \in G,\ \mathrm{dist}(g \cdot \mathbf{T}^{(1)}, \mathbf{T}^{(1)}) < d_{\mathrm{max}}\}$ where $\mathrm{dist}$ denotes the minimum pairwise residue distance between the candidate phantom copy and the canonical chain.
Concretely, we enumerate group elements in order of increasing minimum distance and include them until the cutoff is reached, yielding a finite phantom-copy set of size $\kappa$ even when $G$ is infinite.
For lattice and helical groups acting with bounded density on $\mathbb{R}^3$, $\kappa = O(1)$ independent of the (formal) group order, with the constant determined by $d_{\mathrm{max}}$ and the geometric density of the orbit.



\section{Visualization of Block Symmetric Attention}
\label{app:block}
We empirically validate Theorem~\ref{thm:attention_block} by visualizing the attention matrices that arise during full-complex symmetric inference. We run full-complex (non-\ours{}) symmetric diffusion on FrameDiff, FoldFlow, and RFdiffusion under $C_8$ and tetrahedral symmetry. For FrameDiff and FoldFlow we use an inter-chain offset of $1000$ residue positions.

\begin{figure}[p]
    \centering
    \includegraphics[width=1.0\linewidth]{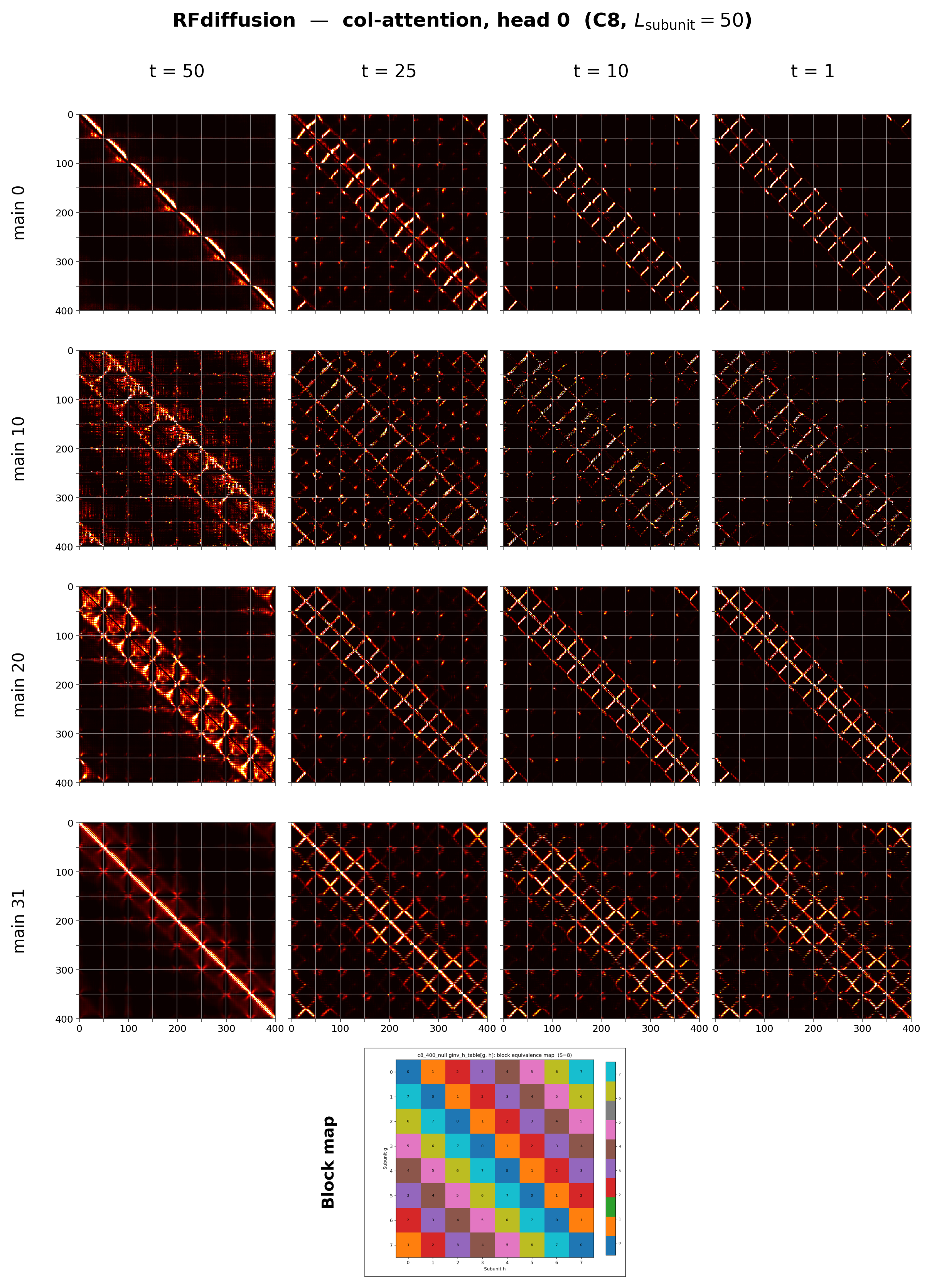}
    \caption{RFdiffusion attention logits under $C_8$ symmetry, at attention head $0$, across a grid of network layers and denoising timesteps. The analytical block structure predicted by Theorem~\ref{thm:attention_block} for $G = C_8$ is shown for reference.}
    \label{fig:attn_grid_rfdiffusion_c8}
\end{figure}

\begin{figure}[p]
    \centering
    \includegraphics[width=1.0\linewidth]{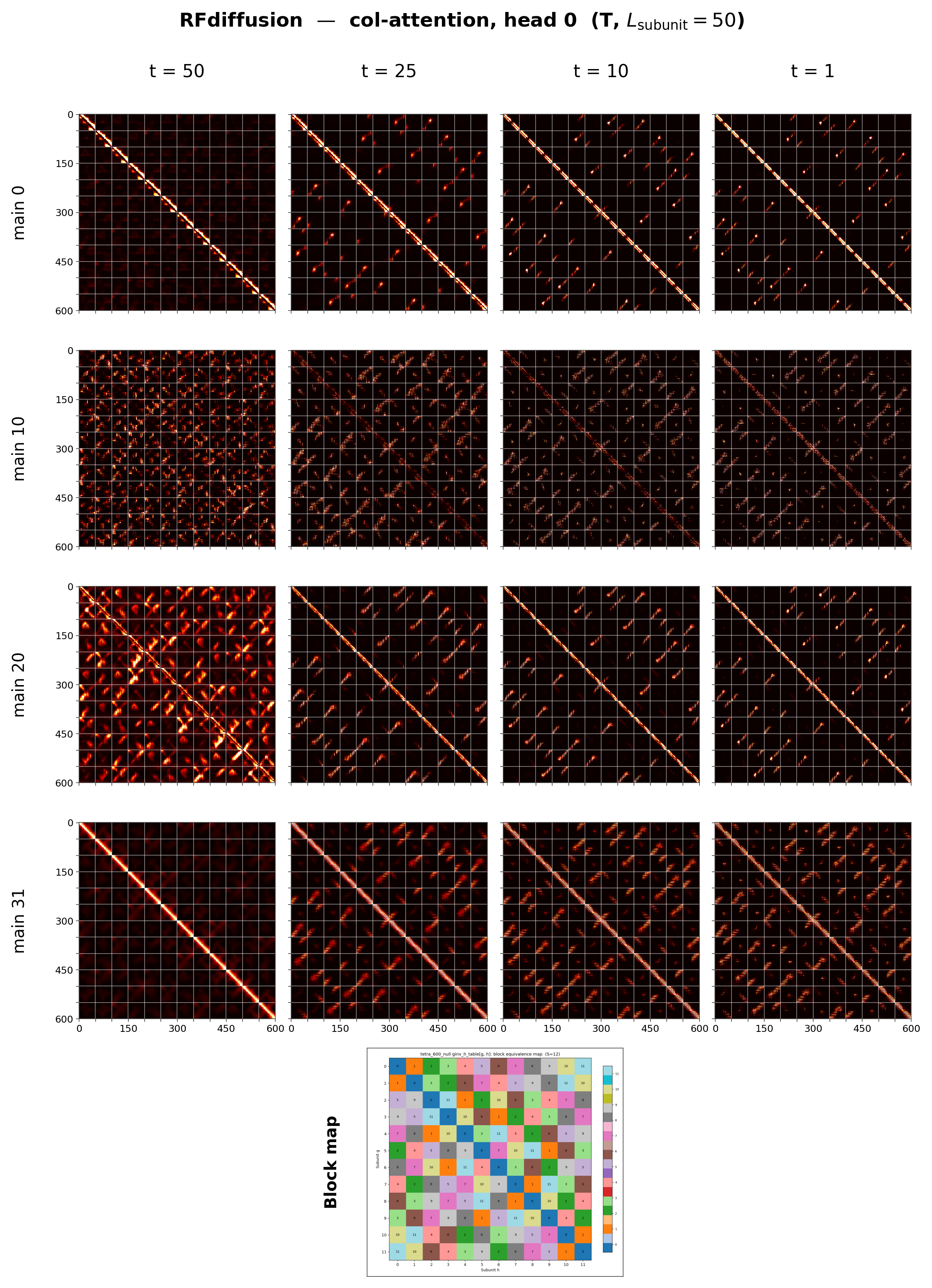}
    \caption{RFdiffusion attention logits under tetrahedral ($T$) symmetry, at attention head $0$, across a grid of network layers and denoising timesteps. The analytical block structure predicted by Theorem~\ref{thm:attention_block} for $G = T$ is shown for reference.}
    \label{fig:attn_grid_rfdiffusion_t}
\end{figure}

\begin{figure}[p]
    \centering
    \includegraphics[width=1.0\linewidth]{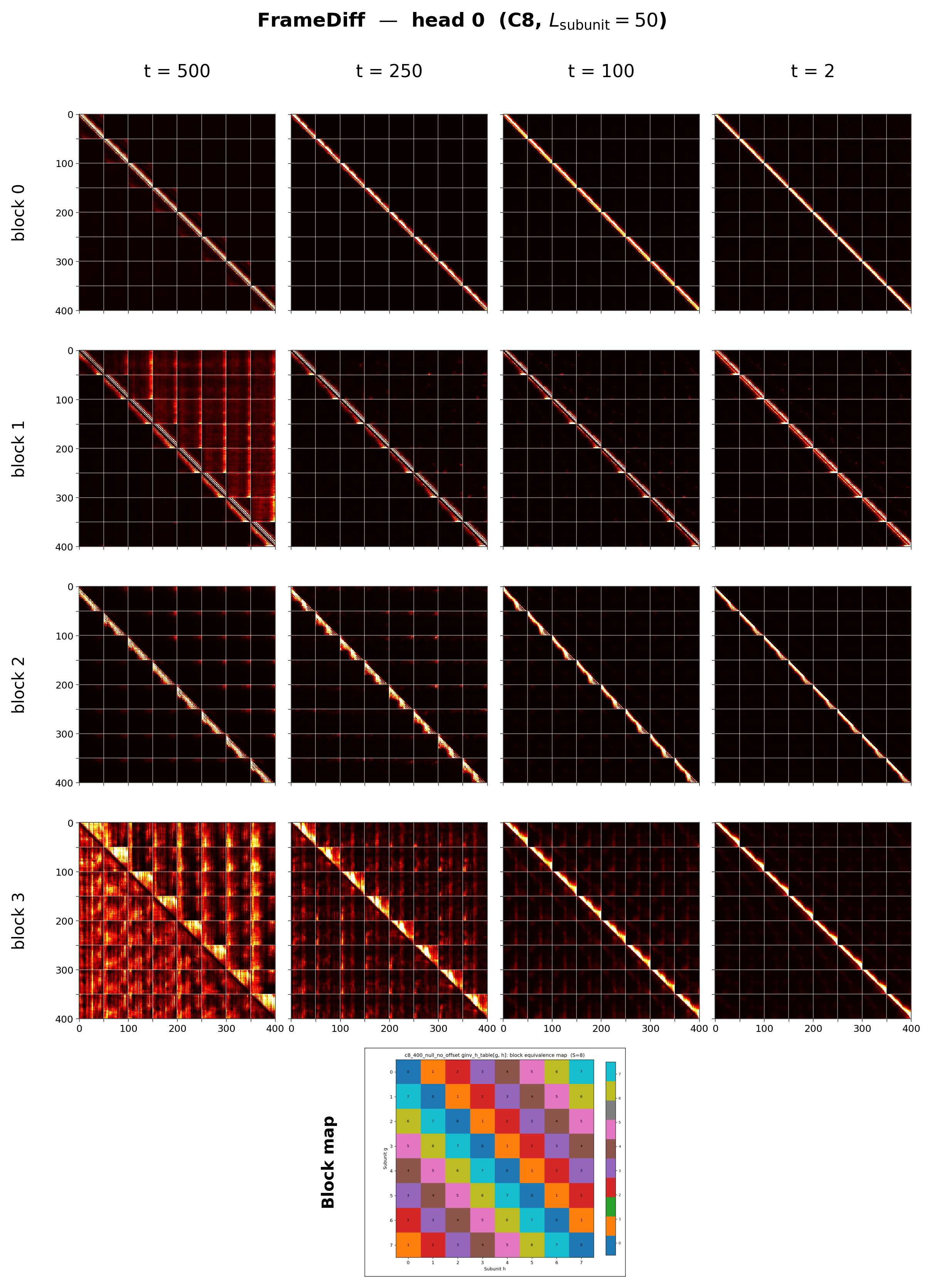}
    \caption{FrameDiff attention logits under $C_8$ symmetry, at attention head $0$, across a grid of network layers and denoising timesteps. The analytical block structure predicted by Theorem~\ref{thm:attention_block} for $G = C_8$ is shown for reference.}
    \label{fig:attn_grid_framediff_c8}
\end{figure}

\begin{figure}[p]
    \centering
    \includegraphics[width=1.0\linewidth]{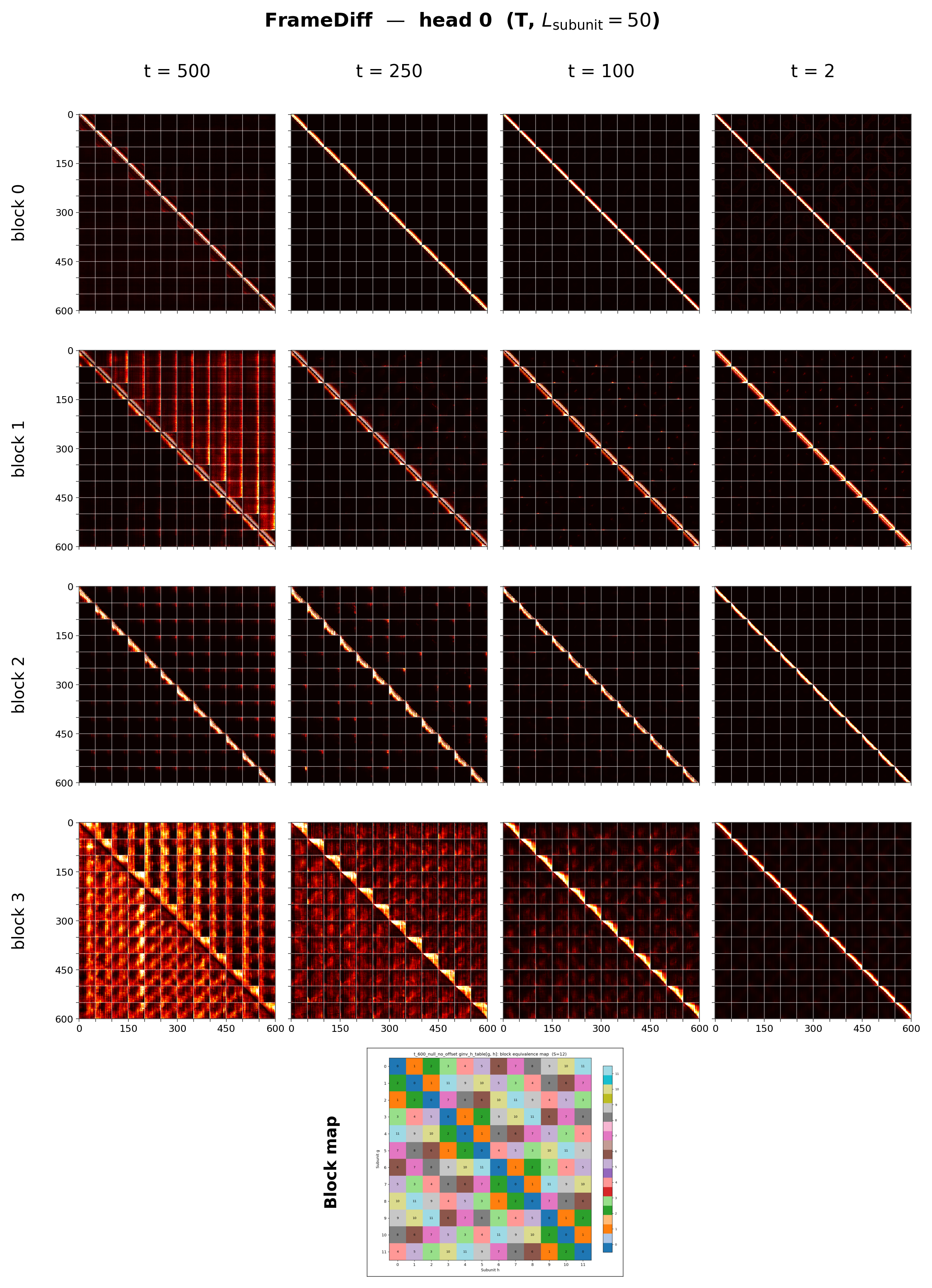}
    \caption{FrameDiff attention logits under tetrahedral ($T$) symmetry, at attention head $0$, across a grid of network layers and denoising timesteps. The analytical block structure predicted by Theorem~\ref{thm:attention_block} for $G = T$ is shown for reference.}
    \label{fig:attn_grid_framediff_t}
\end{figure}

\begin{figure}[p]
    \centering
    \includegraphics[width=1.0\linewidth]{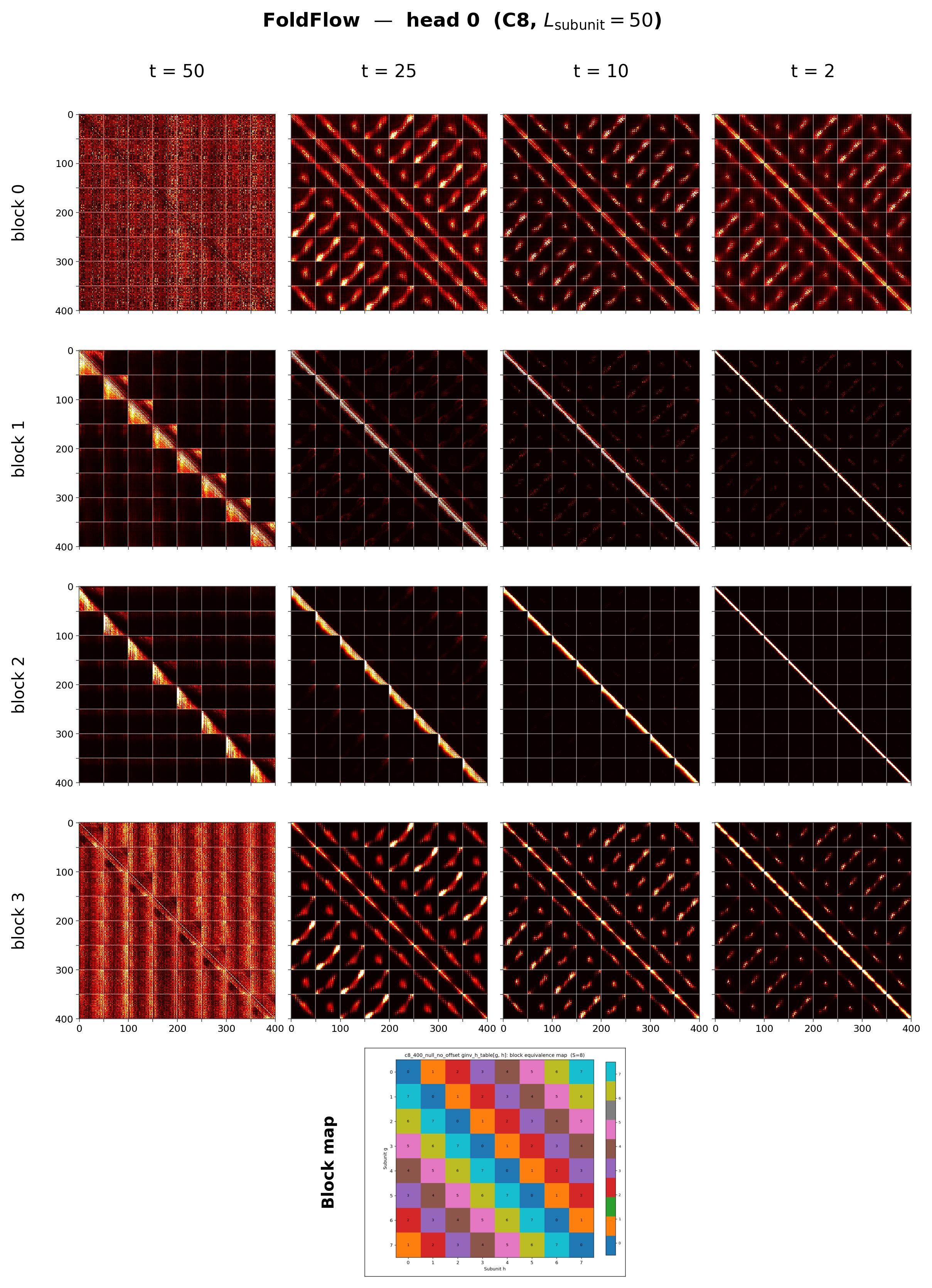}
    \caption{FoldFlow attention logits under $C_8$ symmetry, at attention head $0$, across a grid of network layers and denoising timesteps. The analytical block structure predicted by Theorem~\ref{thm:attention_block} for $G = C_8$ is shown for reference.}
    \label{fig:attn_grid_foldflow_c8}
\end{figure}

\begin{figure}[p]
    \centering
    \includegraphics[width=1.0\linewidth]{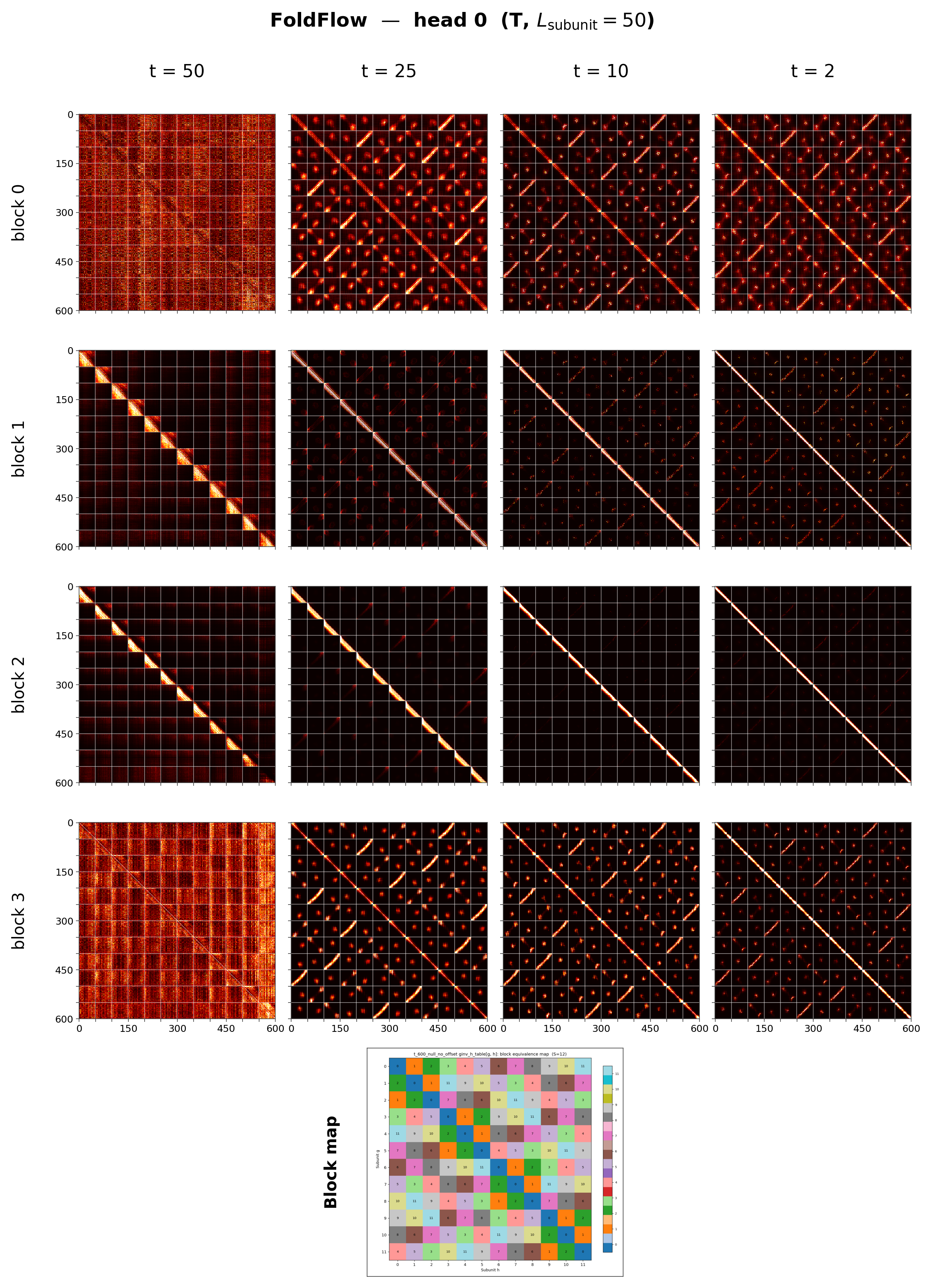}
    \caption{FoldFlow attention logits under tetrahedral ($T$) symmetry, at attention head $0$, across a grid of network layers and denoising timesteps. The analytical block structure predicted by Theorem~\ref{thm:attention_block} for $G = T$ is shown for reference.}
    \label{fig:attn_grid_foldflow_t}
\end{figure}

For each model-symmetry pair, Figures~\ref{fig:attn_grid_rfdiffusion_c8}--\ref{fig:attn_grid_foldflow_t} display attention logits at head $0$ across layers and denoising timesteps. Also displayed is the analytical block structure predicted by the symmetry group $G$. The predicted block structure emerges robustly across all three architectures, all sampled layers, and all sampled timesteps.

\newpage
\section{Algorithm Implementation}
\label{app:algorithms}
\subsection{Symmetric Invariant Point Attention for Framediff and FoldFlow}
Algorithm~\ref{alg:symmetric_ipa} describes our symmetric aware IPA implementation, used in the Framediff and FoldFlow Architecture.
\begin{algorithm}[H]
\label{alg:framediff}
\caption{Framediff SE(3) Symmetric Diffusion Block (single block $\ell \to \ell+1$)}
\label{alg:se3_block}
\begin{algorithmic}[1]
\Statex \textbf{Input:} 
\Statex \quad Node features: $\mathbf{h}_\ell \in \mathbf{R}^{L_{can} \times d_h}$
\Statex \quad Edge features: $\mathbf{z}_\ell \in \mathbf{R}^{L_{can} \times L \times d_z}$, \Statex \quad Backbone frames $\mathbf{T}_\ell \in \text{SE}(3)^{L_{can}}$
\Statex \quad Initial node features $\mathbf{h}_0$
\Statex \quad Symmetry Group $G \in \text{SE}(3)$
\Statex
\Statex \quad Index helpers ($j \in \{1,\dots,L\}$):
\Statex \quad\quad\quad subunit:\ $\sigma(j) = \lceil j / L_{\mathrm{can}} \rceil$ \hfill // map residue idx to group idx
\Statex \quad\quad\quad canonical residue:\ $\rho(j) = ((j-1) \bmod L_{\mathrm{can}}) + 1$ \hfill // map residue idx to canonical idx
\Statex
\Statex \textbf{Output:} 
\Statex \quad Updated Node features: $\mathbf{h}_{\ell+1}$, 
\Statex \quad Updated Edge features: $\mathbf{z}_{\ell+1}$, 
\Statex \quad Updated Frames: $\mathbf{T}_{\ell+1}$

\Statex
\Statex \textbf{NodeUpdate:}

\State $\mathbf{h}' \gets \mathbf{h}_\ell + \text{SymmetricInvariantPointAttention}(\mathbf{h}_\ell, \mathbf{z}_\ell, \mathbf{T}_\ell, G)$
\State $\mathbf{h}'' \gets \text{Transformer}(\text{Concat}(\mathbf{h}', \text{Linear}(\mathbf{h}_0)))$
\State $\mathbf{h}_{\ell+1} \gets \text{MLP}(\mathbf{h}'' + \mathbf{h}')$

\Statex
\Statex \textbf{EdgeUpdate:}
\For{each edge $(n \in L_{can}, m \in L)$}
    \State $e \gets \text{Concat}(\text{Linear}(\mathbf{h}_{\ell+1}^n), \text{Linear}(\mathbf{h}_{\ell+1}^{\rho(m)}), \mathbf{z}_\ell^{nm})$
    \State $e' \gets \text{MLP}(e) + e$
    \State $\mathbf{z}_{\ell+1}^{nm} \gets \text{Linear}(e')$
\EndFor
\Statex
\Statex \textbf{BackboneUpdate:}
\State $\mathbf{T}_{\ell+1} \gets \text{BackboneUpdate}(\mathbf{T}_\ell, \mathbf{h}_{\ell+1})$

\State \Return $\mathbf{h}_{\ell+1}, \mathbf{z}_{\ell+1}, \mathbf{T}_{\ell+1}$
\end{algorithmic}
\end{algorithm}
\begin{algorithm}[H]

\label{alg:symmetric_ipa}
\caption{Efficient Symmetry-Aware Invariant Point Attention (IPA)}
\begin{algorithmic}[1]
\Statex \textbf{Input:}
\Statex \quad Canonical node features: $\mathbf{s} \in \mathbf R^{L_{\mathrm{can}} \times d_s}$
\Statex \quad Block-row Edge features: $\mathbf{z} \in \mathbf R^{L_{\mathrm{can}} \times L \times d_z}$,\ with $L = k \cdot L_{\mathrm{can}}$

\Statex \quad Canonical backbone frames: $\{T_i\}_{i=1}^{L_{\mathrm{can}}} \in \mathrm{SE}(3)$
\Statex \quad Symmetry group: $\{g_1,\, g_2,\, \dots,\, g_k\} \subset G$
\Statex
\Statex \quad Index helpers ($j \in \{1,\dots,L\}$):
\Statex \quad\quad\quad subunit:\ $\sigma(j) = \lceil j / L_{\mathrm{can}} \rceil$ \hfill // map residue idx to group idx
\Statex \quad\quad\quad canonical residue:\ $\rho(j) = ((j-1) \bmod L_{\mathrm{can}}) + 1$ \hfill // map residue idx to canonical idx
\Statex
\Statex \textbf{Output:} Updated Canonical single representations $\tilde{\mathbf{s}} \in \mathbf R^{L_{\mathrm{can}} \times d_s}$
\Statex
\Statex \text{// Convert invariant/equivariant repr. under the symmetry group.}
\State $\mathbf{s}'_j \leftarrow \mathbf{s}_{\rho(j)}$ \hfill \(j \in \{1,\dots,L\}\)
\State $T'_j \leftarrow g_{\sigma(j)} \circ T_{\rho(j)}$ \hfill \(j \in \{1,\dots,L\}\)
\Statex
\Statex \text{// Queries from canonical residues, keys/values from the phantoms.}
\State $\mathbf{q}_i^h \leftarrow \mathrm{LinearNoBias}(\mathbf{s}_i)$ \hfill $i \in \{1,\dots,L_{\mathrm{can}}\}$
\State $\mathbf{k}_j^h,\ \mathbf{v}_j^h \leftarrow \mathrm{LinearNoBias}(\mathbf{s}'_j)$ \hfill $j \in \{1,\dots,L\}$
\State $\vec{\mathbf{q}}_i^{hp} \leftarrow T_i \circ \mathrm{LinearNoBias}(\mathbf{s}_i)$ \hfill $\in \mathbf R^3,\ p \in \{1,\dots,N_{\text{qp}}\}$
\State $\vec{\mathbf{k}}_j^{hp} \leftarrow T'_j \circ \mathrm{LinearNoBias}(\mathbf{s}'_j)$ \hfill $\in \mathbf R^3,\ p \in \{1,\dots,N_{\text{qp}}\}$
\State $\vec{\mathbf{v}}_j^{hp} \leftarrow T'_j \circ \mathrm{LinearNoBias}(\mathbf{s}'_j)$ \hfill $\in \mathbf R^3,\ p \in \{1,\dots,N_{\text{vp}}\}$

\State $b_{ij}^h \leftarrow \mathrm{LinearNoBias}(\mathbf{z}_{ij})$ \hfill \(i\in\{1,\dots,L_{\mathrm{can}}\},\ j\in\{1,\dots,L\}\)
\State $w_C \leftarrow \sqrt{\tfrac{2}{9\,N_{\mathrm{qp}}}}$;\quad $w_L \leftarrow \sqrt{\tfrac{1}{3}}$
\Statex
\Statex \text{// Compute attention weights for phantom $\rightarrow$ canonical only.}
\State $a_{ij}^h \leftarrow \mathrm{Softmax}_{j \in \{1,\dots,L\}}\!\left(
   w_L\!\left(
     \tfrac{1}{\sqrt{c}}\,\mathbf{q}_i^{h\top}\mathbf{k}_j^h
     \,+\, b_{ij}^h
     \,-\, \tfrac{\gamma^h w_C}{2} \sum_{p=1}^{N_{\mathrm{qp}}}
        \bigl\| T_i \circ \vec{\mathbf{q}}_i^{hp} \,-\, T'_j \circ \vec{\mathbf{k}}_j^{hp} \bigr\|^2
   \right)\right)$
\Statex
\State $\tilde{\mathbf{o}}_i^h \leftarrow \sum_{j=1}^{L} a_{ij}^h\,\mathbf{z}_{ij}$
\State $\mathbf{o}_i^h \leftarrow \sum_{j=1}^{L} a_{ij}^h\,\mathbf{v}_j^h$
\State $\vec{\mathbf{o}}_i^{hp} \leftarrow T_i^{-1} \circ \sum_{j=1}^{L} a_{ij}^h\,\bigl(T'_j \circ \vec{\mathbf{v}}_j^{hp}\bigr)$

\State $\tilde{\mathbf{s}}_i \leftarrow \mathrm{Linear}\!\left(
   \mathrm{Concat}_{h,p}\bigl(
     \tilde{\mathbf{o}}_i^h,\ \mathbf{o}_i^h,\ \vec{\mathbf{o}}_i^{hp},\ \|\vec{\mathbf{o}}_i^{hp}\|
   \bigr)\right)$
\State \Return $\tilde{\mathbf{s}} \in \mathbf R^{L_{\mathrm{can}} \times d_s}$
\end{algorithmic}
\end{algorithm}

\newpage
\subsection{Symmetric BiasedAxialAttention for RFdiffusion}
Algorithm~\ref{alg:biasedaxialattention} describes the native Axial Attention used in RFdiffusion and RoseTTAFold. We include it to help the reader visualize \ours{}' implementation differences. 

\begin{algorithm}[H]
\label{alg:biasedaxialattention}
\caption{BiasedAxialAttention (RFdiffusion/RoseTTAFold-native axial attention)}
\begin{algorithmic}[1]
\Statex \textbf{Input:} $\mathbf{z}\in\mathbf R^{L\times L\times d_z}$, bias $\mathbf{b}\in\mathbf R^{L\times L}$
\Statex \textbf{Output:} Updated pair $z'$ of shape $(L,L,d_z)$
\Statex
\If{\textsc{IsRow}}
    \Statex $z\leftarrow z^{\top}$,\quad $b\leftarrow b^{\top}$
\EndIf
\Statex
\State $z\leftarrow \mathrm{LayerNorm}(z)$; \quad $b\leftarrow \mathrm{LayerNorm}(b)$\;
\State $Q\leftarrow W_q z,\ K\leftarrow W_k z,\ V\leftarrow W_v z$, $B\leftarrow W_b b$ \quad\Comment{shapes $(L,L,h,d)$}
\State $g\leftarrow \mathrm{sigm}(W_g z)$
\State $Q\leftarrow Q/\sqrt{d}$; \quad $K\leftarrow K/\sqrt{L}$
\State $$A[:,:,h] = \sum_{k=1}^{d} {Q[:,:,h,k]}^{\top}K[:,:,h,k], \qquad Q[:,:,h,k],K[:,:,h,k] \in \mathbf R^{L\times L}$$
\State $A\leftarrow \mathrm{Softmax}(A+B)$
\State $$O[:,:,h,k] = A[:,:,h]\,{V[:,:,h,k]}^{\top}, \qquad A[:,:,h],V[:,:,h,k] \in \mathbf R^{L\times L},\ \ O[:,:,h,k] \in \mathbf R^{L\times L}$$
\State $O\leftarrow g\odot O$
\State $z'\leftarrow W_o O$
\Statex
\If{\textsc{IsRow}}
    \Statex $z'\leftarrow {z'}^{\top}$
\EndIf
\Statex
\Statex
\Return $z'$\;
\end{algorithmic}
\end{algorithm}
\begin{algorithm}[H]
\label{alg:symmetricbiasedaxialattention}
\caption{SymmetricBiasedAxialAttention (symmetry-aware variant of BiasedAxialAttention)}
\begin{algorithmic}[1]
\Statex \textbf{Input:}
\Statex \quad Block-row pair $\mathbf{z}\in\mathbf R^{L_{can}\times L\times d_z}$, bias $\mathbf{b}\in\mathbf R^{L_{can}\times L}$
\Statex \quad Group order $S$ with $L = S\,L_{can}$
\Statex \quad Permutation table $P\in\mathbf Z^{S\times S}$, where $P[g,h] = \mathrm{idx}(g^{-1}h)$
\Statex \textbf{Output:} Updated restricted pair $z'\in\mathbf R^{L_{can}\times L\times d_z}$
\Statex
\Statex \emph{// Per-block transpose for row attention.}
\If{\textsc{IsRow}}
    \For{$\beta = 0,\ldots,S-1$}
        \State $z[:,\,\beta L_{can}{:}(\beta{+}1)L_{can},\,:] \leftarrow z[:,\,P[\beta,0]\,L_{can}{:}(P[\beta,0]{+}1)L_{can},\,:]^{\top}$
        \State $b[:,\,\beta L_{can}{:}(\beta{+}1)L_{can}] \leftarrow b[:,\,P[\beta,0]\,L_{can}{:}(P[\beta,0]{+}1)L_{can}]^{\top}$
    \EndFor
\EndIf
\Statex
\State $z \leftarrow \mathrm{LayerNorm}(z)$;\quad $b \leftarrow \mathrm{LayerNorm}(b)$
\State $Q \leftarrow W_q z,\ K \leftarrow W_k z,\ V \leftarrow W_v z,\ B \leftarrow W_b b$ \quad\Comment{shapes $(L_{can}, L, h, d)$}
\State $g \leftarrow \mathrm{sigm}(W_g z)$
\State $Q \leftarrow Q/\sqrt{d}$;\quad $K \leftarrow K/\sqrt{L}$
\Statex
\Statex \emph{// Split along the column axis into $S$ group-blocks.}
\State $\{Q^{(s)}\}_{s=0}^{S-1} \leftarrow \mathrm{SplitAlongColumns}(Q,\,n_{\text{splits}} = S)$;
\State $\{K^{(s)}\}_{s=0}^{S-1} \leftarrow \mathrm{SplitAlongColumns}(K,\,n_{\text{splits}} = S)$;
\State $\{V^{(s)}\}_{s=0}^{S-1} \leftarrow \mathrm{SplitAlongColumns}(V,\,n_{\text{splits}} = S)$;
\Statex \Comment{each block $\in\mathbf R^{L_{can}\times L_{can}\times h\times d}$}
\Statex
\Statex \emph{// Compute attention blockwise with permutation table for phantoms $\rightarrow$ canonical only.}
\State $A^{(s)}[:,:,h] \leftarrow \sum_{k=1}^{d}\sum_{g=0}^{S-1}\, Q^{(P[g,0])}[:,:,h,k]^{\top}\, K^{(P[g,s])}[:,:,h,k],\quad s = 0,\ldots,S-1$
\State $A[:,:,h] \leftarrow \bigl[\, A^{(0)}[:,:,h]\,\Vert\,\cdots\,\Vert\,A^{(S-1)}[:,:,h]\,\bigr] \in \mathbf R^{L_{can}\times L}$
\State $A \leftarrow \mathrm{Softmax}(A + B)$
\State $\{A^{(s)}\}_{s=0}^{S-1} \leftarrow \mathrm{SplitAlongColumns}(A,\,n_{\text{splits}} = S)$ \quad\Comment{each $A^{(s)}[:,:,h] \in \mathbf R^{L_{can}\times L_{can}}$}
\Statex
\Statex \emph{// Assemble Output by gathering V blocks.}
\If{\textsc{IsRow}}
    \State $O^{(b_i)}[:,:,h,k] \leftarrow \sum_{k_j=0}^{S-1} A^{(k_j)}[:,:,h]\,V^{(P[P[b_i,0],\,k_j])}[:,:,h,k]^{\top},\quad b_i = 0,\ldots,S-1$
\Else
    \State $O^{(c_b)}[:,:,h,k] \leftarrow \sum_{k_j=0}^{S-1} A^{(k_j)}[:,:,h]\,V^{(P[c_b,\,k_j])}[:,:,h,k]^{\top},\quad c_b = 0,\ldots,S-1$
\EndIf
\State $O[:,:,h,k] \leftarrow \bigl[\, O^{(0)}[:,:,h,k]\,\Vert\,\cdots\,\Vert\,O^{(S-1)}[:,:,h,k]\,\bigr] \in \mathbf R^{L_{can}\times L}$
\Statex
\State $O \leftarrow g \odot O$
\State $z' \leftarrow W_o\,O$
\Statex
\State \Return $z'$
\end{algorithmic}
\end{algorithm}

\section{Implementations of Symmetry Transformations}

We apply the group symmetry operator during generation as a left matrix multiplication with the group elements' rotation matrices:
\begin{equation}
G\cdot T_n = \left[(R_{g_1}\cdot r_n, R_{g_1} \cdot \mathbf{x}_n),\ldots, (R_{g_k}\cdot r_n, R_{g_k} \cdot \mathbf{x}_n)\right] 
\label{eq:standard-symm}
\end{equation} 

As denoted before, let $|G|=k$. For cyclic symmetries ($C_n$, $k=n$) the rotation matrices $R_{g_{i}}$ are rotations of $\frac{2\pi i}{k}$ about $z$.

For dihedral ($D_n$, $k=2n$) they are the cyclic rotations above, composed with a $\pi$ rotation about $x$.

For tetrahedral ($T$, $k=12$) the rotation matrices $R_{g_i}$ comprise
the identity; eight rotations of $\pm 2\pi/3$ about the four body-diagonal
axes $(\pm 1, \pm 1, \pm 1)/\sqrt{3}$ (with an even number of minus signs);
and three rotations of $\pi$ about the coordinate axes.

For octahedral ($O$, $k=24$) the rotation matrices $R_{g_i}$ comprise
the identity; nine rotations of $\pm \pi/2$ and $\pi$ about the three
coordinate axes; eight rotations of $\pm 2\pi/3$ about the four
body-diagonal axes $(\pm 1, \pm 1, \pm 1)/\sqrt{3}$; and six rotations of
$\pi$ about the face-diagonal axes $(\pm 1, \pm 1, 0)/\sqrt{2}$ and cyclic
permutations.

For icosahedral ($I$, $k=60$) the rotation matrices $R_{g_i}$ comprise
the identity, twenty-four rotations of $\pm 2\pi/5$ and $\pm 4\pi/5$ about
the six axes through opposite vertices, located at cyclic permutations of
$(0, \pm 1, \pm \varphi)$ where $\varphi = (1+\sqrt 5)/2$, twenty rotations of $\pm 2\pi/3$ about the ten axes through opposite
face centers; and fifteen rotations of $\pi$ about the fifteen axes through
opposite edge midpoints.

In practice, we find an alternative symmetric operation that parameterizes the symmetry operation as a rotation about an offset center to yield more valid results for Framediff and Foldflow.  

\begin{equation}
G\cdot T_n = \left[(R_{g_1}\cdot r_n, R_{g_1} \cdot (\mathbf{x}_n-t)+t),\ldots, (R_{g_k}\cdot r_n, R_{g_k} \cdot (\mathbf{x}_n-t)+t)\right] 
\label{eq:offset-symm}
\end{equation} 

where $t \in\mathbb{R}^3$ is a translation vector placing the canonical subunit at distance $\|t\|$ from the symmetry axis. For the $C_{k}$ case this causes the the $k$ subunits to initialize on a ring of radius $\|t\|$ rather than all at the origin. 

We choose $T$ so that adjacent subunits at $t=0$ are separated by the 
characteristic spread of the initialization distribution. Two subunits 
related by $R_k$ and placed at distance $r$ from the origin are separated 
by chord length $2r\sin(\pi/k)$; setting this equal to $4\sigma$ and applying a cube-root correction for subunit 
volume gives
\begin{equation}
\|t\| = \frac{2\sigma}{sin(\frac{\pi}{k})} \cdot 
        \sqrt[3]{\tfrac{N}{50}}
\label{eq:prior-radius}
\end{equation}
where $\sigma$ is the standard deviation of the translational frame 
initialization and $N$ is the subunit length.

\textbf{Justification.}
For our $\mathrm{SE}(3)$ diffusion models, at $t=0$, frame 
translations are drawn from an isotropic Gaussian centered at the origin with 
scale $\sigma$. Applying \ref{eq:standard-symm} results in all $k$ symmetric copies 
of the canonical subunit initializing on top of one another. The 
denoising process must push these copies apart over the course of generation. Empirically, we find this 
manifests as a sharp drop in designability for $k \geq 6$: the model exhausts 
its denoising budget on spatial separation, and in some cases, cannot separate chains at all, leading to highly clashing designs. This is especially pronounced for models not pretrained on multi-chain generation such as Framediff and FoldFlow. The application of this offset symmeterization constrains the trajectory's center of mass to lie at radius $\|t\|$ from the axis, freeing the model to spend its capacity on local geometry.

\section{Experimental details}
\label{app:experimental}

We benchmark on cyclic symmetries $C_2, C_3, C_4, C_6, C_8, C_{12}$ with subunit length $n = 50$, yielding total complex sizes from $100$ to $600$ residues. The choice of $n = 50$ balances designability against known out of distribution behavior of FrameDiff and FoldFlow above $\sim 300$ residues at the full-complex level. All experiments are run on a single NVIDIA A100 (40\,GB VRAM).

We use the FrameDiff paper checkpoint at \url{https://github.com/Immortals-33/Framediff/tree/master}, FoldFlow-Base \url{https://github.com/DreamFold/FoldFlow/releases/download/0.1.0/foldflow-base.pth}, and the RFdiffusion base checkpoint \url{http://files.ipd.uw.edu/pub/RFdiffusion/6f5902ac237024bdd0c176cb93063dc4/Base_ckpt.pt}. 

Each is run with the published default hyperparameters. FrameDiff with $500$ diffusion steps, translation schedule $0.1 \to 20$, and IGSO(3) schedule $0.1 \to 1.5$; FoldFlow with translation schedule $0.01 \to 20$ and IGSO(3) schedule $0.01 \to 1.5$; and RFdiffusion with $50$ diffusion steps and $\beta$ schedule $0.01 \to 0.07$.

For FrameDiff and FoldFlow full-complex inference, we preserve within-chain residue indexes and assign an inter-chain separation offset of $1000$. For RFdiffusion full-complex inference we maintain the native inter-chain separation offsets of $\pm L$ where $L$ is the total number of residues in the complex. By definition, any sign on the chain offset disrupts chain-permutation equivariance. 
For RFdiffusion ablation and \ours{} inference, all chains instead share identical sequential indices. Interchain offsets are set to $L$ for RFdiffusion.

For RFdiffusion we centre the diffusion process about the whole-complex. At each step the model receives the entire $kN$-residue complex; we extract the predicted $p(\mathbf{x}_0)$ for the canonical subunit, apply the reverse step, and propagate the update to all phantom subunits via the standard rotational symmetry operation~\ref{eq:standard-symm}, with the reverse trajectory centered on the full complex.

For Framediff and FoldFlow we use the offset symmetry parameterization~\ref{eq:offset-symm} with prior radius determined using equation~\ref{eq:prior-radius}.

Each generated design's sequence is predicted with ProteinMPNN at sampling temperature $0.1$, using tied positions across symmetric subunits, followed by refolding prediction with ESMFold v1 in multimer mode. 
For FrameDiff and FoldFlow, ESMFold prediction is restricted to the canonical subunit plus its two nearest phantom neighbors. 

Novelty against the PDB is computed with Foldseek~\citep{van2024fast} with \texttt{foldseek easy-search} at the subunit level and \texttt{foldseek easy-multimersearch -e inf} at the complex level.

\section{Additional Experimental Metrics}
\label{app:additionalmetrics}
\begin{table}[t]
\centering
\scriptsize
\setlength{\tabcolsep}{3pt}
\renewcommand{\arraystretch}{1.15}
\caption{Designability (frac. designable, scRMSD), Diversity ($TM_{complex}$), and Novelty ($pdbTM_{complex}$) results for cyclic generation for FrameDiff, FoldFlow, and RFdiffusion. Row labels $C_k$ denote symmetry group with subunit length $n = 50$. For each metric, paired columns report the full-complex baseline and \ours{}. All reported uncertainties are 1-sigma standard deviations.}
\label{tab:all_combined}
\begin{tabular}{ll cc cc cc cc}
\toprule
 & & \multicolumn{2}{c}{Designability ($\uparrow$)} & \multicolumn{2}{c}{mean best scRMSD ($\downarrow$)} & \multicolumn{2}{c}{$\mathrm{diversityTM}_{\text{complex}}$ ($\downarrow$)} & \multicolumn{2}{c}{$\mathrm{pdbTM}_{\text{complex}}$ ($\downarrow$)} \\
\cmidrule(lr){3-4}\cmidrule(lr){5-6}\cmidrule(lr){7-8}\cmidrule(lr){9-10}
model & symm & full & \ours{} & full & \ours{} & full & \ours{} & full & \ours{} \\
\midrule
\multirow{6}{*}{FrameDiff}
 & $C_2$    & 0.680 & 0.760 & $2.026_{\pm 1.037}$ & $2.060_{\pm 0.891}$ & $0.223_{\pm 0.048}$ & $0.231_{\pm 0.054}$ & $0.157_{\pm 0.060}$ & $0.132_{\pm 0.054}$ \\
 & $C_3$    & 0.280 & 0.200 & $3.471_{\pm 0.894}$ & $3.356_{\pm 1.050}$ & $0.226_{\pm 0.052}$ & $0.202_{\pm 0.042}$ & $0.097_{\pm 0.028}$ & $0.078_{\pm 0.027}$ \\
 & $C_4$    & 0.160 & 0.220 & $3.348_{\pm 0.881}$ & $3.427_{\pm 0.935}$ & $0.237_{\pm 0.061}$ & $0.232_{\pm 0.062}$ & $0.103_{\pm 0.058}$ & $0.068_{\pm 0.048}$ \\
 & $C_6$    & 0.080 & 0.140 & $3.310_{\pm 0.892}$ & $3.413_{\pm 0.999}$ & $0.260_{\pm 0.067}$ & $0.286_{\pm 0.069}$ & $0.059_{\pm 0.044}$ & $0.044_{\pm 0.021}$ \\
 & $C_8$    & 0.200 & 0.140 & $3.028_{\pm 0.904}$ & $3.119_{\pm 0.938}$ & $0.299_{\pm 0.075}$ & $0.292_{\pm 0.091}$ & $0.047_{\pm 0.029}$ & $0.026_{\pm 0.008}$ \\
 & $C_{12}$ & 0.220 & 0.020 & $3.013_{\pm 0.954}$ & $4.154_{\pm 1.071}$ & $0.374_{\pm 0.091}$ & --                  & $0.033_{\pm 0.015}$ & 0.020 \\
\midrule
\multirow{6}{*}{FoldFlow}
 & $C_2$    & 0.840 & 0.820 & $1.838_{\pm 0.753}$ & $1.686_{\pm 0.739}$ & $0.249_{\pm 0.051}$ & $0.247_{\pm 0.041}$ & $0.126_{\pm 0.060}$ & $0.130_{\pm 0.064}$ \\
 & $C_3$    & 0.220 & 0.280 & $3.336_{\pm 0.924}$ & $3.029_{\pm 0.951}$ & $0.214_{\pm 0.053}$ & $0.215_{\pm 0.047}$ & $0.069_{\pm 0.025}$ & $0.075_{\pm 0.019}$ \\
 & $C_4$    & 0.180 & 0.120 & $2.983_{\pm 0.899}$ & $3.285_{\pm 0.712}$ & $0.219_{\pm 0.042}$ & $0.220_{\pm 0.049}$ & $0.058_{\pm 0.025}$ & $0.046_{\pm 0.022}$ \\
 & $C_6$    & 0.140 & 0.160 & $3.202_{\pm 0.777}$ & $2.964_{\pm 0.890}$ & $0.263_{\pm 0.064}$ & $0.263_{\pm 0.077}$ & $0.054_{\pm 0.038}$ & $0.040_{\pm 0.018}$ \\
 & $C_8$    & 0.140 & 0.120 & $2.957_{\pm 0.714}$ & $3.062_{\pm 0.734}$ & $0.320_{\pm 0.084}$ & $0.283_{\pm 0.068}$ & $0.021_{\pm 0.007}$ & $0.035_{\pm 0.019}$ \\
 & $C_{12}$ & 0.140 & 0.160 & $2.922_{\pm 0.863}$ & $2.947_{\pm 0.654}$ & $0.395_{\pm 0.079}$ & $0.368_{\pm 0.108}$ & $0.017_{\pm 0.012}$ & $0.018_{\pm 0.006}$ \\
\midrule
\multirow{6}{*}{RFdiffusion}
 & $C_2$ & 0.970 & 0.955 & $1.036_{\pm 0.776}$ & $0.889_{\pm 0.679}$ & $0.252_{\pm 0.077}$ & $0.260_{\pm 0.080}$ & $0.037_{\pm 0.019}$ & $0.038_{\pm 0.018}$ \\
 & $C_3$ & 0.820 & 0.760 & $1.694_{\pm 1.367}$ & $1.758_{\pm 1.445}$ & $0.229_{\pm 0.081}$ & $0.225_{\pm 0.073}$ & $0.034_{\pm 0.019}$ & $0.031_{\pm 0.015}$ \\
 & $C_4$ & 0.670 & 0.545 & $2.429_{\pm 1.712}$ & $2.722_{\pm 1.591}$ & $0.255_{\pm 0.096}$ & $0.243_{\pm 0.088}$ & $0.024_{\pm 0.016}$ & $0.026_{\pm 0.019}$ \\
 & $C_6$ & 0.570 & 0.472 & $3.534_{\pm 2.046}$ & $3.984_{\pm 1.828}$ & $0.259_{\pm 0.106}$ & $0.244_{\pm 0.090}$ & $0.017_{\pm 0.011}$ & $0.017_{\pm 0.012}$ \\
 & $C_8$ & 0.485 & 0.465 & $4.703_{\pm 1.975}$ & $5.105_{\pm 1.747}$ & $0.271_{\pm 0.098}$ & $0.268_{\pm 0.103}$ & $0.013_{\pm 0.007}$ & $0.012_{\pm 0.007}$ \\
 & $C_{12}$ & 0.510 & 0.400 & $5.681_{\pm 1.865}$ & $5.914_{\pm 1.921}$ & $0.304_{\pm 0.114}$ & $0.306_{\pm 0.123}$ & $0.010_{\pm 0.008}$ & $0.006_{\pm 0.003}$ \\
\bottomrule
\end{tabular}
\end{table}

\begin{figure}[H]
    \centering
    \includegraphics[width=1.0\linewidth]{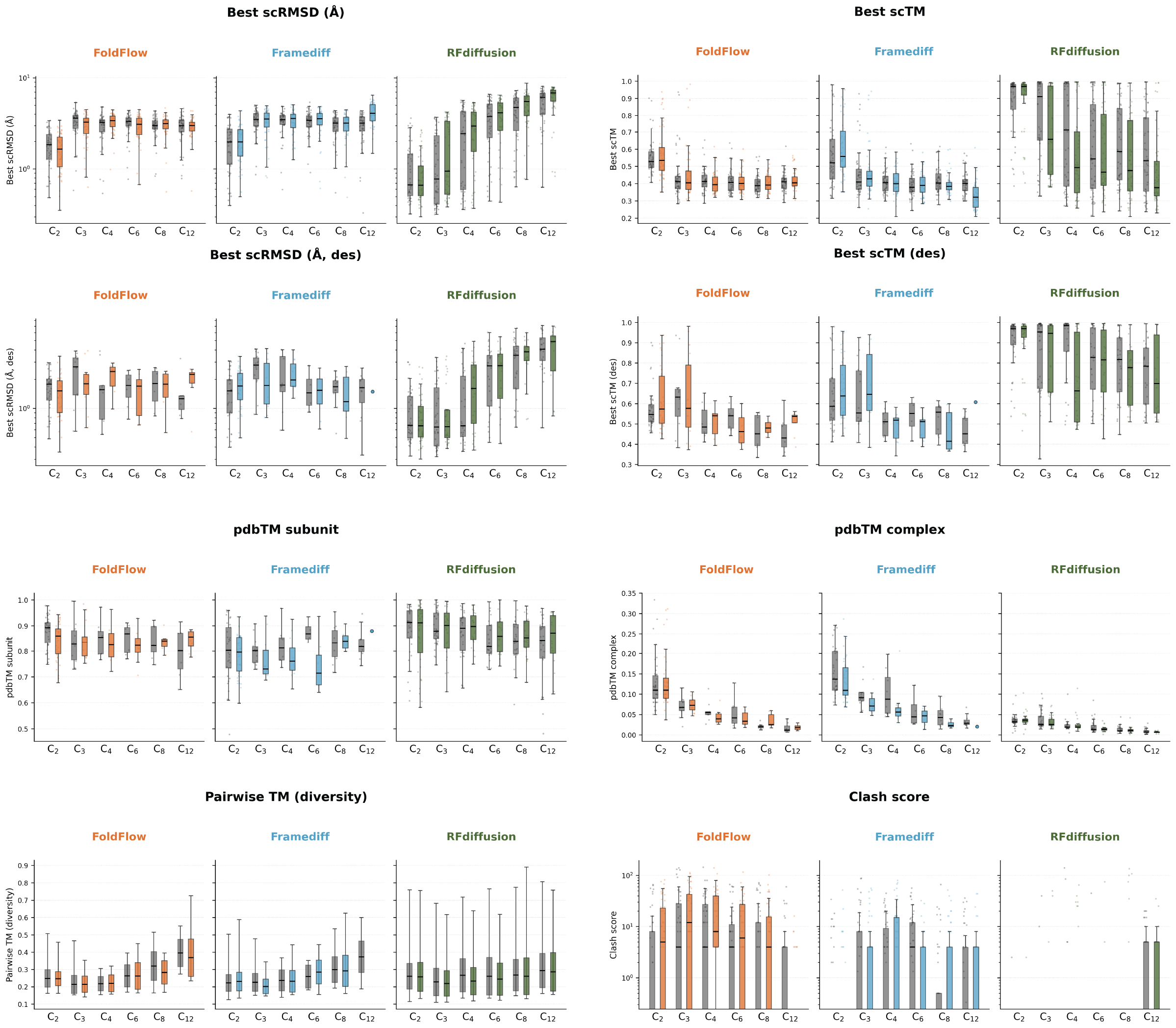}
    \caption{Boxplot visualization of Table~\ref{tab:all_combined}. Distributional metrics for Cyclic generation with subunit length=50 with ZeUS, for all models. Grey box is the baseline full-complex inference, while coloured is \ours{}. \ours{} achieves equivalent performance across all metrics.}
    \label{fig:cyclic_metrics}
\end{figure}

\begin{table}[H]
\centering
\scriptsize
\setlength{\tabcolsep}{3pt}
\renewcommand{\arraystretch}{1.15}
\caption{FoldFlow -- DSSP secondary-structure fractions. Each cell shows full-complex / \ours{} as mean $\pm$ std.}
\label{tab:foldflow_ss}
\begin{tabular}{llll}
\toprule
setting & helix & sheet & coil \\
\midrule
C2\_100 & 0.863 $\pm$ 0.036 / 0.859 $\pm$ 0.041 & 0 $\pm$ 0 / 0 $\pm$ 0 & 0.137 $\pm$ 0.036 / 0.141 $\pm$ 0.041 \\
C3\_150 & 0.870 $\pm$ 0.064 / 0.852 $\pm$ 0.049 & 0.006 $\pm$ 0.042 / 0 $\pm$ 0 & 0.124 $\pm$ 0.033 / 0.148 $\pm$ 0.049 \\
C4\_200 & 0.871 $\pm$ 0.031 / 0.864 $\pm$ 0.039 & 0 $\pm$ 0 / 0 $\pm$ 0 & 0.129 $\pm$ 0.031 / 0.136 $\pm$ 0.039 \\
C6\_300 & 0.870 $\pm$ 0.034 / 0.851 $\pm$ 0.042 & 0 $\pm$ 0 / 0 $\pm$ 0 & 0.130 $\pm$ 0.034 / 0.149 $\pm$ 0.042 \\
C8\_400 & 0.858 $\pm$ 0.088 / 0.864 $\pm$ 0.040 & 0.007 $\pm$ 0.048 / 0 $\pm$ 0 & 0.135 $\pm$ 0.053 / 0.136 $\pm$ 0.040 \\
C12\_600 & 0.872 $\pm$ 0.046 / 0.860 $\pm$ 0.046 & 0 $\pm$ 0 / 0.002 $\pm$ 0.011 & 0.128 $\pm$ 0.046 / 0.138 $\pm$ 0.045 \\
T\_600 & 0.883 $\pm$ 0.030 / 0.866 $\pm$ 0.041 & 0 $\pm$ 0 / 0 $\pm$ 0 & 0.117 $\pm$ 0.030 / 0.134 $\pm$ 0.041 \\
O\_1200 & 0.879 $\pm$ 0.042 / 0.870 $\pm$ 0.040 & 0 $\pm$ 0 / 0 $\pm$ 0 & 0.121 $\pm$ 0.042 / 0.130 $\pm$ 0.040 \\
\bottomrule
\end{tabular}
\end{table}
\begin{table}[H]
\centering
\scriptsize
\setlength{\tabcolsep}{3pt}
\renewcommand{\arraystretch}{1.15}
\caption{Framediff -- DSSP secondary-structure fractions. Each cell shows full-complex / \ours{} as mean $\pm$ std.}
\label{tab:framediff_ss}
\begin{tabular}{llll}
\toprule
setting & helix & sheet & coil \\
\midrule
C2\_100 & 0.679 $\pm$ 0.251 / 0.736 $\pm$ 0.174 & 0.070 $\pm$ 0.155 / 0.039 $\pm$ 0.121 & 0.251 $\pm$ 0.142 / 0.225 $\pm$ 0.086 \\
C3\_150 & 0.735 $\pm$ 0.206 / 0.723 $\pm$ 0.204 & 0.059 $\pm$ 0.131 / 0.057 $\pm$ 0.123 & 0.206 $\pm$ 0.106 / 0.220 $\pm$ 0.120 \\
C4\_200 & 0.785 $\pm$ 0.172 / 0.714 $\pm$ 0.219 & 0.030 $\pm$ 0.098 / 0.021 $\pm$ 0.067 & 0.184 $\pm$ 0.091 / 0.265 $\pm$ 0.196 \\
C6\_300 & 0.700 $\pm$ 0.254 / 0.719 $\pm$ 0.178 & 0.059 $\pm$ 0.136 / 0.036 $\pm$ 0.086 & 0.242 $\pm$ 0.183 / 0.245 $\pm$ 0.141 \\
C8\_400 & 0.770 $\pm$ 0.201 / 0.712 $\pm$ 0.237 & 0.049 $\pm$ 0.123 / 0.067 $\pm$ 0.159 & 0.180 $\pm$ 0.096 / 0.221 $\pm$ 0.108 \\
C12\_600 & 0.747 $\pm$ 0.217 / 0.705 $\pm$ 0.199 & 0.049 $\pm$ 0.123 / 0.044 $\pm$ 0.124 & 0.204 $\pm$ 0.111 / 0.251 $\pm$ 0.125 \\
T\_600 & 0.836 $\pm$ 0.140 / 0.712 $\pm$ 0.213 & 0.019 $\pm$ 0.099 / 0.049 $\pm$ 0.135 & 0.145 $\pm$ 0.065 / 0.239 $\pm$ 0.123 \\
O\_1200 & 0.846 $\pm$ 0.081 / 0.713 $\pm$ 0.176 & 0.004 $\pm$ 0.031 / 0.042 $\pm$ 0.114 & 0.150 $\pm$ 0.069 / 0.244 $\pm$ 0.101 \\
\bottomrule
\end{tabular}
\end{table}
\begin{table}[H]
\centering
\scriptsize
\setlength{\tabcolsep}{3pt}
\renewcommand{\arraystretch}{1.15}
\caption{RFdiffusion -- DSSP secondary-structure fractions. Each cell shows full-complex / \ours{} as mean $\pm$ std.}
\label{tab:rfdiffusion_ss}
\begin{tabular}{llll}
\toprule
setting & helix & sheet & coil \\
\midrule
C2\_100 & 0.773 $\pm$ 0.214 / 0.770 $\pm$ 0.208 & 0.082 $\pm$ 0.152 / 0.088 $\pm$ 0.154 & 0.145 $\pm$ 0.079 / 0.142 $\pm$ 0.065 \\
C3\_150 & 0.665 $\pm$ 0.303 / 0.656 $\pm$ 0.270 & 0.160 $\pm$ 0.219 / 0.163 $\pm$ 0.204 & 0.175 $\pm$ 0.100 / 0.180 $\pm$ 0.116 \\
C4\_200 & 0.642 $\pm$ 0.252 / 0.619 $\pm$ 0.315 & 0.163 $\pm$ 0.189 / 0.178 $\pm$ 0.215 & 0.196 $\pm$ 0.116 / 0.204 $\pm$ 0.134 \\
C6\_300 & 0.432 $\pm$ 0.248 / 0.550 $\pm$ 0.273 & 0.294 $\pm$ 0.191 / 0.216 $\pm$ 0.165 & 0.274 $\pm$ 0.111 / 0.234 $\pm$ 0.153 \\
C8\_400 & 0.459 $\pm$ 0.283 / 0.441 $\pm$ 0.273 & 0.310 $\pm$ 0.212 / 0.315 $\pm$ 0.189 & 0.231 $\pm$ 0.118 / 0.244 $\pm$ 0.136 \\
C12\_600 & 0.346 $\pm$ 0.273 / 0.297 $\pm$ 0.273 & 0.382 $\pm$ 0.210 / 0.444 $\pm$ 0.193 & 0.273 $\pm$ 0.123 / 0.259 $\pm$ 0.124 \\
T\_600 & 0.723 $\pm$ 0.261 / 0.670 $\pm$ 0.299 & 0.119 $\pm$ 0.178 / 0.131 $\pm$ 0.231 & 0.158 $\pm$ 0.110 / 0.199 $\pm$ 0.117 \\
O\_1200 & 0.810 $\pm$ 0.107 / 0.750 $\pm$ 0.139 & 0.020 $\pm$ 0.043 / 0.041 $\pm$ 0.075 & 0.170 $\pm$ 0.088 / 0.209 $\pm$ 0.117 \\
\bottomrule
\end{tabular}
\end{table}

\section{Polyhedral Generation Experiments}
\label{app:polyhedral}

The self-consistency evaluation is uninformative for polyhedral symmetries (we restrict ourselves to tetrahedral, octahedral, and icosahedral when we mention polyhedral as they are the representative set.) Folding models routinely fail on these complexes even at small sizes~\citep{bryant2022predicting}. 
A failure to refold cannot distinguish between poor design from folding-model failure. Therefore we restrict the analysis of polyhedral generations to distributional comparisons only. We seek to determine equivalence of validity for \ours{} using proxy metrics for designability.  

\paragraph{Setup.}
For tetrahedral ($T$) and octahedral ($O$) symmetries, we generate $50$ designs at subunit length $n = 50$ for each model (FrameDiff, FoldFlow, RFdiffusion), using full-complex baseline and \ours{} inference. Icosahedral ($I$) is omitted from this comparison because full-complex inference at $n = 50$ does not fit on a single A100. Reducing the icosahedral subunit lengths even further is uninformative and biologically implausible~\citep{lovsdorfer2013statistical} and we do not consider them. All reported uncertainties are 1-sigma standard deviations.

\paragraph{Metrics.}
Since designability cannot be directly evaluated, all metrics below are computed on the full set of $50$ generated designs rather than restricted to a designable subset as in the cyclic evaluation.

\begin{itemize}[leftmargin=*, topsep=2pt, itemsep=2pt]
  \item \textbf{Diversity and novelty.} Defined as in the main experiments (Section~\ref{sec:experiments}), but computed across all $50$ generated designs.
  \item \textbf{Radius of gyration (subunit and complex).} $\mathrm{R_g}$ is an RMS distance of a protein's atoms from its centre of mass. It is a measure of spatial extent and compactness: 
      \[
    R_g^{subunit} = \sqrt{\sum_{i\in\mathcal{C}}\|\mathbf{r}_i - \bar{\mathbf{r}}\|^2}
    \]
    where $\mathcal{C}$ is the set of backbone atoms in the canonical chain and $\bar{\mathbf{r}}$ the chain centre of mass. The complex-level $R_g^{\text{complex}}$ is computed analogously over the union of all backbone atoms with respect to the assembly center of mass.

  \item \textbf{Clash score.} Number of atom pairs separated by less than $1.5\,\text{\AA}$ and at least three residues apart in sequence.

    \item \textbf{Shell thickness.} We measure the radial extent of the cage wall between the polyhedron's inner cavity and outer surface. For backbone atoms at coordinates $\{\mathbf{r}_i\}$ in the assembly, with assembly centre of mass $\bar{\mathbf{r}}$, let $r_i = \|\mathbf{r}_i - \bar{\mathbf{r}}\|$. Then
    \[
    \text{shell thickness} = P_{99}\bigl(\{r_i\}\bigr) - P_{1}\bigl(\{r_i\}\bigr),
    \]
    where $P_q$ denotes the $q$-th percentile.
    
    \item \textbf{Buried surface area per chain.} The solvent-accessible surface area (SASA) of a structure is the total area accessible to a $1.4\,\text{\AA}$ solvent probe rolled over its van der Waals surface~\citep{shrake1973environment}. Per-chain BSA is the area of a chain's surface occluded upon assembly: $\mathrm{SASA}_{\mathrm{isolated}} - \mathrm{SASA}_{\mathrm{complex}}$, with both terms restricted to backbone atoms and computed with \texttt{biotite.structure.sasa}. Since all subunits are identical by symmetry, we compute on chain A only.
    \item \textbf{Normalized BSA.} The per-chain BSA divided by $\mathrm{SASA}_{\mathrm{isolated}}$, giving the fraction of a subunit's surface buried at the assembly interface.
    \item \textbf{pdbTM (subunit and complex).} Maximum TM-score against the PDB, returned by Foldseek at both the subunit and full-complex level.

\end{itemize}

\begin{figure}
    \centering
    \includegraphics[width=1.0\linewidth]{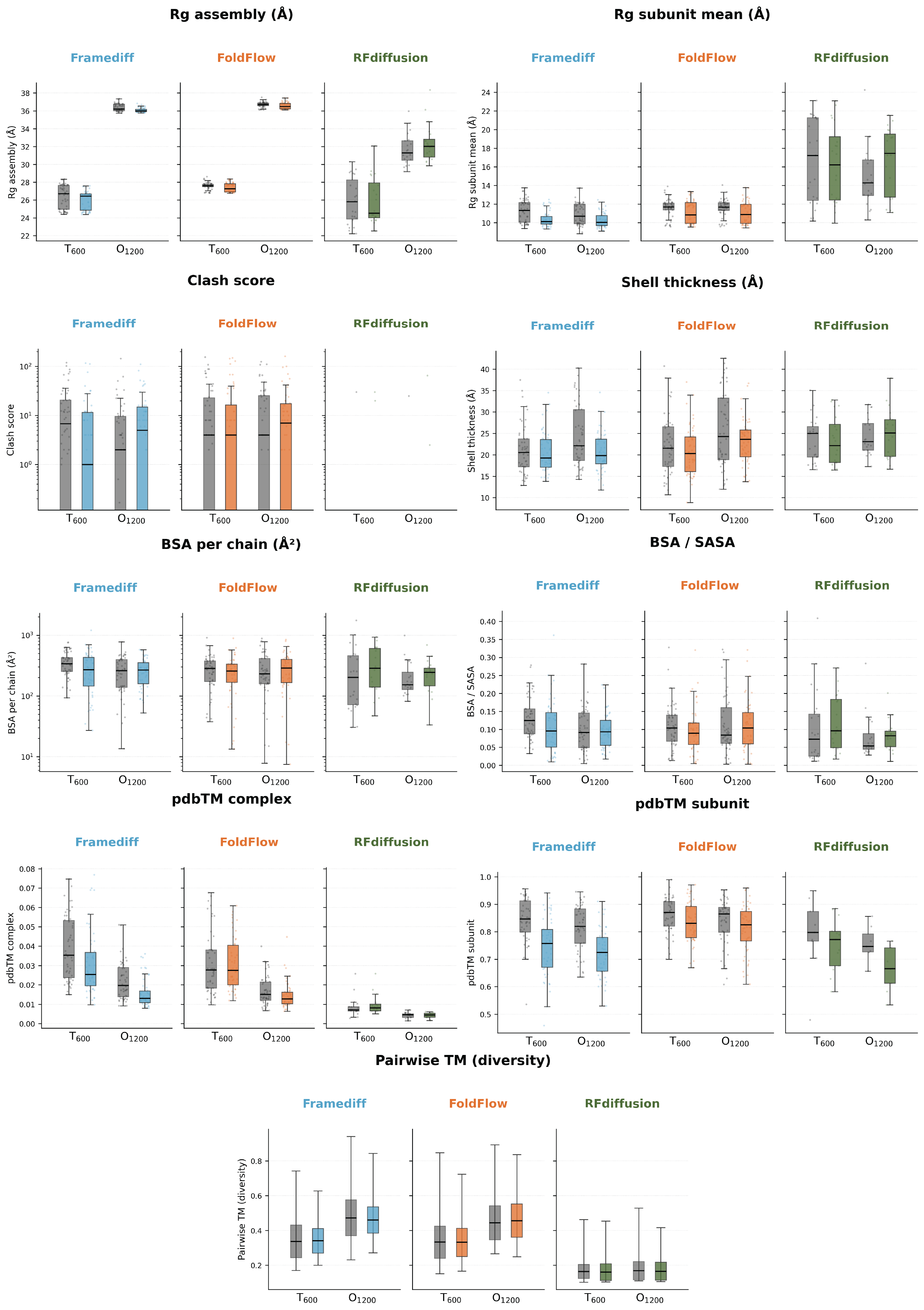}
    \caption{Distributional metrics for tetrahedral and octahedral generations under full-complex baseline and \ours{}, across all three backbones. Grey boxplot is full-complex inference, coloured is \ours{}. Symmetries are labelled with total $N_{res}$. \ours{} is distributionally equivalent.}
    \label{fig:polyhedral_metrics}
\end{figure}

Figure~\ref{fig:polyhedral_metrics} compares the full-complex baseline against \ours{} across $T$ and $O$ symmetries for all three models. On global shape metrics such as shell thickness, per-chain BSA, BSA/SASA ratio, and radius of gyration, the two methods agree closely across all three models. Novelty shows a visible distributional shift: \ours{}-generated assemblies are slightly more PDB-distinct than the baseline. This direction is favorable for de novo design. Diversity distributions are nearly identical, with a maximum between-pathway difference of $0.0124$ in TM-score (FoldFlow, octahedral). All remaining metrics are distributionally similar. We thus claim that the distributional equivalence established for cyclic symmetries in Section~\ref{sec:experiments} extends to polyhedral symmetries.

\newpage
\section{Symmetric Generation Samples}
\label{app:samples}

\begin{figure}[h!]
    \centering
    \includegraphics[width=1.0\linewidth]{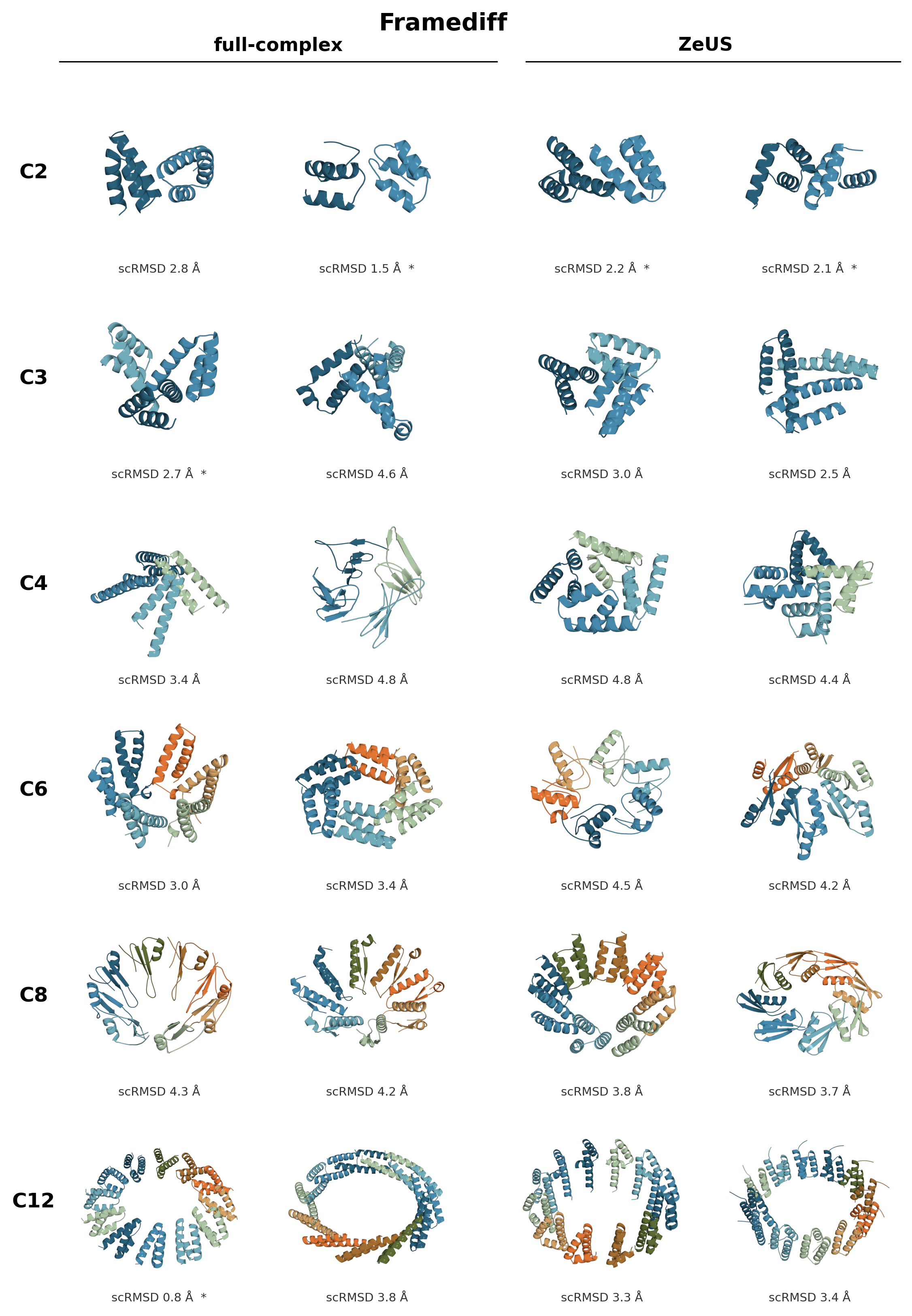}
    \caption{Cyclic samples using FrameDiff with $L_{subunit}=50$ aligned by symmetry type (row) and method (2 samples per col) comparing full-complex and \ours{} inference.}
    \label{fig:framediff}
\end{figure}

\begin{figure}[p]
    \centering
    \includegraphics[width=1.0\linewidth]{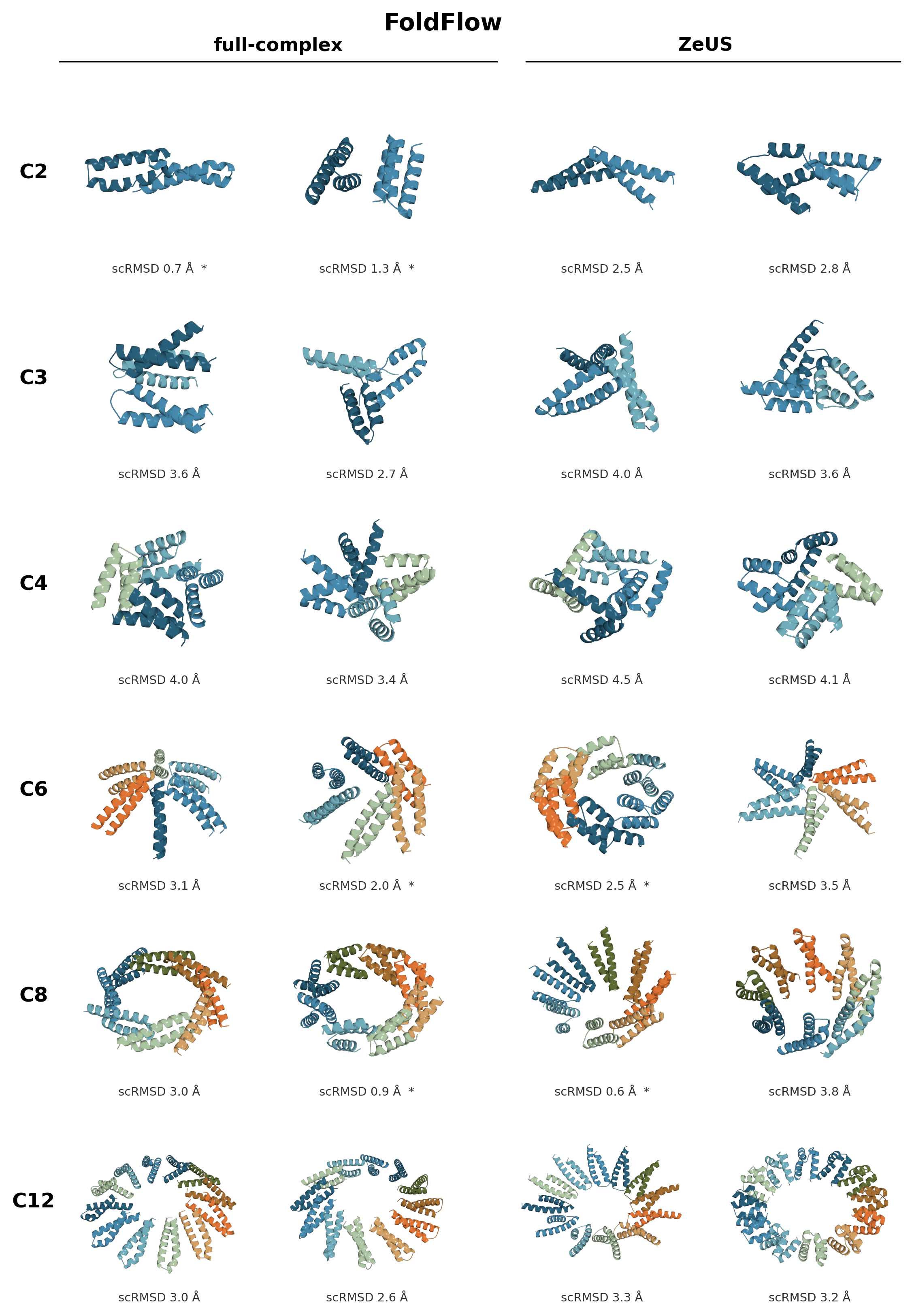}
    \caption{Cyclic samples using FoldFlow with $L_{subunit}=50$ aligned by symmetry type (row) and method (2 samples per col) comparing full-complex and \ours{} inference.}
    \label{fig:foldflow}
\end{figure}

\begin{figure}[p]
    \centering
    \includegraphics[width=1.0\linewidth]{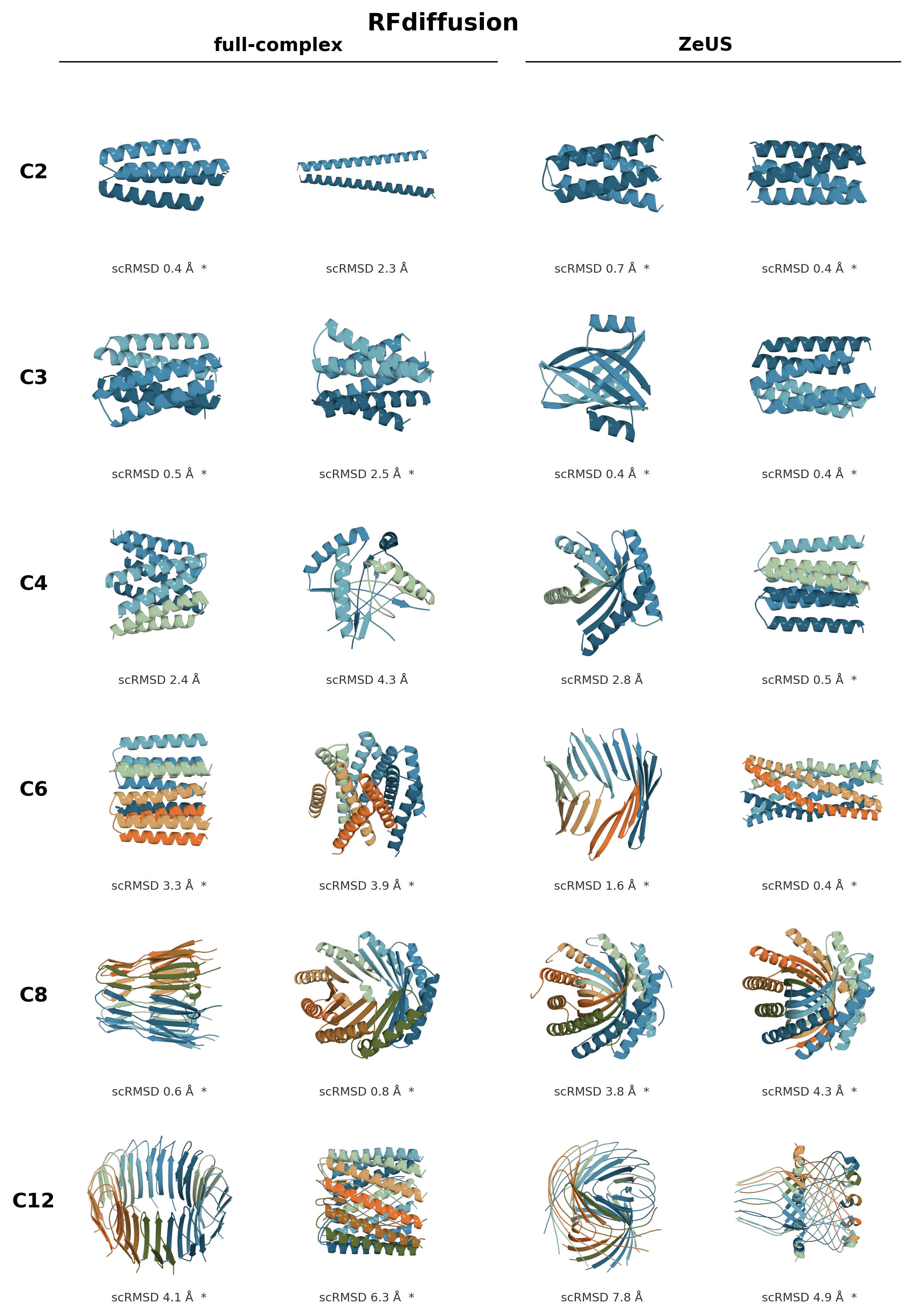}
    \caption{Cyclic samples using RFdiffusion with $L_{subunit}=50$ aligned by symmetry type (row) and method (2 samples per col) comparing full-complex and \ours{} inference.}
    \label{fig:rfdiffusion}
\end{figure}

\begin{figure}[p]
    \centering
    \includegraphics[width=1.0\linewidth]{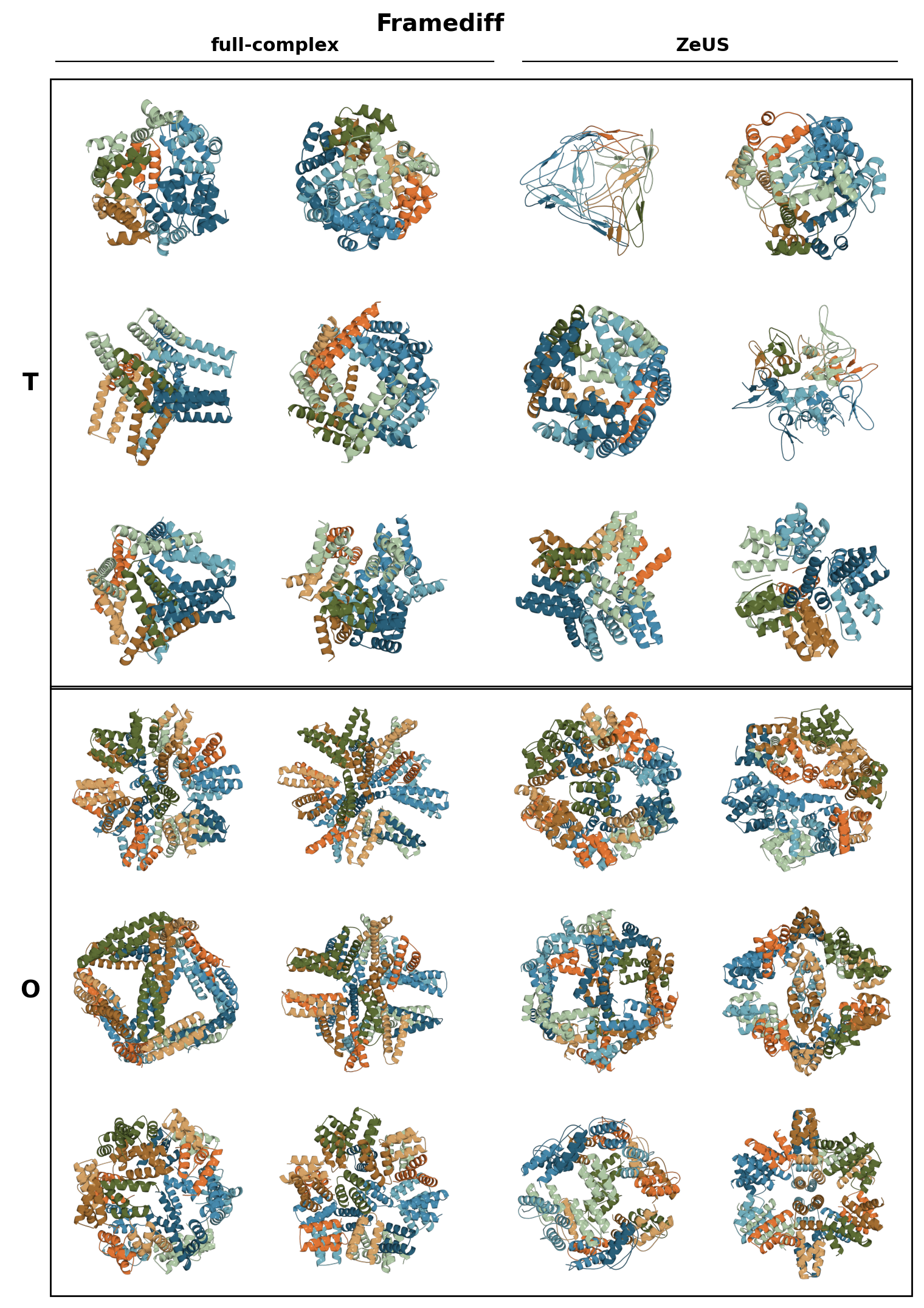}
    \caption{Polyhedral samples using FrameDiff with $L_{subunit}=50$ aligned by symmetry type (row) and method (2 samples per col) comparing full-complex and \ours{} inference.}
    \label{fig:framediff_poly}
\end{figure}

\begin{figure}[p]
    \centering
    \includegraphics[width=1.0\linewidth]{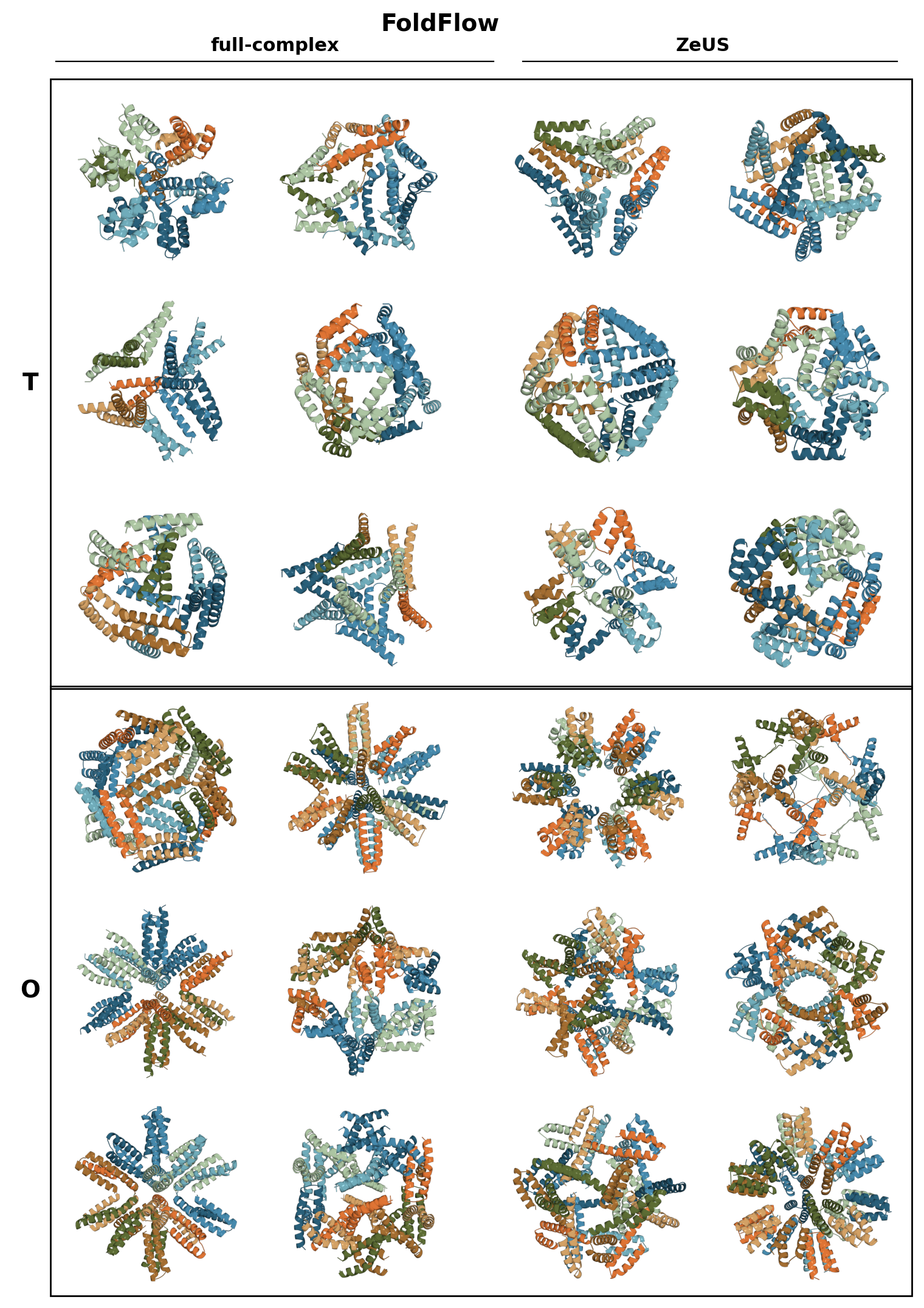}
    \caption{Polyhedral samples using FoldFlow with $L_{subunit}=50$ aligned by symmetry type (row) and method (2 samples per col) comparing full-complex and \ours{} inference.}
    \label{fig:foldflow_poly}
\end{figure}

\begin{figure}[p]
    \centering
    \includegraphics[width=1.0\linewidth]{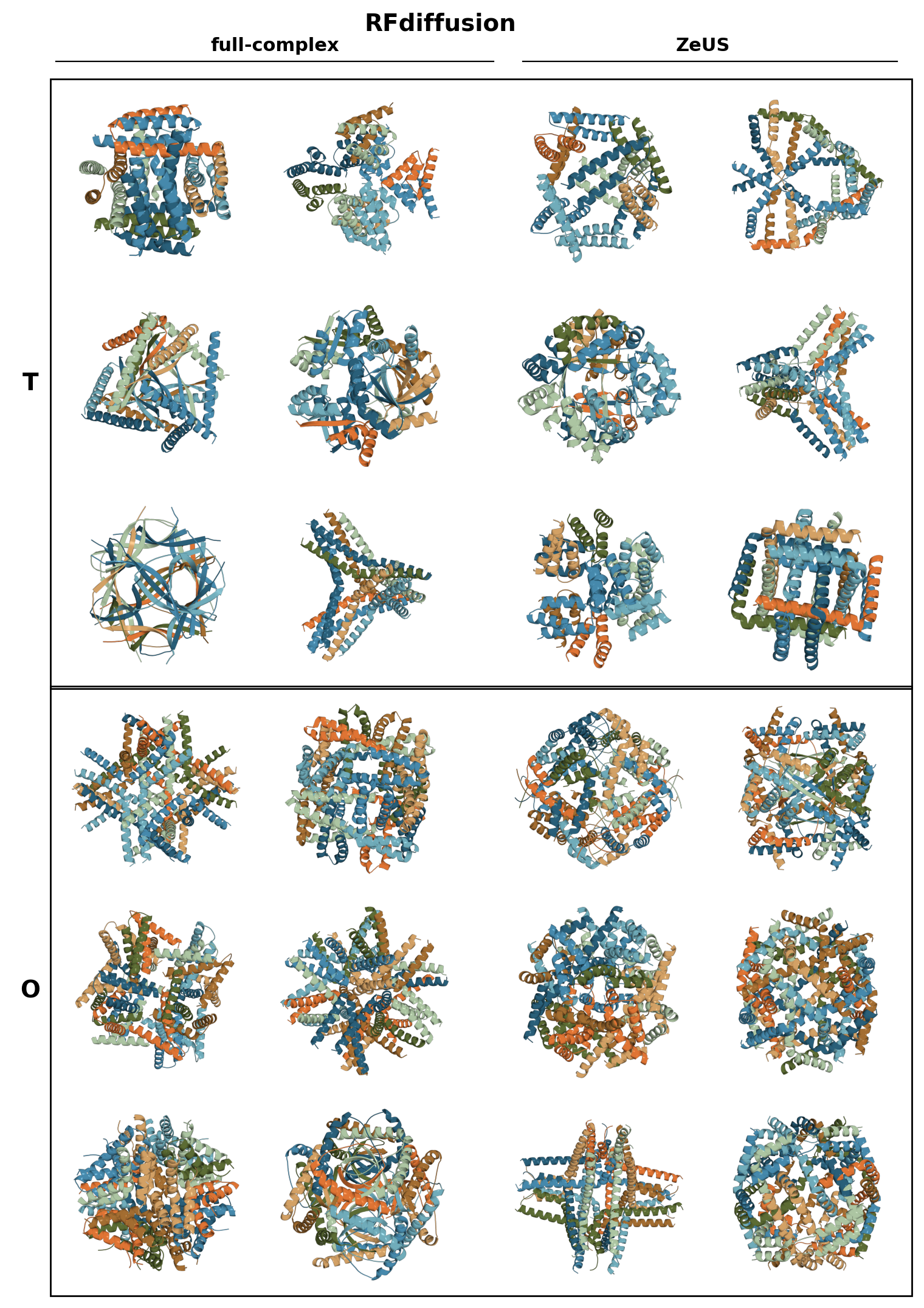}
    \caption{Polyhedral samples using RFdiffusion with $L_{subunit}=50$ aligned by symmetry type (3 samples per row) and method (2 samples per col) comparing full-complex and \ours{} inference.}
    \label{fig:rfdiffusion_poly}
\end{figure}

\newpage
\section{Large Scale Generations}
\label{app:closest_pdb}

We run \ours{} on FrameDiff with the following configurations: $C_{50}$ at subunit length $200$ and radius $225.3\,\text{\AA}$; $D_{50}$ at subunit length $100$, radius $225.3\,\text{\AA}$, and ring offset $28\,\text{\AA}$; helical ($H$) at subunit length $100$, radius $79.875\,\text{\AA}$, twist $0.31$, and rise $1.30\,\text{\AA}$; and icosahedral ($I$) at subunit length $100$, outer radius $57\,\text{\AA}$ with $C_5$-axis radius $20\,\text{\AA}$. Radius is defined as the median $\mathrm{C}_\alpha$ distance to the symmetry axis.

For each generated complex, we retrieve the closest matching natural assemblies via the RCSB Search v2 API, restricted to homomers of the matching symmetry class and filtered by minimum chain count: $\geq 20$ chains for cyclic ($n=268$), $\geq 40$ for dihedral ($n=40$), $\geq 30$ for helical ($n=267$), and $\geq 60$ for icosahedral ($n=797$).

Backbone-only buried surface area is computed as $\mathrm{SASA}_{\mathrm{isolated}} - \mathrm{SASA}_{\mathrm{complex}}$, with both terms restricted to backbone atoms. We compute distributions with $n=10$ generations for each symmetry. 

\begin{figure}[h!]
    \centering
    \includegraphics[width=1.0\linewidth]{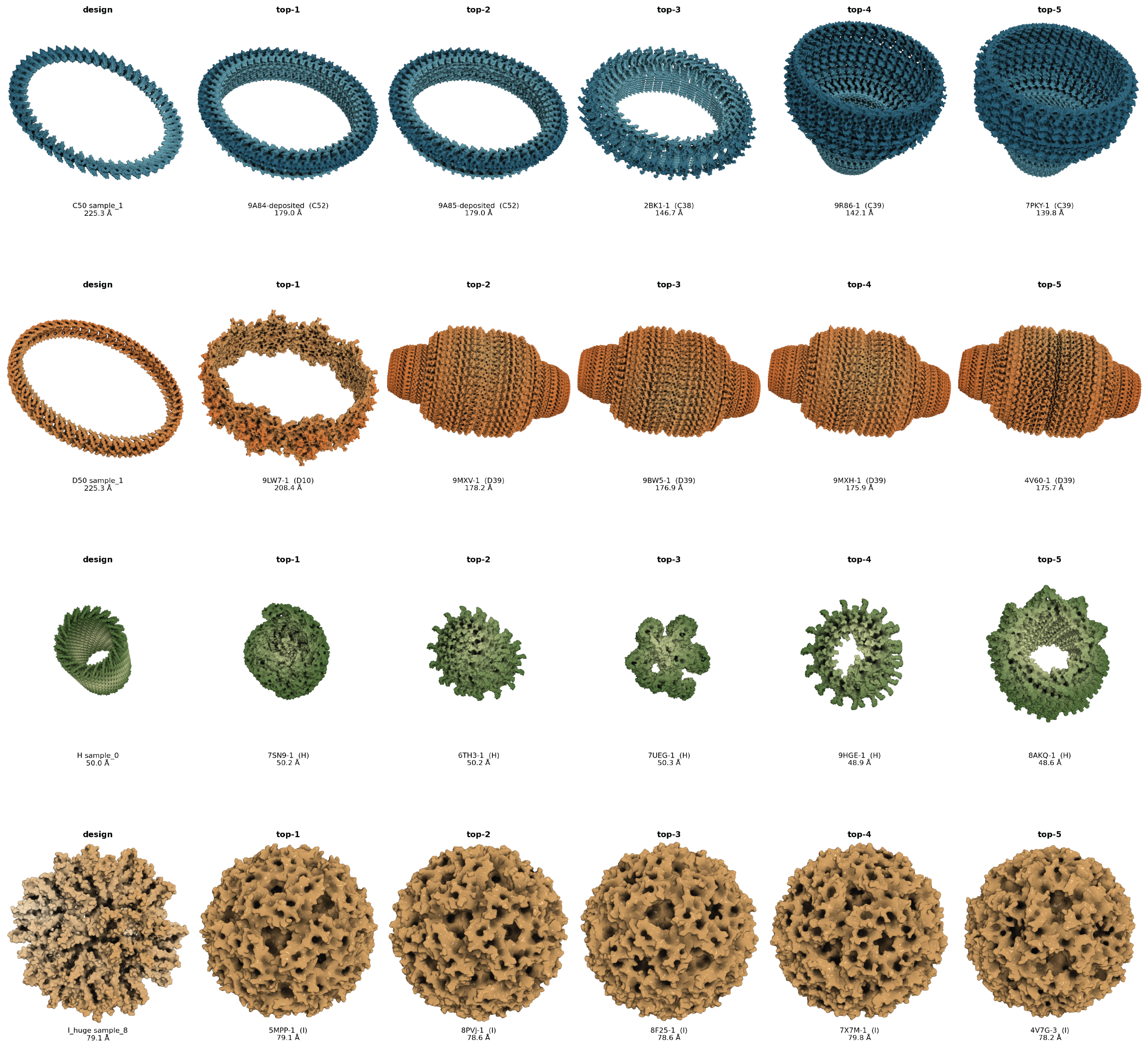}
    \caption{Top $k = 5$ most similar symmetric complexes retrieved from the PDB compared against \ours{} generations (left column).}
    \label{fig:pdb_matched}
\end{figure}

Figure~\ref{fig:pdb_matched} shows the top-$5$ retrieved assemblies for each generated complex. \ours{} generated structures are radially distinct from existing proteins and present plausible alternative large-scale geometries.

\section{Ethics Statement}
\label{app:ethics}
We acknowledge that because our work enables large-scale symmetric protein design, if modified, it could potentially be used for the design of harmful viral capsid proteins by malicious actors. In order to mitigate this, in this existing work, we do not train or fine-tune on any symmetric complexes, and we expect that this will result in model outputs being unlikely to be biologically expressible as of now. In our future work, we will mitigate risks that could arise from misuse of our method for viral capsid design by carefully selecting our fine-tuning corpus to avoid particularly harmful viral strains. On the other hand, despite the potential risks that could arise from viral capsid generation, we expect that our method will also enable a number of potential benefits. For example, the ability to generate viral capsid proteins could prove useful in the study of viral mechanisms, in vaccine development, and in developing new methods for targeted drug delivery.

\newpage

\end{document}